\documentclass[%
 reprint,
superscriptaddress,
 amsmath,amssymb,
 aps,
]{revtex4-2}

\usepackage{graphicx}
\usepackage{dcolumn}
\usepackage{bm}

\begin{document}


\title{Reverse-time analysis uncovers universality classes in directional biological dynamics}

\author{Nicolas Lenner}
 \altaffiliation[Currently at ]{Simons Center for Systems Biology, School of Natural Sciences, Institute for Advanced Study, Princeton, New Jersey, USA.}
  \email{Lenner@ias.edu}
 \affiliation{Max Planck Institute for Dynamics and Self-Organization, Göttingen, Germany}

\author{Stephan Eule}%
\affiliation{Max Planck Institute for Dynamics and Self-Organization, Göttingen, Germany}
\affiliation{German Primate Center—Leibniz Institute for Primate Research, Goettingen, Germany}

\author{Jörg Großhans}
\affiliation{Department of Biology, Philipps University Marburg, Marburg, Germany}
\affiliation{Göttingen Campus Institute for Dynamics of Biological Networks, University of Göttingen, Göttingen, Germany}

\author{Fred Wolf}
 \email{Fred.Wolf@ds.mpg.de}
\affiliation{Max Planck Institute for Dynamics and Self-Organization, Göttingen, Germany}
\affiliation{Göttingen Campus Institute for Dynamics of Biological Networks, University of Göttingen, Göttingen, Germany}
\affiliation{Max Planck Institute for Multidisciplinary Sciences, Göttingen, Germany}
\affiliation{Institute for the Dynamics of Complex Systems, University of Göttingen, Göttingen, Germany}
\affiliation{Center for Biostructural Imaging of Neurodegeneration, Göttingen, Germany}
\affiliation{Bernstein Center for Computational Neuroscience Göttingen, Göttingen, Germany}


\begin{abstract}
Mesoscopic bio-systems typically evolve towards functionally important target states, such as cell-
cycle checkpoints or decision boundaries for the release of specific behaviors. For the data-driven
inference of the underlying directional out-of-equilibrium dynamics, we here develop a theory of
target state aligned (TSA) ensembles. Target state alignment allows to analyze directional dynamics in reverse time, starting from the final conditions of the forward process. Knowledge about the initial conditions of the forward process is not required for the analysis. Our theory reveals whether and when such a system can be represented by a single, effective stochastic equation of motion. We show how, in these effective dynamics, genuine biological forces can be separated from spurious forces, which invariably arise from target state alignment. We apply our inference scheme to the example of cytokinetic ring constriction, and derive the universal low-noise and short-term behavior of TSA ensembles. Our theory establishes a transparent mathematical foundation for the analysis and inference of directed biological dynamics by target state alignment.
\end{abstract}

\maketitle


\section{Introduction}
\begin{figure}
\centering
\includegraphics[width=.8\linewidth]{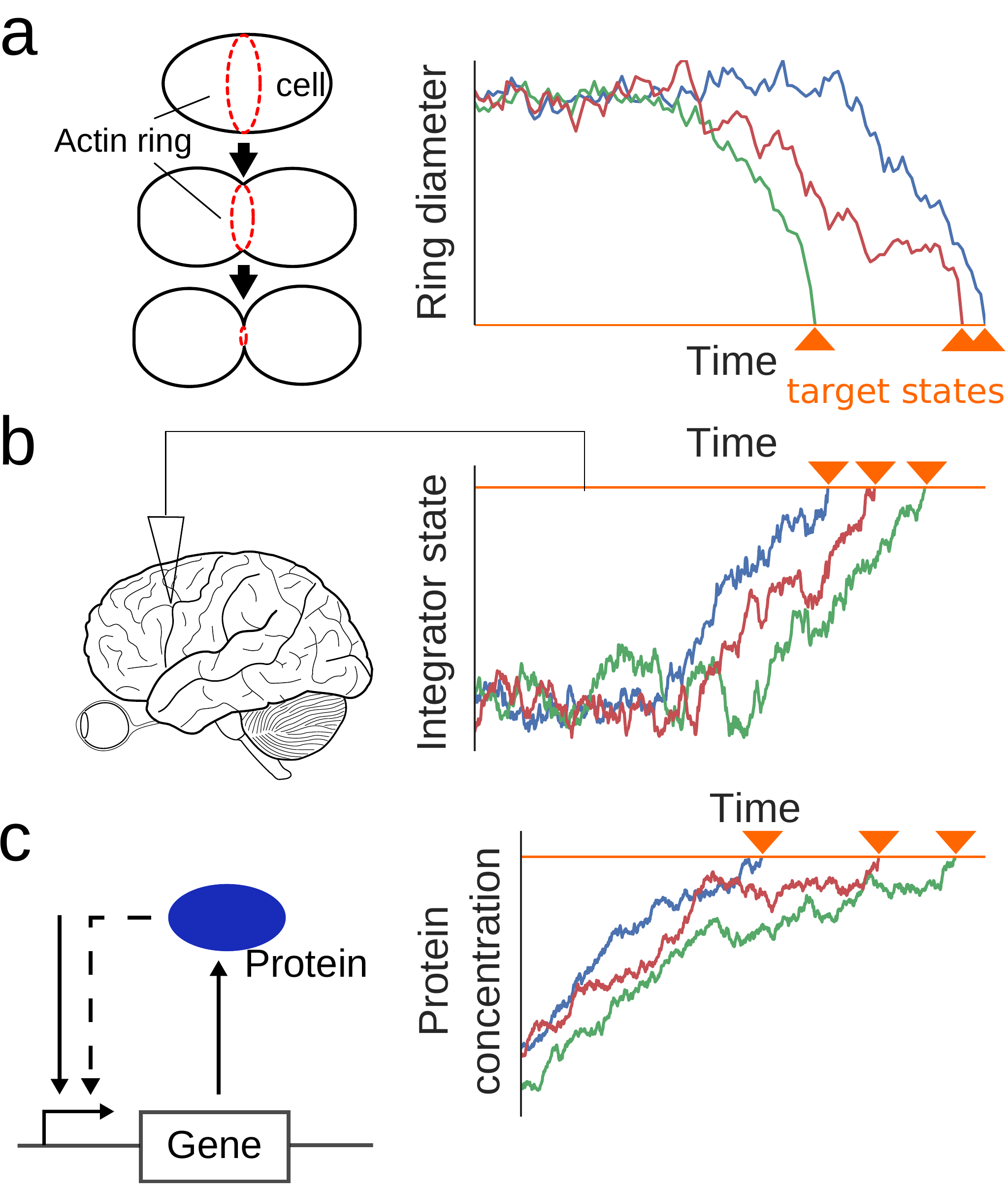}
\caption{
\textbf{Biological processes with target states.} 
\textbf{(a):} Cytokinetic ring constriction towards 
cell separation. 
\textbf{(b):} Neuronal evidence accumulation during decision making.
\textbf{(c):} Activation of a gene regulatory network by concentration threshold crossing
For each process, representative trajectories are shown. The target states (orange line) of cell separation \textbf{(a)}, decision- \textbf{(b)} or concentration-threshold  \textbf{(c)},  and the completion time (orange triangle) of each individual trajectory are marked.
}
\label{chTSAexp_fig:fig1}
\end{figure}

The dynamics of biological systems are often directed towards functionally important target states. Prominent examples of target states (TS) are cell cycle check-points \cite{sha2003hysteresis,pomerening2003building,coudreuse2010driving,rata2018two,schwarz2018precise,domingo2011switches,nachman2007dissecting}, branch-points in cell fate determination \cite{pardee1974restriction,ahrends2014controlling,maamar2007noise,xiong2003positive,losick2008stochasticity,balazsi2011cellular}, or discrete behavioral decisions from the continuous accumulation of sensory evidence \cite{hanes1996neural,hanks2015distinct,ratcliff2016diffusion,ratcliff2008diffusion,brunton2013rats,hanks2015distinct,churchland2011variance,roitman2002response} (Fig.~\ref{chTSAexp_fig:fig1}). Such TS-directed dynamics are crucial for the functioning of biological systems on various scales and considerable effort has been dedicated to clarify their phenomenology and mechanistic underpinnings \cite{coudreuse2010driving,xiong2003positive,brunton2013rats}. For many instances, this research revealed that TSs are not approached deterministically \cite{nachman2007dissecting,balazsi2011cellular,hanes1996neural}. Rather, due to internal and external noise sources, many biological systems generate a diverse set of trajectories all homing in on the same TS. This richness of behavior reflects their capability to buffer noise, achieve TS-arrival robustly and is best captured by a statistical ensemble approach. Biological systems with TS-directed dynamics are in general out-of-equilibrium and non-stationary and thus require large sample sizes for accurate characterization \cite{hoze2017statistical,holcman2015analysis,michalet2012optimal,vestergaard2014optimal,fogelmark2018fitting,burov2011single,meroz2015toolbox,hofling2013anomalous}. Fortunately, technological progress in recording from cells, systems and entire organisms \cite{chalfie1994green,lippincott2003development} and in multi channel high-speed data acquisition \cite{tomer2012quantitative,krzic2012multiview,power2017guide} has opened exciting prospects to efficiently generate such extensive datasets.

Ultimately, the large-scale datasets that can now be acquired in many biological systems should enable understanding the mechanisms of TS-convergence through the data-driven inference of stochastic dynamical models. However, beyond the scale of individual biomolecules \cite{qian1991single,lee2017unraveling,manzo2015review,shen2017single,liphardt2002equilibrium,collin2005verification,hummer2001free,de2012pulling}, the power and limitations of model inference from out-of equilibrium non-stationary ensembles are not well understood. Key to the analysis of such ensembles is the seemingly innocent step of trajectory temporal alignment. In principle, one might choose to either align individual trajectories at process onset or at the time of TS-arrival. In practice, however, these two reference times often cannot be defined with equal certainty. Firstly, stochastic behavior makes it hard in general to exactly determine the onset time of a specific behavior for each individual sample. Secondly, complex biological systems typically assume the TS-directed dynamics only during a particular functional stage and the underlying mechanisms gradually fade-in as this stage is entered. Thirdly and in particular during fade-in and onset, systems may exhibit substantial sample-to-sample heterogeneity. By contrast, near TS-arrival one expects the relevant dynamical behavior to be expressed most purely and most uniformly. In addition TSs, such as daughter cell separation in cytokinesis, often are unique and can be localized in time quite precisely. For all of these reasons, analyzing TS-directed dynamics not in forward time, measure from process onset, but in reverse time, measured from TS-arrival, may offer many advantages in terms of temporal precision, ensemble purity and thus quality of inference. So far, however, TS-alignment (TSA) has not been employed frequently and - perhaps for this reason - its theoretical foundations have not been systematically analyzed. Below we first expose a couple of surprising intricacies mathematically associated with the very procedure of TSA. We then develop a theory for analyzing directed biological dynamics in reverse time and demonstrate the use of this approach in the inference of dynamical models for cytokinetic ring constriction.

\section{Terminal pseudo forces and the mixed nature of TSA ensembles}
\begin{figure*}
\centering
\includegraphics[width=.99\linewidth]
{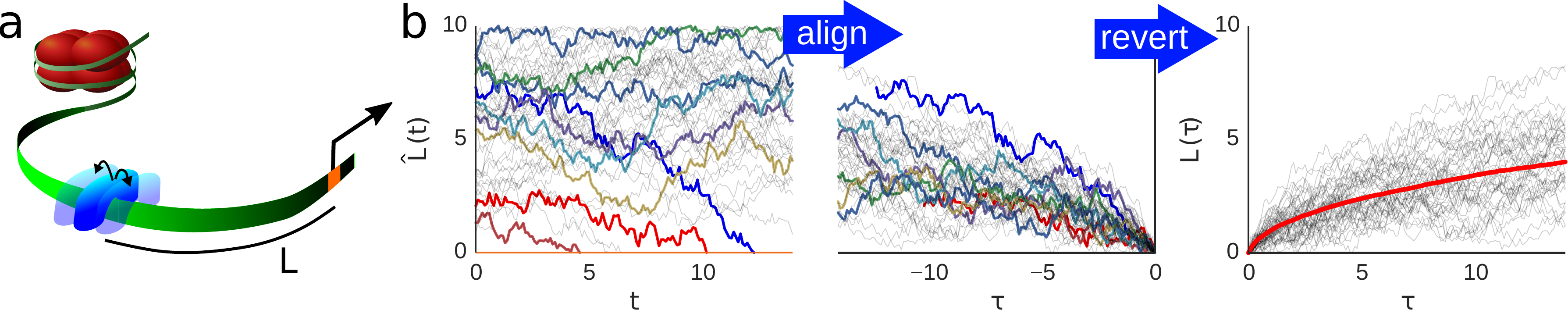}
\caption{
\textbf{Target state alignment creates pseudo forces for 1d random target 
search.}
\textbf{(a):} Random walk like sliding of transcription factor confined between inaccessible DNA and promoter binding site (orange). 
\textbf{(b):} 
(Left): Sample path realizations of a random walk with one reflecting and one absorbing 
boundary (orange). 
(Middle): Target state (promoter binding site) aligned sample paths. 
(Right): Aligned and time-reversed ensemble of sample paths. The mean, growing with $\propto \tau^\frac{1}{2}$  (red line), indicates the presence of a non-linear alignment force.
}
\label{chTSAexp_fig:pureDiffusion}
\end{figure*}

The intricacies arising from target state alignment become apparent already for the simplest case, i.e.~a random search process for a target site \cite{berg1981diffusion,gorman2012single,schotz2011target,hettich2018transcription} (Fig.~\ref{chTSAexp_fig:pureDiffusion}a). 
The TSA ensemble resulting from undirected random motion terminated at a target site is depicted in Fig.~\ref{chTSAexp_fig:pureDiffusion}.
By construction all sample paths concentrate near the ensemble mean close to the target state.
If the TSA ensemble had an effective stochastic differential equation (SDE) with conventional low-noise behavior,
a low-noise approximation could be used close to the target state such that the ensemble mean would be the zero noise solution and thus directly reveal the drift term. This however must obviously be wrong as the mean, which grows $\propto\tau^{\frac12}$, would indicate a diverging force $\propto \frac{1}{L}$ although the forward dynamics are force free.
%

We analysed how such spurious pseudo forces arise for processes satisfying a Langevin equation of the form
\begin{equation}
\label{chTSAexp_fwd_sde}
 d\widehat{L}(t) = f(\widehat{L}) \, dt + \sqrt{D} \, dW_t \, .
\end{equation}
Here $f(\widehat{L})$ denotes a deterministic drift term and $\sqrt{D}$ the strength of the fluctuations $dW_{t}$, with
$dW_t$ the Wiener process increment with zero mean $\langle dW_t \rangle = 0$ and delta covariance  $\langle \, dW_t  \, dW_t'\rangle = \delta(t-t')$.
Each observation $i$ consists of a trajectory 
$\widehat{L}_i(t)$ with wall clock time $t$ and lifetime $T_i$.

The reverse-time TSA ensemble is described by a time-dependent distribution $R(L, \tau)$ 
with $L_i(\tau)= \widehat{L}_i(T_i - \tau)$ as a function of the reverse time $\tau=T_i-t$.
%
%
%
%
%
To construct $R(L,\tau)$ two aspects must be taken into account: 
i.) the underlying dynamics evolve in reverse time; ii.) the lifetime of the trajectories is itself a random variable. Over time, less and less trajectories remain in the ensemble until eventually all trajectories have reached their lifetime. $R(L,\tau)$ is thus not normalized and decays with $\tau$.
In absence of noise ($D=0$), time-reversal of Eq.~\eqref{chTSAexp_fwd_sde} is trivial. 
All trajectories starting at $\widehat{L}_{0}$ end at
$\widehat{L}_{f}$ at time $t=T$. 
The time-reversed dynamical equation is $dL(\tau)=-f(L) d\tau$.

In a stochastic system however, 
changing the sign in front of the time derivative does not yield the correct time-reversed equation. For instance for a stationary Ornstein-Uhlenbeck processes $f(\widehat{L}) = - \widehat{L}$, 
inverting the sign of the drift term $-L \to L$ results in exponentially diverging trajectories -- a completely different behavior as in forward time.

Assuming that all sample paths are of the same lifetime $T$, the correct time-reversal of the SDE Eq.~\eqref{chTSAexp_fwd_sde} is \cite{anderson1982reverse}
\begin{align}
\label{chTSAexp_tr_lv}
  dL(\tau) &= 
      \left(
      -
       f(L) 
       +
       f^G(L)
       \right)
         d\tau
      +
      \sqrt{D} \ d W_\tau \, ,
\intertext{with}
\label{chTSAexp_exact_guidingForce}
f^G(L)
&=
       D
       \frac{\partial }{\partial L} 
       \log\left(P^{\mathrm{fw}}(L,T-\tau) \right) 
\end{align}
a guiding force that depends on the solution $P^{\mathrm{fw}}(\widehat{L},t)$ of the forward Fokker-Planck equation (FPE)
\begin{align}
    \label{fwd_FPE}
 \frac{\partial}{\partial t} P(\widehat{L},t)
 =
 -
 \frac{\partial}{\partial \widehat{L} }
 f(\widehat{L})  P(\widehat{L},t)
  +
 \frac{D}{2}
 \frac{\partial^2}{\partial \widehat{L}^2}
 P(\widehat{L},t)
\end{align}
with absorbing boundary conditions imposed at the target state $L_\mathrm{ts}$.
The guiding force $f^G(L)$ ensures that the forward- and reverse-time dynamics are identical.
Hence stating the time-reversed SDE in principle requires knowledge of the solution of the forward process and its initial distribution.
Although time reversed dynamics might appear as a simple initial value problem it is not.

\begin{figure}
\centering
\includegraphics[width=.99\linewidth]
{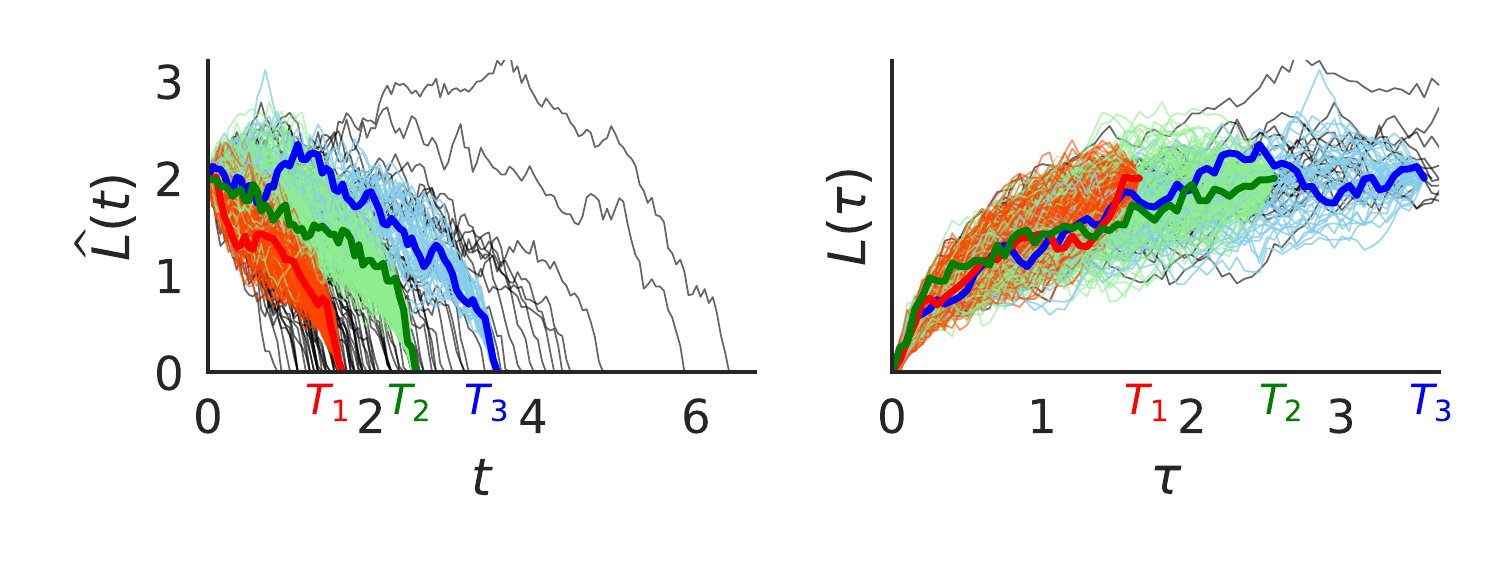}
\caption{
\textbf{
Construction of the full aligned time reversed ensemble from sub-ensembles of 
different lifetimes.} 
(Left): The full forward ensemble is split into sub-ensembles with different completion times $T_i$. 
We show three exemplary cases in red ($T_1$), green ($T_2$) and blue ($T_3$). To guide the eye, one sample path per sub-ensemble is highlighted. 
(Right): After target state alignment and time reversal all sub-ensembles together form the new ensemble 
$R(L,\tau)$.
}
\label{chTSAexp_fig:demoTimereversal}
\end{figure}

In a natural setting the lifetime $T_i$ is not fixed but is itself a random variable.
To assemble the ensemble in reverse time, we partition 
the trajectories into sub-ensembles $R(L, \tau| T;L_f)$ of fixed lifetime $T_{i}$. In the example depicted in Fig.\ref{chTSAexp_fig:demoTimereversal} the initial condition of each of the time-reversed 
sub-ensembles is a delta-function $R(L,\tau=0 |T_{i};L_f)= 
\delta(L-L_\mathrm{ts})$. 
Because the guiding force Eq.~\eqref{chTSAexp_exact_guidingForce} depends on the full forward 
distribution $P^{\mathrm{fw}}(L, T_{i}-\tau)$ up to time $T_i$ according to Eq.~\eqref{chTSAexp_tr_lv} and Eq.~\eqref{chTSAexp_exact_guidingForce}, each of these sub-ensembles experiences a different force.

To assemble a complete TSA ensemble, sub-ensembles satisfying Eq.~\eqref{chTSAexp_tr_lv} must be superimposed (Fig.~\ref{chTSAexp_fig:demoTimereversal}). Three aspects are noteworthy. First, each sub-ensemble of lifetime $T$ only contributes to the complete ensemble up to this time. Second, the relative weight of trajectories of this lifetime is given by the hitting time distribution of the forward process $\rho(T|L_f)$, which in turn depends on the forward initial condition.
Third, the relative weight of each sub-ensemble depends on the initial conditions of the forward process $P^\mathrm{in}(L_f)$.
Hence, $R(L, \tau)$ for the aligned time-reversed ensemble is
\begin{align}
\label{chTSAexp_prob_full_tr_ens}
 R(L,\tau) = 
\int_{L_\mathrm{ts}}^\infty
dL_f
 P^\mathrm{in}(L_f)
 \int_{\tau}^{\infty}
 dT 
 \
 R(L,\tau|T;L_f)
 \rho(T|L_f)
 \
 \ .
\end{align}
The lower integration limit of the inner integral accounts for the fact that only sub-ensembles that have at least
a length of $\tau$ contribute to the full ensemble. The outer integral accounts for the distribution of initial values of the forward process (see SI II).

Let us briefly consider the implications of the above result.
First, for fixed lifetime sub-ensembles, we can 
state the general form of the guiding force close to the target state. With $P^{\mathrm{fw}}(L,T-\tau)$ vanishing close to the absorbing state $L_\mathrm{ts}$, the generic form of the density near $L_\mathrm{ts}$ is to leading order $P^{\mathrm{fw}}(L)\propto (L-L_\mathrm{ts})^\delta$ with $\delta > 0$. The guiding force close to the target state then evaluates to $f^G(L) \propto \frac{1}{L}$, explaining the behavior of the mean of the TSA random walk (Fig.~\ref{chTSAexp_fig:pureDiffusion}).
Second, the construction of the TSA ensemble $R(L,\tau)$ from sub-ensembles of varying lifetimes $T_i$ (and thus varying guiding forces) suggests that in general there is no unique SDE which describes the full reverse-time dynamics. Different force laws seem to be active at the same state $(L,\tau)$. 
At first sight both of these observations should raise substantial caution. With reverse time dynamics generically "contaminated" by a diverging guiding force and TSA ensembles in principle a mixture of different effective dynamics, can TSA really provide a strategy to infer directed biophysical dynamics? To asses the power and limitations of TSA what is needed is a general understanding of  the form of guiding forces of the complete ensemble and under which conditions the TSA ensemble follows a unique SDE as intuition would suggest.

\begin{figure}[h]
\centerline{\includegraphics[width=0.99\linewidth]
{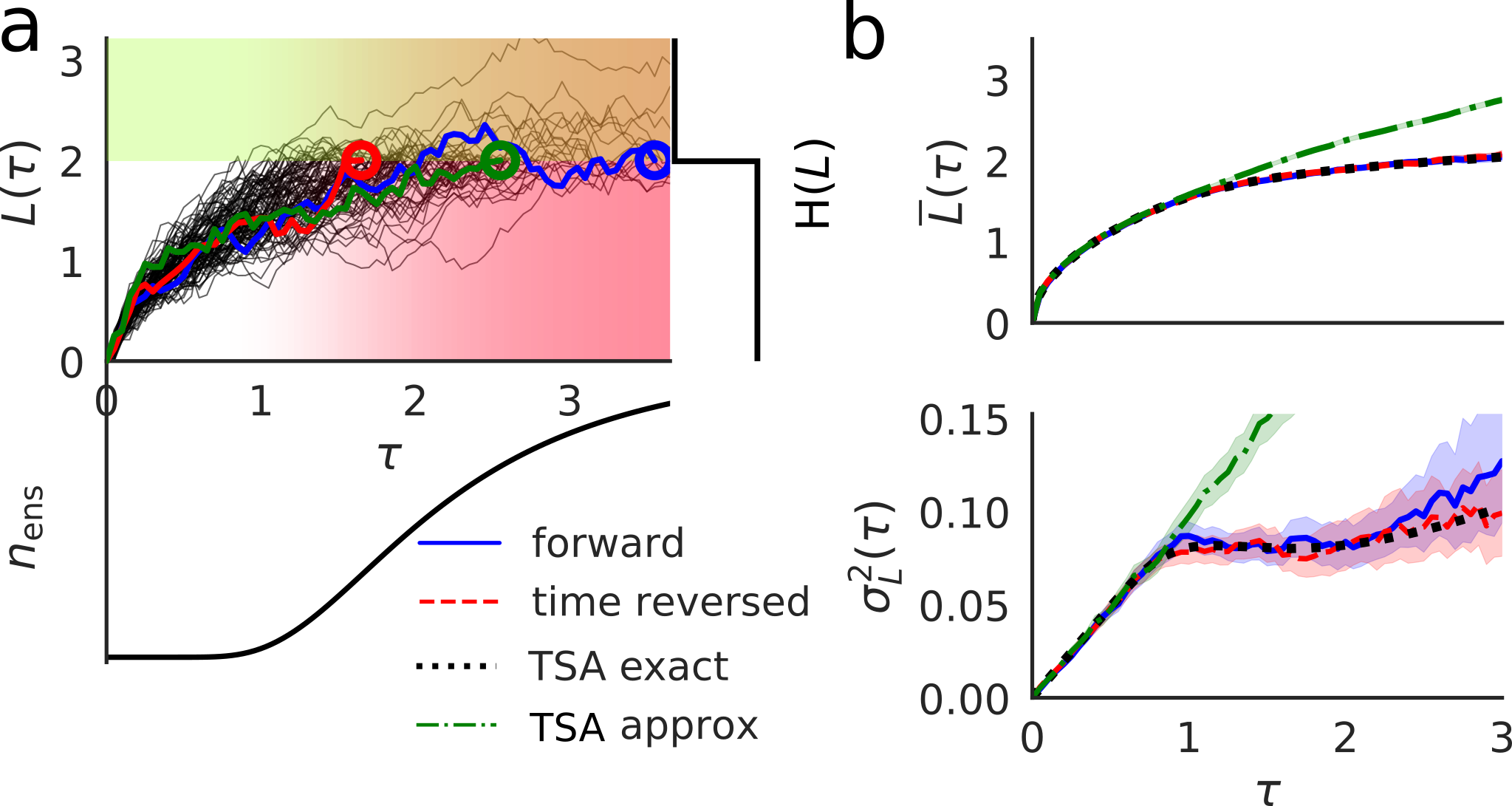}}
\caption{\textbf{ TSA FPE and SDE exactly describe the target state aligned ensemble.}
\textbf{(a):} Schematic depiction of the dependency of the TSA dynamics Eq.~\eqref{chTSAexp_eq:fp_tsa} and Eq.~\eqref{chTSAexp_tr_lv_freeE} on the forward initial condition $H(L)$ and the killing measure $k(L,\tau)$. In the green region, i.e. above the bulk of the initial distribution measured by the sigmoidal $H(L)$, we find $f^\mathcal{F}(L)\approx 0$. Below $f^\mathcal{F}(L)$ contributes in full strength. In $\tau$ direction, an increase in the red gradient indicates, that more and more trajectories are killed.
The residual white $L-\tau$ plane close to the target state defines the region where
$k(L,\tau)\approx0$ and $H(L)=1$ holds, and the approximation of well separated initial and final states holds. Circles mark the killing of example trajectories.
\textbf{(b):} Comparison of the forward (blue), sub-ensemble based (red), exact (black)  and approximate (green) reverse-time dynamics for $f(L)=-\gamma/L$.
Shown are the mean (Top) and variance (Bottom) of all four cases. 
95\% bootstrap confidence intervals are shown for the cases involving sampling. To exclude numerical 
inaccuracies due to rarely visited tails of the distribution of completion 
times $\rho(T|L_f)$, we directly sampled $T_i$ from the numerically 
obtained hitting time distribution of the forward process. Results 
were obtained using each 1000 sample path realizations with 
parameter settings $\gamma = 1$, $D =0.2$, $\widehat{L}_\mathrm{init}=2$.
}\label{chTSAexp_driftCompExact2}
\end{figure}

\section{Reverse-time FPE and SDE for the TSA ensemble}
Starting from Eq.~\eqref{chTSAexp_prob_full_tr_ens}, we show in the SI II that the dynamics underlying the evolution of $R(L,\tau)$ can be cast into the form of a generalized FPE 
\begin{align}
\label{chTSAexp_eq:fp_tsa}
 \frac{\partial}{\partial \tau} R(L,\tau)
 =
 &-
 \frac{\partial}{\partial L}
 \left(
 \left(
 f(L) + f^\mathcal{F}(L)
 \right) R(L,\tau)
 \right)
  \notag \\
 &+
 \frac{D}{2}
 \frac{\partial^2}{\partial L^2}
 R(L,\tau)
-
 P^\mathrm{in}(L) \rho(\tau|L)
 \end{align}
with a time-dependent sink term $-P^\mathrm{in}(L) \rho(\tau|L)$. Here $P^\mathrm{in}(L)$ denotes the distribution of initial states for the forward dynamics.
This term ensures that the distribution of sample path lifetimes in the reverse-time ensemble is the same as in the forward dynamics. Unlike for the sub-ensemble dynamics Eq.~\eqref{chTSAexp_tr_lv}, $f(L)$ is not sign inverted compared to the forward dynamics.
The ``free energy force''
\begin{align}
\label{chTSAexp_eq:def_freeE_force}
 f^\mathcal{F}(L)=  D
  \frac{\partial}{\partial L}
  \log \left(
 \int_{L_\mathrm{ts}}^{L} dL' \;
   e^{-
   \frac{2 \Phi(L')}{D}
 }
 H(L')
 \right)
\end{align}
captures the combined effect of the total entropy production of all sample paths and the time-reversion of the dynamics up to a position $L$. The potential $\Phi(L) = \int^{L} f(L') dL'$ is defined with respect to the sign inverted drift term. 
$
 H(L)=  
 1 - \int_{L_\mathrm{ts}}^{L} P^{\mathrm{in}}(L') dL'
$
is a sigmoidal function which continuously changes from one to zero. 
Above the bulk of $P^{\mathrm{in}}(L)$ the free energy force $f^\mathcal{F}(L)$ therefore vanishes, and forward and TSA dynamics are indistinguishable. Below however the free energy force not only reverses the forward dynamics but also adds additional terms, for instance the guiding force term $\propto \frac{1}{L}$.

From the theory of reaction-diffusion systems \cite{wilemski1973general} we know that a Fokker-Planck equation with sink proportional to the density can be cast into a SDE with killing measure $k(L,\tau)$ proportional to the rate of degradation 
 \cite{holcman2005survival,schuss2015brownian,erban2007reactive}. 
Analogously for the reverse-time FPE for TSA ensembles the corresponding SDE reads
\begin{align}
 \label{chTSAexp_tr_lv_freeE}
  dL(\tau) &= 
      \left(
       f(L) 
       +
       f^\mathcal{F}(L)
       \right)
         d\tau
      +
      \sqrt{D} \ d W_\tau \, ,
\end{align}
equipped with a killing measure
$
k(L,\tau) d\tau
=
\frac{\rho(\tau|L) P^\mathrm{in}(L)}{R(L,\tau)} d\tau \ .
$
For finite TSA ensembles originally comprised of $n_\mathrm{ens}$ sample paths, this implies 
that the killing measure terminates $n_\mathrm{ens} \rho(\tau|L) d\tau$ of the remaining sample paths at each timestep $d\tau$ as determined by the weight $\frac{P^\mathrm{in}(L)}{R(L,\tau)}$.
The explicit lifetime dependency of the sub-ensemble based construction Eq.~\eqref{chTSAexp_prob_full_tr_ens} can thus be moved to a time- and ensemble-dependent boundary condition in the form of a killing measure.

\section{The TSA ensemble close to the target state}
Expressing the dynamics of $R(L,\tau)$ in terms of a single SDE allows to 
separate the contribution of the forward initial conditions from the pure TSA dynamics close to the target state. This can be seen by inspecting $f^\mathcal{F}(L)$ defined in Eq.~\eqref{chTSAexp_eq:def_freeE_force}. For $L$ sufficiently below the bulk of the forward initial distribution $P^\mathrm{in}(L)$ the term $H(L)$,
which describes the influence of the forward initial conditions on the reverse time TSA dynamics, evaluates to approximately one. If most of the sample paths have not yet reached the sink at $P^\mathrm{in}(L)$, this in turn implies that almost no sample path has terminated in reverse time. The killing measure $k(L,\tau)$ evaluates to zero and $H(L)\approx 0$.
With $k(L,\tau) = 0$ and $H(L)=1$ the remaining unique SDE describes TSA dynamics temporally and spatially close to target states and irrespective of initial conditions. Fig.~\ref{chTSAexp_driftCompExact2} depicts one example in which the ensemble mean and variance in this approximation agrees excellently with the exact solution. In the SI V we discuss further examples that illustrate the accuracy of this approximation.

\subsection{Separating genuine and alignment induced forces}
Next we examined, using this approximation, the general mapping between forward forces and aligned reverse-time dynamics close to the target state.
Using $H(L)=1$ we calculated a mapping for power-law forces
\begin{equation}
\label{chTSAexp_alpha_force_law}
 f(\widehat{L}) = - \gamma \widehat{L}^\alpha
 \ ,
\end{equation}
with $\gamma >0$, $\alpha \in \mathbb{R}$, target state at $\widehat{L}_\mathrm{tp}=0$ and $\widehat{L}_\mathrm{in} \to \infty$. 
The free energy force is given by
\begin{align}
\label{chTSAexp_tr_lv_ness_alpha_force_general_case}
 f^\mathcal{F}(L)&=\frac{(D\alpha+D) \left(-\frac{2 \gamma }{D
   \alpha+D}\right)^{\frac{1}{\alpha+1}} e^{\frac{2 \gamma  L^{\alpha+1}}{D
   \alpha+D}}}{\Theta(\alpha+1) \Gamma \left(\frac{1}{\alpha+1}\right)-\Gamma
   \left(\frac{1}{\alpha+1},-\frac{2 L^{\alpha+1} \gamma }{\alpha
   D+D}\right)}
   \ .
\end{align}
Eq.~\eqref{chTSAexp_tr_lv_ness_alpha_force_general_case} is valid for $\alpha \ne -1$, with the connecting form $f^\mathcal{F}(L) = \frac{2\gamma + D}{L}$ at $\alpha=-1$. Here $\Gamma(L)$ is the gamma-function, $\Gamma \left( n,L \right)$ the upper incomplete gamma-function and $\Theta(n)$ the Heaviside-function. The derivation is presented in the SI V.

\begin{figure*}
\centering
\includegraphics[width=.99\linewidth]
{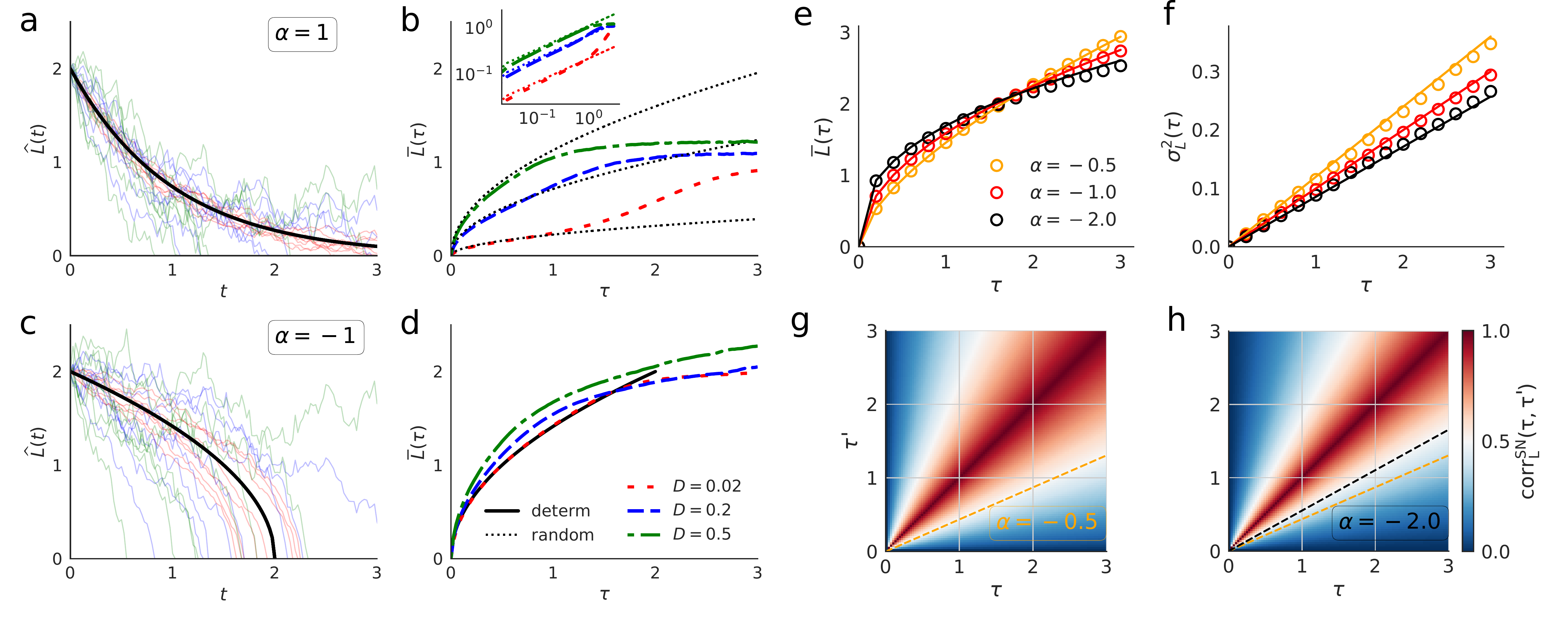}
\caption{
\textbf{TSA ensembles are noise and parameter sensitive, and qualitatively differ for diffusion or drift dominated target state approaches.}
\textbf{(a):}
For $\alpha >0$ only noise induced irreversible transitions occur in finite time. 
\textbf{(b):} Close to their target state reverse time statistics of noise induced transitions ($\alpha > 0$) are indistinguishable from random fluctuations and thus $\alpha$-independent. 
The mean grows $\sim \tau^\frac{1}{2}$ (random-walk-like, Eq.~\eqref{chTSAexp_diffusion_mean}, dotted) independent of the noise strength $D$ as shown in both the log-log inset and main plot. 
\textbf{(c):}
The hitting time of force driven irreversible transitions ($\alpha < 0$) scatters around the deterministic solution (black line).
\textbf{(d):}
The mean of the aligned reverse time ensemble (dashed line) for $\alpha < 0$ deviates positively from the deterministic solution with increasing noise strength $D$.
Together \textbf{(c)} and \textbf{(d)} motivate
a small noise moments expansion.
\textbf{(e),(f):} The small noise expansions of mean Eq.~\eqref{chTSAexp_sup_sn_mean} and variance Eq.~\eqref{chTSAexp_sup_sn_var} (lines) 
agree with the numeric evaluation of the reverse-time TSA dynamics with $f^\mathcal{F}(L)$ Eq.~\eqref{chTSAexp_tr_lv_ness_alpha_force_general_case} (circles). 
\textbf{(g),(h):} The small noise two-time correlation Eq.~\eqref{chTSAexp_sup_sn_corr} solely depends on the power-law exponent $\alpha$. 
Linear equi-correlation lines (dashed) allow for an easy distinction.
For all plots we chose $\gamma=1$ and $D=0.2$ if not stated differently.
}
\label{chTSAexp_fig:fig6}
\end{figure*}

\subsection{Noise induced vs.~force induced target state arrival}
For power-law forces, the behavior near arrival falls into two generic classes.
For $\alpha \ge 0$ the forward force vanishes at the target state and the system typically
terminates \textit{noise induced} (Fig.~\ref{chTSAexp_fig:fig6}a). 
Aligned, in reverse time and close to the target state $\widehat{L}_\mathrm{ts}=0$, such an ensemble is indistinguishable from the dynamics of a random walk $(\alpha=0,\gamma=0)$ with $f^\mathcal{F}(L)=\frac{D}{L}$ (SI IV). 
We tested this behavior in Fig.~\ref{chTSAexp_fig:fig6}b for three different noise strengths $D$. The analytic expression for the mean
\begin{align}
\label{chTSAexp_diffusion_mean}
   \overline{L}(\tau)  = \sqrt{\frac{8 D}{\pi } \tau} \ ,
\end{align}
of the aligned reverse-time random walk almost perfectly matches the mean of the aligned reverse-time data for the case $\alpha=1$ and close to $\widehat{L}_\mathrm{ts}=0$. 
It starts deviating for $L f(L) >D$ where the approximation is expected to break down (SI V). Note that Eq.~\eqref{chTSAexp_diffusion_mean} also explains the result found for the random target search (Fig.~\ref{chTSAexp_fig:pureDiffusion}).

For $\alpha < 0$ the termination of the forward dynamics is \textit{force induced} as 
$f(\widehat{L})$ diverges at the absorbing boundary. 
The 
corresponding reverse time dynamics exhibit the genuine (sign inverted) force 
law dependence plus corrections proportional to the noise strength $D$. 
Fig.~\ref{chTSAexp_fig:fig6}c,d shows exemplary sample paths for $\alpha=-1$. It demonstrates the gradual deviation of the mean from the deterministic solution with increasing noise strength $D$.

Further insight can be obtained by examining the small $L$ weak noise regime of the free energy force Eq.~\eqref{chTSAexp_tr_lv_ness_alpha_force_general_case}. 
In this regime the reverse time SDE for $\alpha<0$ simplifies to
\begin{equation}
\label{chTSAexp_tr_lv_sn}
 dL(\tau) = 
 \left(
 \gamma L^\alpha - \frac{\alpha D}{L}
 \right) d\tau
 + 
 \mathcal{O}\left( \frac{ D^2 }{L^{2 + \alpha}} \right)
 +
\sqrt{D} \;dW_\tau 
\ ,
\end{equation}
see SI VI.
This shows that for all $\alpha<0$ the leading order correction to the time reversed genuine force is random-walk like ($\sim D/L$), modulated in its strength by the power-law exponent $\alpha$. 
Despite its simplicity Eq.~\eqref{chTSAexp_tr_lv_sn} can not be solved analytically for general $\alpha <0$. 
Adopting the low noise expansion for moments of ordinary SDEs \cite{gardiner1985handbook},
we expand Eq.~\eqref{chTSAexp_tr_lv_sn} around its deterministic solution  $(D \to 0)$ for small $D$
and obtain 
\begin{align}
\label{chTSAexp_sup_sn_mean}
 &\overline{L}(\tau)
 =
 \left( (1-\alpha) \gamma \tau \right)^{\frac{1}{1-\alpha}}
+
 D 
  \frac{(7 \alpha -3) ((1 - \alpha ) \gamma \tau)^{\frac{\alpha
   }{\alpha -1}}}{4 (3 \alpha -1) \gamma }
\\
   \label{chTSAexp_sup_sn_var}
&\sigma_L^2(\tau)
=
   D \frac{1 - \alpha}{1 - 3 \alpha} \tau
\\
\label{chTSAexp_sup_sn_corr}
 &\mathrm{corr}_L(\tau,\tau')
  =
  \left(\frac{\min[\tau,\tau']}{\max[\tau,\tau']}\right)^{\frac{3}{2} + 
\frac{1}{\alpha -1}}
\end{align}
for mean, variance and two-time correlation function up to order $D$ (SI VI). 
In Fig.~\ref{chTSAexp_fig:fig6}e,f we compare this approximation to simulations and find excellent agreement. 

Motivated by their high accuracy we propose Eqs.~\eqref{chTSAexp_sup_sn_mean}-\eqref{chTSAexp_sup_sn_corr} as basis for fast inference schemes: Given sufficiently good statistics, $\alpha$ can be read of directly from a 2d correlation plot using Eq.~\eqref{chTSAexp_sup_sn_corr} (Fig.~\ref{chTSAexp_fig:fig6}g,h). Once $\alpha$ is identified, $D$ can be extracted from the variance Eq.~\eqref{chTSAexp_sup_sn_var}. The force strength $\gamma$ follows directly from a $\alpha$ and $D$ constrained fit of the mean Eq.~\eqref{chTSAexp_sup_sn_mean}. Notably this is quite distinct from a normal low noise approximation, in which the mean would be the zero noise solution and thus only depend on $\gamma$.


\section{Inferring actomyosin turnover from the terminal dynamics of cytokinesis}
To demonstrate the inference of biological dynamics from TSA ensembles we set up a biophysical model of cytokinetic ring constriction which has all ingredients that one expects to complicate the correct identification of the directional dynamics of cytokinesis.
The force driving cytokinesis is mediated by a contractile ring of myosin motors and actin filaments. The contractile behavior of this ring can exhibit very different types of concentration-tension relationships depending on the regime of molecular turnover. As a single myosin minifilament needs to bind two actin fibers to exert a force the tension can be assume $\propto c_\mathrm{myo} c_\mathrm{act}^2$.
Three qualitatively different types of dynamics are conceivable. First, the effective constriction force is proportional to the myosin concentration along the cable of length $L$. The shorter the cable gets, the more myosin per ring perimeter accumulates. The constriction force increases
$\propto 1/L$ ($\alpha=-1$), assuming a low myosin turnover rate.

Second, with strong myosin turnover
the effective constriction force reflexts the actin fiber concentration and thus is $\propto 1/L^2$ ($\alpha=-2$).
The third mechanism is favored for high turnover rates of both force generating
molecules. Independent of $L$, force molecules are under this scenario
present at a constant concentration.
The constriction force is constant ($\alpha=0$).

To test our inference scheme and to generate artificial cytokinetic sample paths we use a model by Zumdieck et.~al.~ \cite{zumdieck2007stress}, which covers all three scenarios as limiting cases. The details of the model and the analytic limits are covered in the SI VII.
In our simulations the full process of constriction starts from a equilibrium perimeter which initially is stable. The dynamics then transition to the final regime which leads to ring constriction and separation. Internal fluctuations of the ring perimeter are modeled as white noise. 
After target state alignment we obtain a TSA ensemble with contributions from both biological and guiding forces. Its target and equilibrium state are well separated
and the dynamics should assume a pure force law short before cell separation.

\begin{figure*}
\centering
\includegraphics[width=.85\linewidth]
{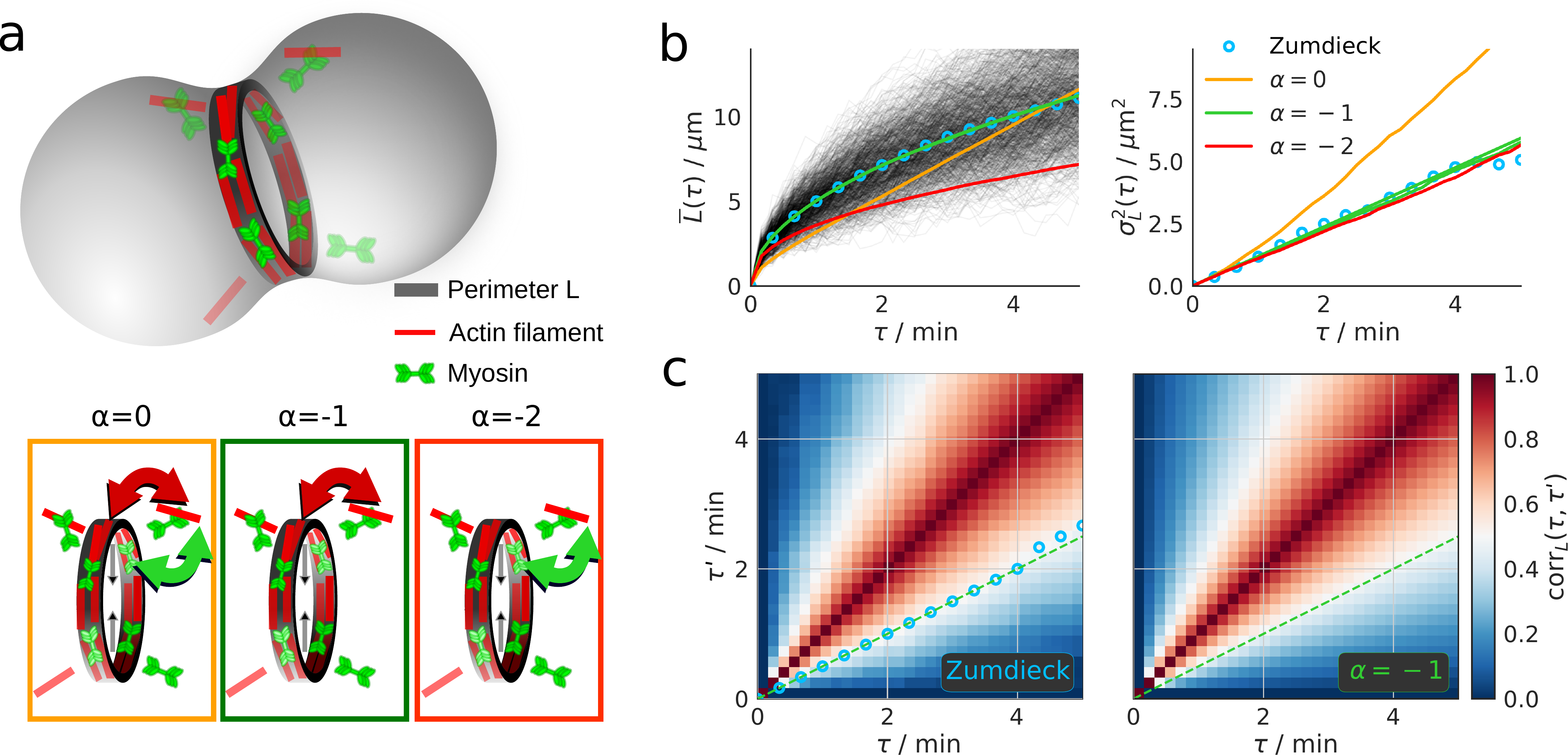}
\caption{
\textbf{Accurate inference of actomyosin turnover from dynamics in the noise
extended Zumdieck model \cite{zumdieck2007stress} of cytokinetic ring
constriction}. 
\textbf{(a):}
Schematic representation 
of three different cytokinesis scenarios assuming constriction-forces proportional to molecule concentrations along the ring:
Constant shrinkage due to high myosin turnover ($\alpha=0$, orange), $1/L$-shrinkage with stalled myosin turnover ($\alpha=-1$, green) and $1/L^2$-shrinkage for actin-pair driven constriction with stalled turnover ($\alpha=-2$, red). 
\textbf{(b):}
Using path-ensemble maximum likelihood inference (SI VIII) 
the dynamical law leading to constriction 
can unambiguously be inferred from realizations of the Zumdieck model (grey lines, blue circles). 
We visually confirm the inferred underlying force law 
($\alpha=-1$) by numerically evaluating mean and variance for the reverse-time TSA dynamics for each of the three inferred maximum likelihood parameter sets $(\gamma_{\mathrm{ML}}^\alpha,D_{\mathrm{ML}}^\alpha)$. 
\textbf{(c):} The two-time correlation of the Zumdieck model and the 
correlations of the inferred dynamics for $\alpha=-1$ are in perfect agreement. To facilitate the comparison we show one equi-correlation line for both (Zumdieck: blue circles, $\alpha=-1$: green dashes). 
}\label{chTSAexp_fig:cytokin}
\end{figure*}

For the quantitative inference of the most likely constriction scenario, we evaluated the path-ensemble likelihood for the TSA ensemble forces Eq.~\eqref{chTSAexp_eq:def_freeE_force} 
with $H(L)=1$ and for each of the three scenarios with $\alpha= 0,-1,-2$ (for details see SI VII,VIII). The correct case
$\alpha=-1$ was clearly singled out by the highest maximum likelihood value (Fig.~\ref{chTSAexp_fig:cytokin}). The other scenarios can also be tested and unambiguously identified (see SI VII).

Without considering the peculiarities of TSA ensembles several mistakes in interpreting the sample data might have occurred. First, a scenario of constant force could erroneously be classified as one driven by a divergent force component which we now understand as the guiding force correction of the form $D/L$. Second, the strength of the force for the myosin dominated case could be overestimated as both genuine and guiding force are proportional to $1/L$. Third, for the actin dominated case an erroneous crossover from $1/L$ to $1/L^2$ might be inferred. In contrast, our TSA method obtains the genuine force law for all three cases.

\section{Discussion}
In this study, we examined the stochastic dynamics problem associated with inferring directed biological processes by target state alignment. We show whether and when such dynamics can be represented by a single SDE and how spurious forces, which inevitably arise due to target state alignment, can be separated from genuine biological forces.
The universal low-noise and short-term behavior of TSA ensembles are derived.
The biophysical applicability is demonstrated for a model of cytokinetic ring constriction, an example for directional dynamics, containing all potential confounders of correct inference.

Previously target state alignment has been used as means of data analysis in various fields. It has been employed to access entropic force differences of DNA in confined spaces \cite{turner2002confinement}, or to determine the dynamics leading to a behavioral decision e.g. represented by the initiation f saccadic eye movements \cite{hanes1996neural}. The inevitable occurrence of spurious forces in TSA ensembles however has so far not been noted or analyzed.

Examining the structure of TSA ensembles we find that they
fall into two distinct classes. For power-law exponents $\alpha >0$, forces vanish at the boundary and directed dynamics become indistinguishable from the dynamics of a target state aligned random search process near termination. For power-law exponents  
$\alpha <0$, target state directed dynamics are force induced and reverse-time statistics can be treated as corrections to the zero noise solution. Expanding on these results, we present small noise expressions for reverse-time mean, variance and two-time correlation function valid for force induced target state convergence. The generic analysis of noise and force induced transitions can serve as guidance when searching for the driving mechanism. For direct high precision inference of the true biological forces, we propose a path ensemble maximum likelihood scheme. Its applicability for the distinction of different scenarios of cytokinetic ring constriction also highlights possible force miss-assignments when ignoring the peculiarities of TSA ensembles.

The TSA approach introduces a new perspective to the design of experiments. It makes the search for suitable initial conditions superfluous and facilitates the investigation of directed dynamics under conditions that are as natural as possible.
Nevertheless, the identification of transition times, when dynamics change e.g. from an equilibrium state into target state directed dynamics such as during cytokinesis, remains a biologically highly relevant question. The here sketched path ensemble framework provides a possible route to this inference problem. A further generalization of the formalism at hand might include different types of fluctuating environments, the inclusion of external variables and an extension of the framework to more than one degree of freedom.

Concluding, the here presented theory provides the mathematical foundations for the inference of TSA dynamics. It offers an intuitive understanding of the characteristics of TSA ensembles and provides a new tool for the study of directed biological dynamics.




\begin{acknowledgments}
We thank Matthias Häring, Erik Schultheis and the Wolf group for stimulating discussions and proofreading of the manuscript. This work was supported by the German Research Foundation (Deutsche Forschungsgemeinschaft, DFG) through FOR 1756, SPP 1782, SFB 1528, SFB 889, SFB 1286, SPP 2205, DFG 436260547 in relation to NeuroNex (National Science Foundation 2015276) \& under Germany’s Excellence Strategy - EXC 2067/1- 390729940; by the Leibniz Association (project K265/2019); and by the Niedersächsisches Vorab of the VolkswagenStiftung through the Göttingen Campus Institute for Dynamics of Biological Networks.
\end{acknowledgments}


\bibliography{literature}

\end{document}



\title{Supplementary Information - Reverse-time analysis uncovers universality classes in directional biological dynamics}

\author{Nicolas Lenner}
 \altaffiliation[Currently at ]{Simons Center for Systems Biology, School of Natural Sciences, Institute for Advanced Study, Princeton, New Jersey, USA.}
  \email{Lenner@ias.edu}
 \affiliation{Max Planck Institute for Dynamics and Self-Organization, Göttingen, Germany}

\author{Stephan Eule}%
\affiliation{Max Planck Institute for Dynamics and Self-Organization, Göttingen, Germany}
\affiliation{German Primate Center—Leibniz Institute for Primate Research, Goettingen, Germany}

\author{Jörg Großhans}
\affiliation{Department of Biology, Philipps University Marburg, Marburg, Germany}
\affiliation{Göttingen Campus Institute for Dynamics of Biological Networks, University of Göttingen, Göttingen, Germany}

\author{Fred Wolf}
 \email{Fred.Wolf@ds.mpg.de}
\affiliation{Max Planck Institute for Dynamics and Self-Organization, Göttingen, Germany}
\affiliation{Göttingen Campus Institute for Dynamics of Biological Networks, University of Göttingen, Göttingen, Germany}
\affiliation{Max Planck Institute for Multidisciplinary Sciences, Göttingen, Germany}
\affiliation{Institute for the Dynamics of Complex Systems, University of Göttingen, Göttingen, Germany}
\affiliation{Center for Biostructural Imaging of Neurodegeneration, Göttingen, Germany}
\affiliation{Bernstein Center for Computational Neuroscience Göttingen, Göttingen, Germany}


\maketitle


%
%
%
%
%
%
%
%
This supplemental material provides the technical background of the theory of TSA dynamics and examples discussed in the main paper.  We present the derivation of the reverse time target state aligned stochastic different equation (TSA SDE), analytically explore the influence of the forward initial conditions on the reverse time dynamics and study the genuine form of TSA dynamics close to the target state.  
For each step in the derivation we provide analytically tractable examples and show how the theory can be applied to infer genuine forces from dynamics of biological examples. We start with a self containing summary of relevant concepts for stochastic processes in reverse time.

\newpage
\tableofcontents
\newpage


\section{Stochastic dynamics setting}
\subsection{Markov processes}
A time dependent stochastic process is defined by its probability to be at its current position $\widehat{L}_{n}$ at time $t_{n}$ and by the transitions it takes to get there starting at $\widehat{L}_{0},t_{0}$. In this chapter $\widehat{L}$ denotes processes in forward time $t$.
Formally such as process can be written as the transition probability
$
P(\widehat{L}_{n},t_{n}|\widehat{L}_{n-1},t_{n-1}, \dots , \widehat{L}_{0},t_{0})
$. 
For simplicity we assume that the current state of a stochastic process does only depend on its last step and not  on its full history, i.e.~that the process is memory free and obeys the Markov property
\begin{align}
  P(\widehat{L}_n,t_n|\widehat{L}_{n-1},t_{n-1}, \dots , \widehat{L}_{0},t_{0}) =
   P(\widehat{L}_n,t_n|\widehat{L}_{n-1},t_{n-1})
   \ .
\end{align}
The Markov property implies that the Chapman Kolmogorov equation
\begin{align}
\label{sup_chapman}
  P(\widehat{L}_{n},t_{n}|\widehat{L}_{n-2},t_{n-2}) =
  \int d \widehat{L}_{n-1}
   P(\widehat{L}_n,t_n|\widehat{L}_{n-1},t_{n-1})
   P(\widehat{L}_{n-1},t_{n-1}|\widehat{L}_{n-2},t_{n-2})
\end{align}
is fulfilled
which we will use repeatedly in the subsequent derivations.

\subsection{SDE and Fokker Planck representations}
The evolution of a dynamical system under the influence of random forcing can often be studied using a  Langevin equation
\begin{equation}
\label{sup_fwd_sde}
 d\widehat{L}(t) = f(\widehat{L}) \, dt + \sqrt{D} \, dW_t
 \ .
\end{equation}
The first term defines the deterministic drift $f(\widehat{L})$ the second the strength $D$ of random fluctuations. Here the term $dW_t$ denotes the Wiener process increment with zero mean $\langle dW_t \rangle = 0$ and delta correlated covariance  $\langle D \, dW_t  \, dW_t'\rangle = D \delta(t-t')$. Throughout this text we adopt the Ito-interpretation of Eq.~\eqref{sup_fwd_sde} which assures that the random forcing is always applied at the beginning of a discrete time interval $dt$.

Using Itos Lemma\cite{gardiner1985handbook}
\begin{align}
\label{itos_lemma}
 dg(\widehat{L}(t)) = 
 \left(
 f(\widehat{L}) \frac{\partial g(\widehat{L})}{\partial \widehat{L}}
 +
 \frac{D}{2} \frac{\partial^2 g(\widehat{L})}{\partial \widehat{L}^2}
 \right) dt
 +
 \sqrt{D} \frac{\partial g(\widehat{L})}{\partial \widehat{L}} \; dW_t
 \ ,
\end{align}
which shows how to apply a change of variables for quantities that are governed by 
Eq.~\eqref{sup_fwd_sde}, we can derive an evolution equation for its probability distribution 
$P(\widehat{L},t|\widehat{L}_0,t_0)$.
This evolution equation is called the (forward) Fokker-Planck equation (FPE) and is an equivalent description of the stochastic dynamics described in Eq.~\eqref{sup_fwd_sde}. A derivation can be found in Gardiner\cite{gardiner1985handbook}. 
Averaging Eq.~\eqref{itos_lemma} with respect to $P(\widehat{L},t|\widehat{L}_0,t_0)$, twice integrating by parts and identifying $g(\widehat{L})$ with a delta-function then leads to the seeked evolution equation\cite{gardiner1985handbook}.
The resulting partial differential equation
\begin{align}
 \label{sup_fwd_fp}
 \frac{\partial}{\partial t} 
P(\widehat{L},t|\widehat{L}_0,t_0)
 =
    - \frac{\partial}{\partial \widehat{L}} f(\widehat{L}) \,
P(\widehat{L},t|\widehat{L}_0,t_0)
    + \frac{D}{2} \frac{\partial^2}{\partial \widehat{L}^2} 
P(\widehat{L},t|\widehat{L}_0,t_0)
\end{align}
is the seeked (forward) Fokker-Planck equation. The adjoint equation is called the backward Fokker-Planck equation
\begin{align}
 \label{sup_bwd_fp}
  \frac{\partial}{\partial t} P(\widehat{L}_f,t_f | \widehat{L},t) 
  =
    - f(\widehat{L}) \frac{\partial}{\partial \widehat{L}} P(\widehat{L}_f,t_f | 
\widehat{L},t)
    - \frac{D}{2}  
      \frac{\partial^2 }{\partial \widehat{L}^2} P(\widehat{L}_f,t_f | 
\widehat{L},t) \ .
\end{align}
and defines the evolution from a initial condition $\widehat{L}$ to a finale state $\widehat{L}_f$. It is important to note that after relabeling both equations yield the same transition probabilities and that it is only a different perspective to solve the same problem.
The term 'backward'-Fokker-Planck equation is somewhat ill-posed and can not be exchanged for 'time-reversed'. In analogy to quantum mechanics, the forward Fokker-Planck equation describes the time evolution of the probability distribution and an average of any dynamical property is taken at a specific time point $t$ like in the Schr{\"o}dinger picture.
Conversely, the backward Fokker-Planck equation corresponds to the Heisenberg picture where the focus lies on the time evolution of a dynamical observable and the averages are taken over initial conditions. 
Like in quantum mechanics both approaches are thus describing exactly the same physics. They are merely offering different perspectives and can be interchanged by a simple transformation. Ultimately both approaches hence refer to the same transition probability.
In the next section we show how the concept of time-reversal is to be introduced into a stochastic framework.

\section{Time reversal and alignment}
\label{Time_reversed_Fokker_Planck_and_Langevin_equation}
The mathematical construction of the aligned and time reversed ensemble 
can be split into three parts.
(i) The general definition of time reversal for SDEs, (ii) its adaptation to 
sample paths $\widehat{L}_i(t)$ that hit a target state
$\widehat{L}_\mathrm{ts}$ after time $T_i$, and (iii) the alignment and time reversal of a complete ensemble of
sample paths $L_i(\tau)$ into one ensemble depending on their ``time to completion'' $\tau=T_i-t$.

\begin{figure}[ht]
\centerline{\includegraphics[width=0.8\linewidth]
{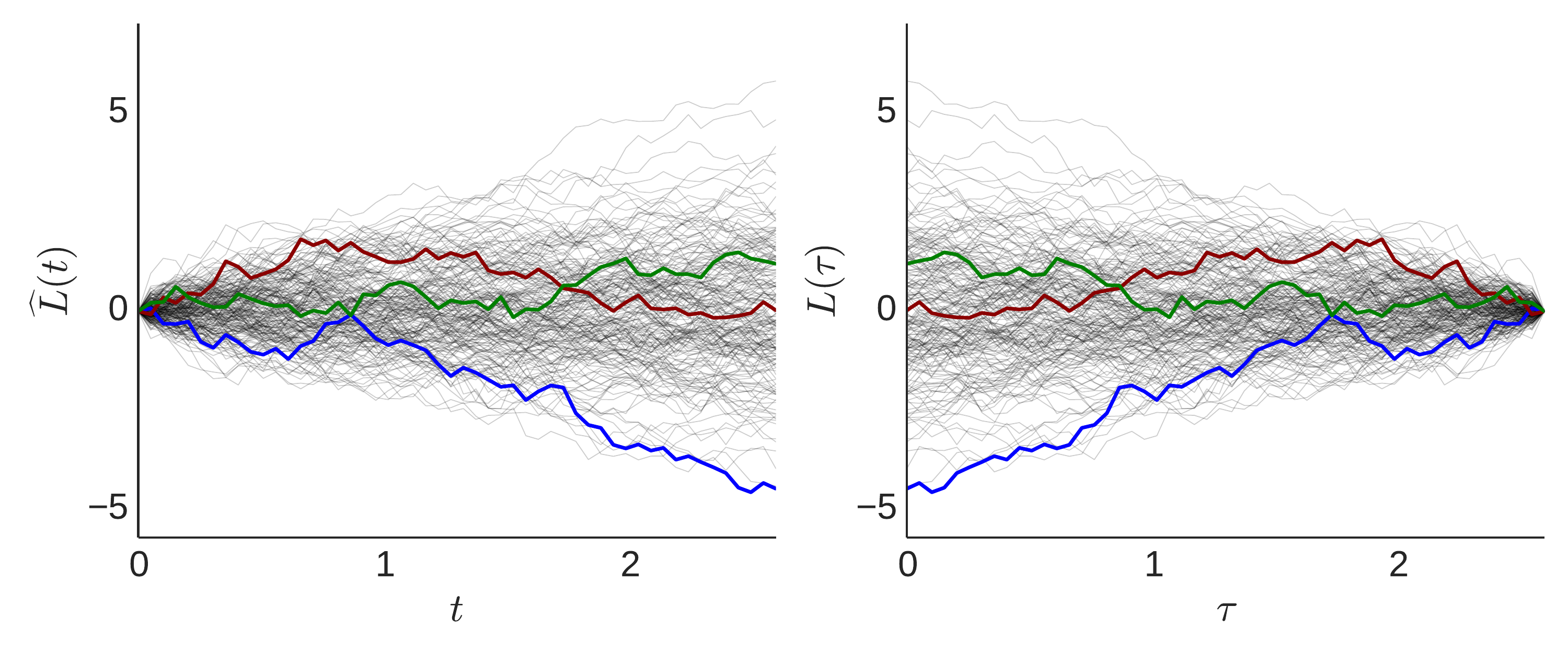}}
\caption{
\textbf{
Time reversal of an ensemble with natural boundary conditions.} (Left): 
Evolution of a set of sample 
paths $\widehat{L}_i(t)$ up to a time $T$. (Right): The set of time reverted 
sample paths $L_i(\tau)$.
Natural boundary conditions are defined as $P(\widehat{L}\to \pm \infty,t)=0$.
}\label{demoGenTimereversal}
\end{figure}

In  the first part we present the SDE for the time reversal of 
sample paths $\widehat{L}_i(t)$  of fixed lifetime $T$. 
The ensemble $P(\widehat{L},t)$ of sample paths $\widehat{L}_i(t)$, evolves freely according to a general stochastic law of motion. 
The reverse time ensemble then consists of sample paths $L_i(\tau) = \widehat{L}_i(T - \tau)$. 
All sample paths $\widehat{L}_i(t)$ start at the same time $t_0$ and 
evolve for the same time interval $T$.
The time reversed SDE describes precisely this ensemble, but in 
reverse time. 
It starts from the final 
distribution of the forward process 
$P(L,\tau_0) = P(\widehat{L},T)$
and evolves into the 
initial distribution 
$P(L,T) = P(\widehat{L},t_0)$.
One illustrative example realization  of both a forward ensemble
$P(\widehat{L},t)$
and time reversed ensemble
$P(L,\tau)$
is shown in 
Fig.~\ref{demoGenTimereversal}.

\begin{figure}[ht]
\centerline{\includegraphics[width=0.8\linewidth]
{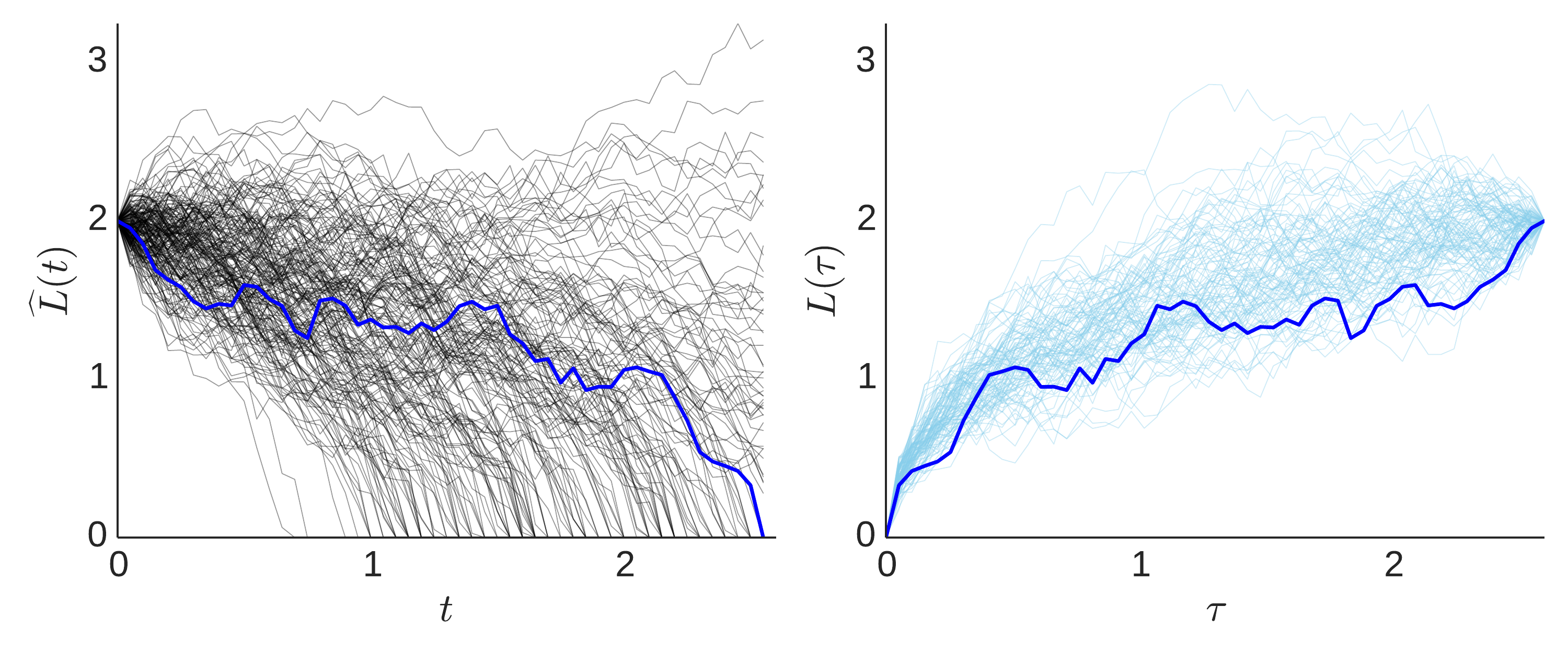}}
\caption{
\textbf{
Time reversal of a terminating sub-ensemble.} (Left): Evolution of a set of sample 
paths $\widehat{L}_i(t)$ up to an irreversible target point. Different sample 
paths $\widehat{L}_i(t)$ hit $\widehat{L}_\mathrm{ts}$ at different times 
$T_i$. (Right): Time reversal of a representative sample path $\widehat{L}_i(t)$ 
that has hit $\widehat{L}_\mathrm{ts}$ after time $T_i$ (dark blue). A 
realization of the sub-ensemble that has terminated after $T_i$ (light blue).
}\label{demoSubTrEnsemble}
\end{figure}

In the second part,
we consider the time reversal of sample paths 
$\widehat{L}_i(t)$ that hit a target state 
$\widehat{L}_\mathrm{ts}$ after a fixed time $T_i=T$.
In this step,
we pick precisely those sample paths $L_i$ of the full forward ensemble 
$P(\widehat{L},t)$ that arrive at a target state after a particular time to completion $T_i$. 
The time reversed SDE for this sub-ensemble is the same as derived in 
the first part,
with the initial condition of the reverse time process specialized as a delta-function 
$P(L,\tau_0) = \delta(\widehat{L}-\widehat{L}_\mathrm{ts})$
localized at the target state.
All sample paths of this sub-ensemble start at 
$\widehat{L}_\mathrm{ts}$ and progress towards the initial conditions of the 
forward process. They are all of the same lifetime $T_i$.
A realization of this time reversed sub-ensemble together with the full forward 
ensemble up to time $T_i$ is shown in Fig.~\ref{demoSubTrEnsemble}.

\begin{figure}[ht]
\centerline{\includegraphics[width=0.8\linewidth]
{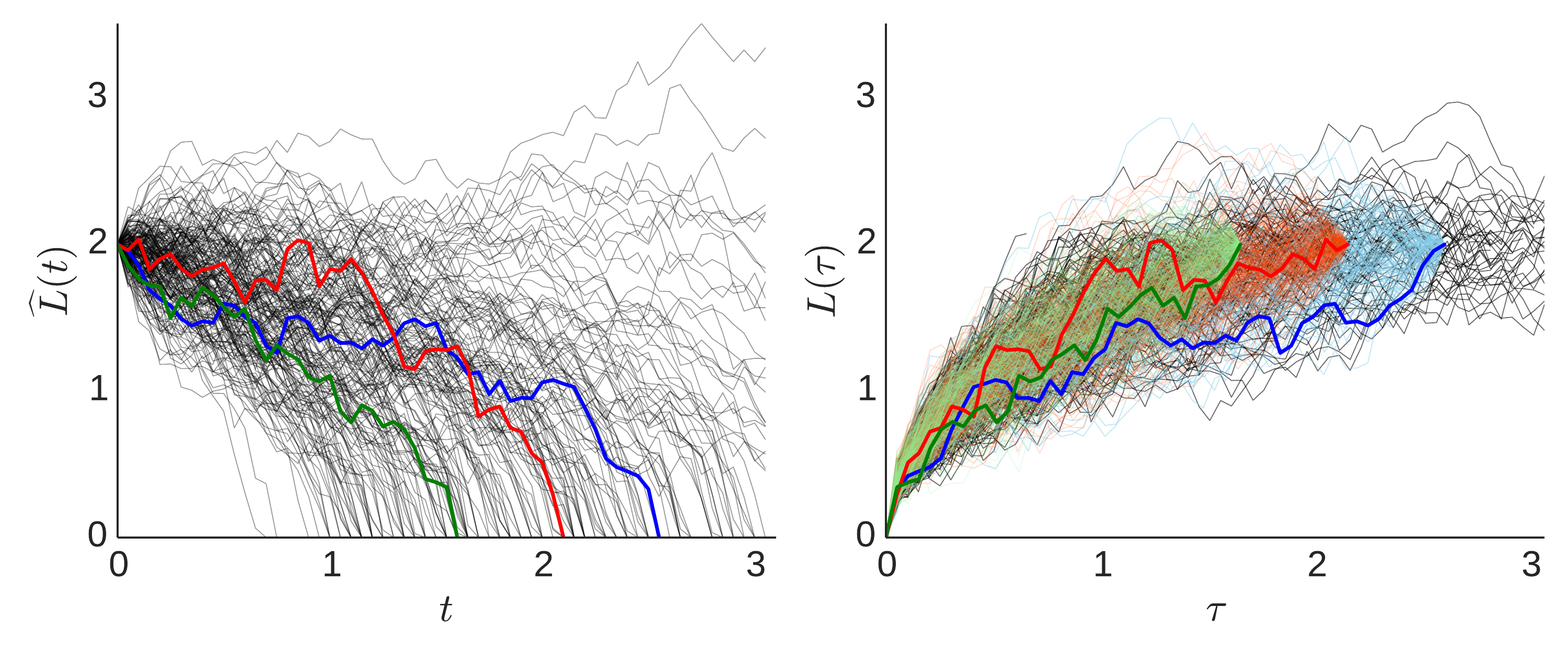}}
\caption{
\textbf{
Time reversal and alignment of the complete terminating ensemble.} (Left): A set of 
realizations of the complete forward ensemble. Three exemplary sample 
paths of different lifetimes $T_i$ are highlighted.
(Right): 
Time reversal of terminating sub-ensembles according to their lifetime $T_i$. The full ensemble is formed of sub-ensembles where three are 
exemplary highlighted.
}\label{demoFullAlignedTrEnsemble}
\end{figure}

Finally, in the last step, we 
construct the complete ensemble of time reversed TSA sample paths 
$L_i(\tau)$, of different lifetimes $T_i$, in a common time to completion reference frame. 
Depending on its noise and 
underlying stochastic law of motion each sample path $\widehat{L}_i(t)$ 
hits $\widehat{L}_\mathrm{ts}$ after a different time $T_i$, which by itself is a random variable. The  
distribution $\rho_{\widehat{L}_\mathrm{ts}}(T|\widehat{L}_0)$ of lifetimes $T$ is 
defined by the forward 
SDE of Eq.~\eqref{sup_fwd_sde} and its initial conditions.
To obtain the complete time reversed ensemble, we first construct
a time reversed sub-ensemble for each lifetime $T_i$.
$\rho_{\widehat{L}_\mathrm{ts}}(T|\widehat{L}_0)$ .
The complete time reversed ensemble is then obtained by
averaging over these sub-ensembles with respect to the distribution of lifetimes
$\rho_{\widehat{L}_\mathrm{ts}}(T|\widehat{L}_0)$ .
This construction of the complete time reversed ensemble out of 
sub-ensembles of different lifetimes is demonstrated in Fig.~\ref{demoFullAlignedTrEnsemble}.
Target state alignment is achieved by constructing each time reversed 
sub-ensemble starting from $\widehat{L}_\mathrm{ts}$ and $\tau =0$. The full 
time reversed ensemble then also starts from this initial state
$(\widehat{L}_\mathrm{ts},\tau 
=0)$.

The following sections give a mathematically detailed account of 
the three steps leading to the construction of an target state aligned and time reversed 
ensemble. We start with the derivation of a general time reversed SDE.

\subsection{The reverse time Fokker-Planck equation}
\label{the_reverse_time_Fokker-Planck_equation}
Time reversal of a stochastic process is a non trivial operation. While a deterministic process is described by an ODE and a single final condition, time reversal of a stochastic process does not simply turn a forward SDE into a reverse time SDE. This finding is rooted in the observation that transition probabilities of the reverse process must depend both on the initial and final conditions of the forward process. We start by noting that a stochastic process constrained by initial and final conditions can be constructed from the product of two transition probabilities 
\begin{align}
\label{sup_bridge_trans_prod}
P(\widehat{L}_f,t_f|\widehat{L},t) P(\widehat{L},t|\widehat{L}_0,t_0)
\ .
\end{align}
To ensure that sample paths start at $(\widehat{L}_0,t_0)$ and end at $(\widehat{L}_f,t_f)$ with probability one, this process is normalized using the Chapman Kolmogorov Eq.~\eqref{sup_chapman}.
The normalized process 
\begin{align}
\label{sup_bridge_trans_nobound}
P(\widehat{L},t|\widehat{L}_f,t_f,\widehat{L}_0,t_0) 
= 
\frac{
P(\widehat{L}_f,t_t|\widehat{L},t) P(\widehat{L},t|\widehat{L}_0,t_0)
}{
P(\widehat{L}_f,t_f|\widehat{L}_0,t_0) 
}
\end{align}
is called a bridge process of duration $T=t_f-t_0$ and defines the evolution of stochastic processes conditioned on both the initial $(\widehat{L}_0,t_0)$ and final conditions $(\widehat{L}_f,t_f)$\cite{majumdar2015effective}. We note that $P(\widehat{L},t|\widehat{L}_f,t_f,\widehat{L}_0,t_0)$  is symmetric under time reversal i.e. relabeling of indices. The initial conditions of the forward process $ (\widehat{L}_0,t_0)$ are then the final conditions of the reverse time process $(L_f,\tau_f)$ and vice versa. The bridge process defined in reverse time $\tau=t_f-t$ and with $\tau_0=0$ then reads
$P(L,\tau|L_0,\tau_0,L_f,\tau_f)$. The SDE that describes this process is known
\cite{anderson1982reverse} 
and assumes the form
\begin{align}
\label{sup_tr_lv}
  dL(\tau) =
      \left(
      -
       f(L) 
       + D
       \frac{\partial }{\partial L} 
       \log\left(P^{\mathrm{fw}}(L,T-\tau|L_f,0) \right)
       \right)
       d\tau
      +
      \sqrt{D} \ d W_\tau
\end{align}
where $P^{\mathrm{fw}}(L,T-\tau|L_f,0)$ denotes the solution of the forward Fokker Planck equation evaluated in reverse time $\tau$.
The initial conditions of this process are given by the final distribution
$P^{\mathrm{fw}}(\widehat{L}_f,t_f|\widehat{L}_0,t_0)$ of the forward 
process.
The guiding force 
$
 D
       \frac{\partial }{\partial L} 
       \log P^{\mathrm{fw}}(L,T-\tau|L_f,0)
$
ensures that every $L_i(\tau)$ synthesized in reverse time $\tau$
is statistically indistinguishable from a forward sample path $\widehat{L}_i$  of lifetime $T$.
Close to $\tau \approx T$, 
$P^{\mathrm{fw}}(L,T-\tau|L_f,0)$ becomes peaked and deviations from 
$P(\widehat{L}_0)=\delta(\widehat{L}-\widehat{L}_0)$
are suppressed by strong gradients of the guiding potential. The system is driven 
towards the initial conditions of the forward process. 
Note that the guiding force takes the form of an entropic force\cite{sokolov2010statistical} which reverses the single trajectory entropy production of the forward process\cite{seifert2005entropy} until the entropy of the forward initial conditions is reinstalled. 
An illustration of both an example forward ensemble and its time reversed 
equivalent is show in Fig.~\ref{demoGenTimereversal}.

The SDE Eq.~\eqref{sup_tr_lv} captures the dynamics of a 
time 
reversed process of fixed duration $T$ and is the first key equation to our approach. Its original derivation\cite{anderson1982reverse} builds on the equivalence of Langevin and Fokker-Planck equations in which Eq.~\eqref{sup_tr_lv} is obtained from the reverse time Fokker-Planck Equation
\begin{align}
\label{sup_tr_fp}
 \frac{\partial P(L,\tau|L_f,\tau_f,L_0,\tau_0)}{\partial 
\tau}  
 &=
     \frac{\partial}{\partial L}
    \left[
      \left(
       f(L) 
       - D
       \frac{\partial }{\partial L} 
       \log P^{\mathrm{fw}}(L,T-\tau|L_f,0)
      \right)
      P(L,\tau|L_f,\tau_f,L_0,\tau_0)
    \right] 
     \notag \\
    &\ \ \ + \frac{D}{2} \frac{\partial^2}{\partial L^2} 
P(L,\tau|L_f,\tau_f,L_0,\tau_0)
     \ .
\end{align}
The probability distribution 
$P(L,\tau|L_f,\tau_f,L_0,\tau_0)$ describes the evolution 
of reverse sample paths $L(\tau)$ of fixed lifetime $T$.
It depends on both, the 
initial and final conditions $\widehat{L}_0$ and $\widehat{L}_f$ of the forward 
process. 
The subsequent derivation of Eq.~\eqref{sup_tr_fp} follows Anderson's\cite{anderson1982reverse}. We here use the notation employed in the main text and add intermediate steps omitted in Anderson\cite{anderson1982reverse}. 
A mathematically more rigorous derivation can be found in F{\"o}llmer\cite{follmer1986time}.

The derivation proceeds in three steps: First, we will derive an evolution equation of 
the joint 
probability
distribution
$P(\widehat{L},t,\widehat{L}_f,t_f|\widehat{L}_0,t_0)$ 
using both the
backward 
and forward Fokker-Planck equation Eq.~\eqref{sup_bwd_fp} and Eq.~\eqref{sup_fwd_fp} for the forward sample paths $\widehat{L}(t)$
that evolve with time $t$.
From this intermediate result 
the evolution equation for the conditional probability
$P(\widehat{L},t|\widehat{L}_f,t_f,\widehat{L}_0,t_0)$
follows directly.
It describes sample paths $\widehat{L}_i(t)$ that are constrained to both 
initial and final conditions.
Finally we transform the evolution equation of
$P(\widehat{L},t|\widehat{L}_f,t_f,\widehat{L}_0,t_0)$ 
into the reverse time Fokker-Planck Equation for
$P(L,\tau|L_f,\tau_f,L_0,\tau_0)$ by time reversal.

Note, that we will omit writing out the explicit dependence on the initial 
conditions $(\widehat{L}_0,t_0)$ during this derivation 
for notational clarity. 
Even if not stated, every probability distribution used in this derivation 
carries
$(\widehat{L}_0,t_0)$ as a condition.

We start with writing the joint probability distribution from Eq.~\eqref{sup_bridge_trans_prod} 
\begin{equation}
\label{sup_joining_probs}
P (\widehat{L},t,\widehat{L}_f,t_f) =  P(\widehat{L}_f,t_f | \widehat{L},t) P(\widehat{L},t)
\end{equation}
as a product of a conditional probability and a probability density assuming $t_f 
\geq t$. Taking the derivative with respect to $t$ on both sides and 
multiplying both sides with $(-1)$, we 
obtain
\begin{equation}
 - \frac{\partial P (\widehat{L},t,\widehat{L}_f,t_f)}{\partial t} 
 = - P(\widehat{L},t) \frac{\partial}{\partial t} P(\widehat{L}_f,t_f | 
\widehat{L},t)
   -  P(\widehat{L}_f,t_f | \widehat{L},t) \frac{\partial}{\partial t} 
P(\widehat{L},t) \qquad .
\end{equation}
Substituting $\frac{\partial}{\partial_t} P(\widehat{L}_f,t_f | \widehat{L},t)$ using the 
backward Fokker-Planck
Eq.~\eqref{sup_bwd_fp} and $\frac{\partial}{\partial t} P(\widehat{L},t)$  using the 
forward 
Fokker-Planck Eq.~\eqref{sup_fwd_fp} yields
\begin{align} 
\label{joint_fpPlugIn}
 - \frac{\partial P (\widehat{L},t,\widehat{L}_f,t_f)}{\partial t}  
 &= P(\widehat{L},t) 
    \left[ 
    f(\widehat{L}) \frac{\partial}{\partial \widehat{L}} P(\widehat{L}_f,t_f | 
\widehat{L},t)
    + \frac{D}{2}  
      \frac{\partial^2}{\partial \widehat{L}^2} P(\widehat{L}_f,t_f | 
\widehat{L},t)
    \right]
 \notag \\
 &\ \ \ + P(\widehat{L}_f,t_f | \widehat{L},t)
    \left[
    \frac{\partial}{\partial \widehat{L}} f(\widehat{L}) \, P(\widehat{L},t)
    - \frac{D}{2} \frac{\partial^2}{\partial \widehat{L}^2}   P(\widehat{L},t)
    \right] \quad .
\end{align}
In the next step we consider the diffusion terms yielding an effective diffusion like 
term of the form 
$\frac{D}{2}\frac{\partial^2}{\partial \widehat{L}^2} 
P(\widehat{L}_f,t_f,\widehat{L},t) $.
Application of the product rule to both terms leads to
\begin{align} 
%
&\frac{D}{2}
\left[
P(\widehat{L},t)
      \frac{\partial^2 }{\partial \widehat{L}^2} P(\widehat{L}_f,t_f | 
\widehat{L},t)
\right]
 -
 \frac{D}{2}
 \left[
 P(\widehat{L}_f,t_f | \widehat{L},t)
      \frac{\partial^2}{\partial \widehat{L}^2}  P(\widehat{L},t)
  \right]
     =
      \notag \\
      &=
 \frac{D}{2} 
 \left[
 \frac{\partial}{\partial \widehat{L}} 
  \left(
  P(\widehat{L},t) \frac{\partial}{\partial \widehat{L}} P(\widehat{L}_f,t_f | 
\widehat{L},t)
 \right)
      -
 \left(
 \frac{\partial}{\partial \widehat{L}}  P(\widehat{L},t)
 \right)
 \left(
      \frac{\partial}{\partial \widehat{L}} P(\widehat{L}_f,t_f | \widehat{L},t)
 \right)
\right]
\notag \\
\label{sub_diffusion_ident_fwdOrbwd}
 & - \frac{D}{2} 
 \left[
  \frac{\partial}{\partial \widehat{L}} 
  \left(
 P(\widehat{L}_f,t_f | \widehat{L},t)
  \frac{\partial}{\partial \widehat{L}}   P(\widehat{L},t)  
  \right)
  -
  \left(
    \frac{\partial}{\partial \widehat{L}} 
 P(\widehat{L}_f,t_f | \widehat{L},t)
 \right)
 \left(
  \frac{\partial}{\partial \widehat{L}}   P(\widehat{L},t) 
 \right)
 \right]
 \ .
\\
      \intertext{Here the product term cancels}
      \label{first_appl_prod_rule_in_bridge}
     &=
 \frac{D}{2} \frac{\partial}{\partial \widehat{L}}
 \left(
  P(\widehat{L},t)
      \frac{\partial}{\partial \widehat{L}} P(\widehat{L}_f,t_f | \widehat{L},t)
 - P(\widehat{L}_f,t_f | \widehat{L},t)
  \frac{\partial}{\partial \widehat{L}}   P(\widehat{L},t)  
 \right)
 \ ,
%
\\
\intertext{using the product rule a second time, but now only for the first term
}
&=
 \frac{D}{2} \frac{\partial}{\partial \widehat{L}}
 \left(
  \frac{\partial}{\partial \widehat{L}}
  \left(
    P(\widehat{L},t) \, P(\widehat{L}_f,t_f | \widehat{L},t)
   \right)
   -
   P(\widehat{L}_f,t_f | \widehat{L},t)
  \frac{\partial}{\partial \widehat{L}}   P(\widehat{L},t)    
 - P(\widehat{L}_f,t_f | \widehat{L},t)
  \frac{\partial}{\partial \widehat{L}}  P(\widehat{L},t)  
 \right)
 \ ,
\\
\intertext{and summing identical terms, yields}
&=
 \frac{D}{2} \frac{\partial^2}{\partial \widehat{L}^2}
 \left(
  P(\widehat{L},t) \, P(\widehat{L}_f,t_f | \widehat{L},t)
 \right)
 - D 
 \frac{\partial}{\partial \widehat{L}}
 \left(
 \ P(\widehat{L}_f,t_f | \widehat{L},t)
  \frac{\partial}{\partial \widehat{L}}  P(\widehat{L},t) 
 \right)
\ .
\end{align}

Next we consider all $f(\widehat{L})$ containing terms in Eq.~\eqref{joint_fpPlugIn}. Application of the product rule here
yields
\begin{align}
    P(\widehat{L},t)
    f(\widehat{L}) \frac{\partial}{\partial \widehat{L}} P(\widehat{L}_f,t_f | 
\widehat{L},t)
+
P(\widehat{L}_f,t_f | &\widehat{L},t)
    \frac{\partial}{\partial \widehat{L}} f(\widehat{L})  P(\widehat{L},t)
=
\notag\\
    &=
    \frac{\partial}{\partial \widehat{L}}
 \left( 
    f(\widehat{L}) P(\widehat{L}_f,t_f | \widehat{L},t) \,  P(\widehat{L},t) 
\right)
\ .
\end{align}
Now we can collect terms and apply Eq.~\eqref{sup_joining_probs} to join the 
probability distributions 
$P(\widehat{L}_f,t_f | \widehat{L},t)$ and $P(\widehat{L},t)$, which allows to 
bring 
Eq.~\eqref{joint_fpPlugIn} into the form
\begin{align}
\label{sup_rt_fp_full_dist_subst}
 - \frac{\partial P (\widehat{L},t,\widehat{L}_f,t_f)}{\partial t}  
 &=
     \frac{\partial}{\partial \widehat{L}}
    \left[
      f(\widehat{L}) P(\widehat{L},t,\widehat{L}_f,t_f) 
      - D \ P(\widehat{L}_f,t_f|\widehat{L},t) \frac{\partial}{\partial 
\widehat{L}}  P(\widehat{L},t)
    \right] 
\notag \\
    &+ \frac{D}{2} \frac{\partial^2}{\partial \widehat{L}^2} 
P(\widehat{L},t,\widehat{L}_f,t_f)
\ .
\end{align}
Note that the term $- D \ P(\widehat{L}_f,t_f|\widehat{L},t) 
\frac{\partial}{\partial \widehat{L}} 
P(\widehat{L},t)$, although depending on the diffusion constant $D$, is now 
assigned to 
the drift part. Rewriting this term, using again 
Eq.~\eqref{sup_joining_probs} in the form 
$
P(\widehat{L}_f,t_f|\widehat{L},t) = P(\widehat{L}_f,t_f,\widehat{L},t) /
P(\widehat{L},t)
$, yields
\begin{align}
 - \frac{\partial P (\widehat{L},t,\widehat{L}_f,t_f)}{\partial t}  
 &=
     \frac{\partial}{\partial \widehat{L}}
    \left[
      \left(
       f(\widehat{L}) 
       - \frac{D}{P(\widehat{L},t)} 
       \frac{\partial P(\widehat{L},t)}{\partial \widehat{L}} 
      \right)
      P(\widehat{L},t,\widehat{L}_f,t_f)
    \right] 
\notag \\
    &+ 
    \frac{D}{2} \frac{\partial^2}{\partial \widehat{L}^2}  
P(\widehat{L},t,\widehat{L}_f,t_f)
     \ .
\end{align}
This is our first intermediate result. The evolution equation for the joined 
probability distribution 
$P(\widehat{L},t,\widehat{L}_f,t_f)$.

Dividing this equation by $P(\widehat{L}_f,t_f)$, the distribution of the final 
state, results in an evolution equation of the forward probability distribution
\begin{align}
 - \frac{\partial P (\widehat{L},t|\widehat{L}_f,t_f)}{\partial 
t}  
 &=
     \frac{\partial}{\partial \widehat{L}}
    \left[
      \left(
       f(\widehat{L}) 
       - \frac{D}{P(\widehat{L},t)} 
       \frac{\partial P(\widehat{L},t)}{\partial \widehat{L}} 
      \right)
      P(\widehat{L},t|\widehat{L}_f,t_f)
    \right] 
\notag \\
    &+ 
    \frac{D}{2} \frac{\partial^2}{\partial \widehat{L}^2}  
P(\widehat{L},t|\widehat{L}_f,t_f)
     \ ,
\end{align}
which in its force term can be simplified further
\begin{align}
 - \frac{\partial P (\widehat{L},t|\widehat{L}_f,t_f)}{\partial 
t}  
 &=
     \frac{\partial}{\partial \widehat{L}}
    \left[
      \left(
       f(\widehat{L}) 
       -
       D \frac{\partial}{\partial \widehat{L}} 
        \log
        P(\widehat{L},t)
%
      \right)
      P(\widehat{L},t|\widehat{L}_f,t_f)
    \right] 
\notag \\
    &+ \frac{D}{2} \frac{\partial^2}{\partial \widehat{L}^2}  
P(\widehat{L},t|\widehat{L}_f,t_f)
     \ .
\end{align}
To make the interpretation of this equation and especially of the term 
$P(\widehat{L},t)$ explicit, the dependence of every probability term on the initial 
conditions $(\widehat{L}_0,t_0)$, suppressed throughout this derivation,
is now restored to yield
\begin{align}
\label{sup_tr_fp_1pre_derived}
 - \frac{\partial P 
(\widehat{L},t|\widehat{L}_0,t_0,\widehat{L}_f,t_f)}{
\partial 
t}  
 &=
     \frac{\partial}{\partial \widehat{L}}
    \left[
      \left(
       f(\widehat{L}) 
       - D
       \frac{\partial}{\partial \widehat{L}} 
       \log P(\widehat{L},t|\widehat{L}_0,t_0)
      \right)
      P(\widehat{L},t|\widehat{L}_0,t_0,\widehat{L}_f,t_f)
    \right] 
     \notag \\
    &\ \ \ 
    + 
    \frac{D}{2} \frac{\partial^2}{\partial \widehat{L}^2}  
P(\widehat{L},t|\widehat{L}_0,t_0,\widehat{L}_f,t_f)
     \ .
\end{align}
The probability distribution
$P(\widehat{L},t|\widehat{L}_0,t_0,\widehat{L}_f,t_f)$ 
describes the evolution of sample paths $\widehat{L}(t)$ in wall time $t$
connecting an initial state $\widehat{L}_0,t_0$ with a 
final state 
$\widehat{L}_f,t_f$. This is our second intermediate result.

The time reversed Fokker-Planck Equation can be obtained from 
Eq.~\eqref{sup_tr_fp_1pre_derived} by replacing the 
wall time $t$ with the reverse time $\tau$. Setting $\tau = t_f - t$, and the duration of the forward process  after which it is reverted to $T=t_f-t_0$, the resulting equation 
reads
\begin{align}
 \frac{\partial P 
(L,\tau|\widehat{L}_0,t_0,\widehat{L}_f,t_f)}{\partial 
\tau}  
 &=
     \frac{\partial}{\partial L}
    \left[
      \left(
       f(L) 
       - D
       \frac{\partial}{\partial L}
       \log P^{\mathrm{fw}}(L,T-\tau|\widehat{L}_f,0)
      \right)
      P(L,\tau|\widehat{L}_0,t_0,\widehat{L}_f,t_f)
    \right] 
     \notag \\
    &\ \ \ + \frac{D}{2} \frac{\partial^2}{\partial L^2}  
P(L,\tau|\widehat{L}_0,t_0,\widehat{L}_f,t_f)
     \ .
\end{align}
The superscript ($\mathrm{fw}$) was added for clarity. In the final step we replace all forward coordinates $\widehat{L}$ by their time reversed equivalent $L$ and obtain the final result already stated in Eq.~\eqref{sup_tr_fp}. The initial conditions of Eq.~\eqref{sup_tr_fp} can be drawn from the final distribution of the forward process $P^\mathrm{fw}(\widehat{L}_f,t_f|\widehat{L}_0,t_0)$. 
Solving or approximating the time reversed Fokker-Planck Equation therefore 
crucially depends on knowledge about the forward process.

\subsection{The time reversal of a terminating sub-ensemble}
\label{The_irreversible_target_point}
Time reversal of a sub-ensemble which ends at a target state $\widehat{L}_\mathrm{ts}$ after time $T=t_f-t_0$ is a special case of the general formulation of the reverse time Fokker-Planck Eq.~\eqref{sup_tr_fp}. From a constructive perspective two specifications are necessary. First, to invoke an irreversible target state, an absorbing boundary condition 
\begin{align}
\label{eq:def_fwd_fp_abs_bound}
 P(\widehat{L},t|\widehat{L}_0,t_0) = 0 \quad \text{for} \quad \widehat{L}=\widehat{L}_\mathrm{ts}
 \end{align}
must be implemented.
Second, with absorbing boundary conditions the product of the two transition probabilities which define the not yet normalized bridge 
Eq.~\eqref{sup_bridge_trans_prod}, is zero for 
$\widehat{L}_f \to \widehat{L}_\mathrm{ts}$. The second transition $P(\widehat{L}_f,t_t|\widehat{L},t)$ in Eq.~\eqref{sup_bridge_trans_prod} must therefore be replaced by the probability to reach the target state $\widehat{L}_\mathrm{ts}$ starting from $\widehat{L}$ and after a time $t_f-t$. This probability flux into the boundary is called a hitting time distribution and is defined as
\begin{align}
\label{eq:def_hit_dist}
\rho_{\widehat{L}_\mathrm{ts}}(t_f|\widehat{L},t) 
=
- \frac{\partial}{\partial t_f}
\int_{\widehat{L}_\mathrm{ts}}^\infty dL_f \ P(\widehat{L}_f,t_f|\widehat{L},t)
\ .
\end{align}
The un-normalized bridge, which starts at $\widehat{L}_0$, ends at $\widehat{L}_f=\widehat{L}_\mathrm{ts}$ and only includes sample paths which reach the target state exactly after a time $T=t_f-t_0$, is then defined as
\begin{align}
\label{sup_bridge_trans_bound_notNorm}
\rho_{\widehat{L}_\mathrm{ts}}(t_f|\widehat{L},t) P(\widehat{L},t|\widehat{L}_0,t_0)
\ .
\end{align}
To ensure that sample paths starting at $\widehat{L}_0$ end at 
$\widehat{L}_f = \widehat{L}_\mathrm{ts}$ with probability one, Eq.~\eqref{sup_bridge_trans_bound_notNorm} is normalized with respect to the hitting time distribution $\rho_{\widehat{L}_\mathrm{ts}}(t_f|\widehat{L}_0,t_0)$. The full bridge ending at a target state then reads
\begin{align}
\label{sup_bridge_trans_bound}
R_{\widehat{L}_\mathrm{ts}}(\widehat{L},t|\widehat{L}_0,t_0;t_f)
:=
 \frac{\rho_{\widehat{L}_\mathrm{ts}}(t_f|\widehat{L},t)
        P(\widehat{L},t|\widehat{L}_0,t_0)
 }{\rho_{\widehat{L}_\mathrm{ts}}(t_f|\widehat{L}_0,t_0)}
 \ .
\end{align}
We will show below, 
that $\rho_{\widehat{L}_\mathrm{ts}}(t_f|\widehat{L},t)$ is a solution of the backward Fokker Planck and that the normalization of this bridge process is again a hitting time distribution. The dynamics of Eq.~\eqref{sup_bridge_trans_bound}  written in reverse time $\tau$ are therefore formally identical to Eq.~\eqref{sup_tr_fp} and read
\begin{align}
\label{sup_tr_fp_derived}
 \frac{\partial 
 R_{L_\mathrm{ts}}(L,\tau|L_f,\tau_f;\tau_0)
 }{\partial 
\tau}  
 &=
     \frac{\partial}{\partial L}
    \left[
      \left(
       f(L) 
       - D 
       \frac{\partial}{\partial L}
       \log P^{\mathrm{fw}}(L,T-\tau|L_f,0)
      \right)
      R_{L_\mathrm{ts}}(L,\tau|L_f,\tau_f;\tau_0)
    \right] 
     \notag \\
    &\ \ \ + \frac{D}{2} \frac{\partial^2}{\partial L^2}  
R_{L_\mathrm{ts}}(L,\tau|L_f,\tau_f;\tau_0)
     \ .
\end{align}
We again added the upper index ($\mathrm{fw}$) for clarity. The corresponding reverse time SDE is thus  identical to Eq.~\eqref{sup_tr_lv} with $P^{\mathrm{fw}}(L,T-\tau|\widehat{L}_0,0)$ assumed with an absorbing boundary at $L_\mathrm{ts}$. 
Assuming absorbing boundary conditions for
$P^{\mathrm{fw}}(\widehat{L},t|\widehat{L}_0,t_0)$, the 
guiding force in Eq.~\eqref{sup_tr_lv} automatically prevents time 
reversed sample paths $L_i(\tau)$ from returning to $\widehat{L}_\mathrm{ts}$ 
for 
$\tau >0$. This can be seen by evaluating the guiding force 
$
D
\frac{\partial}{\partial L}
\log
P^{\mathrm{fw}}(L,T - \tau|\widehat{L}_0,0)
$
 with 
respect to the absorbing boundary condition. Lets assume the 
irreversible target 
state $\widehat{L}_\mathrm{ts}$ is a lower boundary of the one dimensional 
dynamic. As 
$
P^{\mathrm{fw}}(L,T - \tau|\widehat{L}_0,0)
\xrightarrow[L \to \widehat{L}_\mathrm{ts}]{} 0
$
approaches zero from above
close to the absorbing boundary,
$
\log P^{\mathrm{fw}}(L,T-\tau|\widehat{L}_0,0) 
\xrightarrow[L \to \widehat{L}_\mathrm{ts}]{} - \infty
$
diverges.
The guiding force
$
D
\frac{\partial}{\partial L}
\log
P^{\mathrm{fw}}(L,T - \tau|\widehat{L}_0,t_0)
\xrightarrow[L \to \widehat{L}_\mathrm{ts}]{}  \infty
$
therefore diverges as well
and thus behaves as an infinitely large barrier that prevents all $L_i(\tau)$ 
from returning to $\widehat{L}_\mathrm{ts}$.

\subsubsection{The target state bridge normalization}
\label{Chapman_Kolmogorov_and_the_bridge_normalization}
In this sub section we show that $\rho_{\widehat{L}_\mathrm{ts}}(t_f|\widehat{L}_0,t_0)$  is the normalization for the target-state bridge process. One step in this calculation is based on the assumption that the Chapman-Kolmogorov equation Eq.~\eqref{sup_chapman} also holds, when not the full probability space is considered but only those transitions which do not end at a target state, i.e. $P(\widehat{L}',t'|\widehat{L},t)$. 
This assumption is implicit to all calculations of Fokker-Planck equations with absorbing boundary conditions imposed. Nevertheless, as we are not aware of an explicit demonstration of this claim, we will briefly sketch a derivation below.
This sub-section first shows that a Chapman Kolmogorov also holds for this subspace of transitions. In a second step this result is used to determine the normalization of a bridge process with target state at an absorbing boundary.

We first demonstrate that a Chapman-Kolmogorov like equation also holds for 
$P(\widehat{L},t|\widehat{L}_0,t_0)$ and $P(\widehat{L}_f,t_f|\widehat{L},t)$ which are not normalized, as transitions to the absorbing boundary are split off as an extra term.
Splitting the expression into those two terms yields
\begin{align}
\label{sup_full_sol_fwd_fp}
  P^{\mathrm{full}}(\widehat{L},t|\widehat{L}_0,t_0) 
  = 
   \delta 
   \left(
   \widehat{L}-\widehat{L}_\mathrm{ts}
   \right)
   P^{\mathrm{ts}}(t|\widehat{L}_0,t_0) 
   + P(\widehat{L},t|\widehat{L}_0,t_0)
\ ,
\end{align}
where $P^\mathrm{ts}(t|\widehat{L}_0,t_0)$ denotes the over time $t-t_0$ at the target state accumulated probability mass.
For this normalized expression Chapman-Kolmogorov holds
\begin{align}
 P^{\mathrm{full}}(\widehat{L}_f,t_f|\widehat{L}_0,t_0)
 &=
\int_{\widehat{L}_\mathrm{ts}}^\infty 
d\widehat{L} \
 P^{\mathrm{full}}(\widehat{L}_f,t_f|\widehat{L},t)
P^{\mathrm{full}}(\widehat{L},t|\widehat{L}_0,t_0)
\intertext{and can be rewritten using the definition Eq.~\eqref{sup_full_sol_fwd_fp} to yield}
&=
\int_{\widehat{L}_\mathrm{ts}}^\infty 
d\widehat{L} \
\left(
\delta 
\left(
\widehat{L}_f-\widehat{L}_\mathrm{ts}
\right)
P^{\mathrm{ts}}(t_f|\widehat{L},t)
%
+ P(\widehat{L}_f,t_f|\widehat{L},t)
\right)
P^{\mathrm{full}}(\widehat{L},t|\widehat{L}_0,t_0)
\ .
\intertext{Evaluating the first term of the product we find}
&=
\delta 
\left(
\widehat{L}_f-\widehat{L}_\mathrm{ts}
\right)
P^{\mathrm{ts}}(t_f|\widehat{L}_0,t_0)
+
\int_{\widehat{L}_\mathrm{ts}}^\infty 
d\widehat{L} \
P(\widehat{L}_f,t_f|\widehat{L},t)
P^{\mathrm{full}}(\widehat{L},t|\widehat{L}_0,t_0) 
\ .
\intertext{Due to the absorbing boundary condition,
transitions from $\widehat{L}_0 \to \widehat{L} \to \widehat{L}_f$, with $\widehat{L}_f \neq \widehat{L}_\mathrm{ts}$, are only possible if the intermediate step does not end at the absorbing boundary. We can therefore simplify the remaining integral to include only transitions that do not end at the target state. The final equation reads}
&=
\delta 
\left(
\widehat{L}_f-\widehat{L}_\mathrm{ts}
\right)
P^{\mathrm{ts}}(t_f|\widehat{L}_0,t_0)
+
\int_{\widehat{L}_\mathrm{ts}}^\infty 
d\widehat{L} \
P(\widehat{L}_f,t_f|\widehat{L},t)
P(\widehat{L},t|\widehat{L}_0,t_0)
\ .
\end{align}
Comparing this result with the definition of $P^{\mathrm{full}}(\widehat{L}_f,t_f|\widehat{L},t)$ in Eq.~\eqref{sup_full_sol_fwd_fp}, this implies, that the Chapman Kolmogorov like identity
\begin{align}
\label{sup_chapman_for_ts}
P(\widehat{L}_f,t_f|\widehat{L}_0,t_0)
=
  \int_{\widehat{L}_\mathrm{ts}}^\infty 
d\widehat{L} \
 P(\widehat{L}_f,t_f|\widehat{L},t)
P(\widehat{L},t|\widehat{L}_0,t_0)
\end{align}
holds. We next use this identity to obtain the normalization for the target-state bridge Eq.~\eqref{sup_bridge_trans_bound_notNorm} by integration over $\widehat{L}$. Using the definition of the hitting time distribution
\begin{align}
\int_{\widehat{L}_\mathrm{ts}}^\infty 
d\widehat{L} \
 \rho_{\widehat{L}_\mathrm{ts}}(t_f|\widehat{L},t) P(\widehat{L},t|\widehat{L}_0,t_0) 
 &=
-
 \int_{\widehat{L}_\mathrm{ts}}^\infty 
d\widehat{L} \
 P(\widehat{L},t|\widehat{L}_0,t_0) 
 \frac{\partial}{\partial t_f}
 \int_{\widehat{L}_\mathrm{ts}}^\infty 
d\widehat{L}_f \ 
 P(\widehat{L}_f,t_f|\widehat{L},t)
 \ ,
 \intertext{taking both the derivative with respect to $t_f$ and the integral with respect to $\widehat{L}_f$ to the outside, we arrive at}
  &=
-
 \frac{\partial}{\partial t_f}
 \int_{\widehat{L}_\mathrm{ts}}^\infty 
d\widehat{L}_f \ 
 \int_{\widehat{L}_\mathrm{ts}}^\infty 
d\widehat{L} \
 P(\widehat{L},t|\widehat{L}_0,t_0) 
 P(\widehat{L}_f,t_f|\widehat{L},t)
 \ .
 \intertext{We replace the inner integral by the above derived version of the Chapman Kolmogorov equation Eq.~\eqref{sup_chapman_for_ts}}
 &=
 -
 \int_{\widehat{L}_\mathrm{ts}}^\infty 
d\widehat{L}_f \ 
 P(\widehat{L}_f,t_f|\widehat{L}_0,t_0)
 \intertext{and find that the hitting time distribution }
 &=
 \rho_{\widehat{L}_\mathrm{ts}}(t_f|\widehat{L}_0,t_0)
\end{align}
for processes starting at $(\widehat{L}_0,t_0)$ and ending at $(\widehat{L}{_\mathrm{ts}},t_f)$ is the normalization of the bridge process Eq.~\eqref{sup_bridge_trans_bound}.

\subsubsection{Backward Fokker-Planck for hitting time distribution}
\label{Backward_Fokker_Planck_for_hitting_time_distribution}
To conclude the above section on bridge processes with target states we recall that the hitting time distribution $\rho_{\widehat{L}_\mathrm{ts}}(t_f|\widehat{L},t)$ fulfills the backward Fokker-Planck equation Eq.~\eqref{sup_bwd_fp}. This can be easily seen by taking the derivative of Eq.~\eqref{eq:def_hit_dist} with respect to $t$
\begin{align}
 \frac{\partial}{\partial t}\rho_{\widehat{L}_\mathrm{ts}}(t_f|\widehat{L},t)
 &= 
 - \frac{\partial}{\partial t_f} \int_{\widehat{L}_\mathrm{ts}}^\infty dL_f  
\frac{\partial}{\partial t} \ P(\widehat{L}_f,t_f|\widehat{L},t)
\ .
\intertext{Substituting the backward Fokker-Planck equation Eq.~\eqref{sup_bwd_fp}}
&=
 - \frac{\partial}{\partial t_f} \int_{\widehat{L}_\mathrm{ts}}^\infty dL_f
 \left(
    - f(\widehat{L}) \frac{\partial}{\partial \widehat{L}} P(\widehat{L}_f,t_f | 
\widehat{L},t)
    - \frac{D}{2}  
      \frac{\partial^2 }{\partial \widehat{L}^2} P(\widehat{L}_f,t_f | 
\widehat{L},t)
\right)
 \notag
 \intertext{and replacing all $P(\widehat{L}_f,t_f |\widehat{L},t)$ dependent terms by the definition of the hitting time distribution Eq.~\eqref{eq:def_hit_dist}, we obtain the backward Fokker Planck for hitting time distributions}
&=
    - f(\widehat{L}) \frac{\partial}{\partial \widehat{L}} \rho_{\widehat{L}_\mathrm{ts}}(t_f|\widehat{L},t)
    - \frac{D}{2}  
      \frac{\partial^2 }{\partial \widehat{L}^2} \rho_{\widehat{L}_\mathrm{ts}}(t_f|\widehat{L},t)
\ .
\end{align}

\subsection{The time reversal and alignment of the full terminating ensemble}
\label{time_to_completion_as_stochastic_variable}
So far we considered the time reversal of sample paths 
$\widehat{L_i}(t)$ of identical lifetime $T_i=T$. The lifetime $T_i$, however, is a stochastic variable. Different sample paths $\widehat{L_i}(t)$ reach 
the target state after different lifetimes $T_i$.
The distribution of lifetimes is the hitting time distribution
$\rho_{\widehat{L}_\mathrm{ts}}(t_f|\widehat{L}_0,t_0)$.
The full forward ensemble of all sample paths starting at $(\widehat{L}_0,t_0)$ and ending at $\widehat{L}_\mathrm{ts}$ after a random time $t_f$  can therefore be understood as comprised of sub-ensembles 
$
R_{L_\mathrm{ts}}(\widehat{L},t|\widehat{L}_0,t_0;t_f)
$ each of a different lifetime $T_i$. Each of these sub-ensembles then contributes to the full ensemble with a weight given by the hitting time distribution $\rho_{\widehat{L}_\mathrm{ts}}(t_f|\widehat{L}_0,t_0)$

In this perspective, target state alignment consists of detaching all these sub-ensembles from their initiation point $(\widehat{L}_0,t_0)$ and shifting all
sample paths to reach the target state at a common reference time $t_f$. The resulting target state aligned ensemble in forward time
\begin{align}
\label{sup_prob_full_tr_ens}
R_{\widehat{L}_\mathrm{ts}}(\widehat{L},t;t_f|\widehat{L}_0)
=
\int_{-\infty}^t
dt_0
\rho_{\widehat{L}_\mathrm{ts}}(t_f|\widehat{L}_0,t_0)
R_{\widehat{L}_\mathrm{ts}}(\widehat{L},t|\widehat{L}_0,t_0;t_f)
\end{align}
is then given as the hitting time weighted integral of target-state aligned sub-ensembles.
By target state alignment we choose one $t_f$ for all sub-ensembles. 
Keeping $t_f$ fixed, $T_i=t_f-t_0$ or respectively $t_0$ becomes the random variable over which we integrate. Notably, not all sub-ensembles contribute to the full TSA ensemble at all times. The longer the lifetime $t_f-t$, the fewer sub-ensembles have a lifetime $T_i$ long enough to still contribute. The integral bounds therefore enforce that at time $t$ only those sub-ensembles with a minimal lifetime $t_f-t$ contribute. 

In reverse time $\tau$ the expression for the TSA ensemble is
\begin{align}
\label{sup_rtsa_delta}
 R_{L_\mathrm{ts}}(L,\tau;\tau_0|L_f)
=
\int_\tau^\infty d\tau_f
\rho_{L_\mathrm{ts}}(\tau_0|L_f,\tau_f)
\
 R_{L_\mathrm{ts}}(L,\tau|L_f,\tau_f;\tau_0)
 \ .
\end{align}
With $\tau_0$ identically fixed for all sub-ensembles only those sub-ensembles contribute at time $\tau$ with a lifetime $T_i \ge \tau-\tau_0$. The reverse time TSA ensemble $R_{L_\mathrm{ts}}(L,\tau;\tau_0|L_f)$ is therefore not normalized. It decays with the the hitting time distribution $\rho_{L_\mathrm{ts}}(\tau_0|L_f,\tau_f)$.

In the next step we generalize this expression to include sample-paths with their initial position $\widehat{L}_0$ drawn from an input distributions $P^\mathrm{in}(\widehat{L}_0)$. For the reverse time TSA ensemble this generalization simply amounts to integrating  over Eq.~\eqref{sup_rtsa_delta} with respect to  $P^\mathrm{in}(L_f)$ and we arrive at
\begin{align}
\label{sup_rtsa_dist}
 R_{L_\mathrm{ts}}(L,\tau;\tau_0)
=
\int_{L_\mathrm{ts}}^\infty
dL_f
P^\mathrm{in}(L_f)
\int_\tau^\infty d\tau_f
\rho_{L_\mathrm{ts}}(\tau_0|L_f,\tau_f)
\
 R_{L_\mathrm{ts}}(L,\tau|L_f,\tau_f;\tau_0)
 \ .
\end{align}
This expression is exactly the same as stated in the main text, where the concept of TSA ensembles is introduced. There, we set 
$R(L,\tau):=R_{L_\mathrm{ts}}(L,\tau;\tau_0)$ and neglected the for the intuition irrelevant dependency on 
$L_\mathrm{ts}$ and $\tau_0$. 

Independent of its notation,
$R_{L_\mathrm{ts}}(L,\tau;\tau_0)$ satisfies an easy to interpret reverse time Fokker-Planck equation which reads
\begin{align}
\label{eq:revt_TSA_FP}
&\frac{\partial}{\partial \tau}
R_{L_\mathrm{ts}}(L,\tau;\tau_0)
  =
-
  P^\mathrm{in}(L)
 \rho_{L_\mathrm{ts}}(\tau|L) 
\notag \\
 &-
     \frac{\partial}{\partial L}
     \left(
    \left[
      f(L) 
      +D 
      \  
      \frac{\partial}{\partial 
L}
 \log\left(
  \int_{L_\mathrm{ts}}^{L} dL' \;
   e^{-
  \int^{L'}
 \frac{2 f(L'')}{D} dL''
 }
 \left(
 1
 -
 \int_{L_\mathrm{ts}}^{L'} P^{\mathrm{in}}(L'') dL''
 \right) 
 \right)
    \right] 
     R_{L_\mathrm{ts}}(L,\tau;\tau_0)
     \right)
\notag \\    
    &+ \frac{D}{2} \frac{\partial^2}{\partial L^2}  
R_{L_\mathrm{ts}}(L,\tau;\tau_0)
\ .
\end{align}
The exact derivation is presented further below in this section. 
The first term constitutes a sink where probability mass proportional to the hitting time distribution 
is lost. For delta initial conditions $P^\mathrm{in}(L) = \delta(L-L_f)$ the sink is only active for $L = L_f$. For arbitrary input distributions $P^\mathrm{in}(L_f)$ the rate of probability loss is still proportional to the hitting time distribution  but now additionally to the number of sample paths which originated in an interval $P^\mathrm{in}(L) dL$.
This is exactly in accordance with the construction of $ R_{L_\mathrm{ts}}(L,\tau;\tau_0)$ in Eq.~\eqref{sup_rtsa_dist}. The second line represents the drift and the third a simple diffusion term.

This form of the Fokker Planck equation for the reverse time TSA ensemble Eq.~\eqref{eq:revt_TSA_FP}  corresponds to the Langevin equation  
\begin{align}
\label{sup_tr_lv_tsa}
  dL(\tau) =
\left(
      f(L) 
      +D 
      \  
      \frac{\partial}{\partial 
L}
 \log\left(
  \int_{L_\mathrm{ts}}^{L} dL' \;
   e^{-
  \int^{L'}
 \frac{2 f(L'')}{D} dL''
 }
 \left(
 1
 -
 \int_{L_\mathrm{ts}}^{L'} P^{\mathrm{in}}(L'') dL''
 \right) 
 \right)  
 \right)
       d\tau
      +
      \sqrt{D} \ d W_\tau
\end{align}
under Ito interpretation.

For further use, we split Eq.~\eqref{sup_tr_lv_tsa} into conceptually meaningful components. 
We call the total deterministic contributions to the force, the TSA force $f^\mathrm{TSA}(L)$. This allows to rewrite Eq.~\eqref{sup_tr_lv_tsa} as
\begin{align}
  dL(\tau) =
        f^\mathrm{TSA}(L) \: d\tau
      +
      \sqrt{D} \ d W_\tau
     \ .
\end{align}
The TSA force
\begin{align}
 f^\mathrm{TSA}(L)
 =
 f(L) + f^\mathcal{F}(L)
\end{align}
is comprised of two parts. A term which is exactly identical to the drift term of the forward Fokker Planck Eq.~\eqref{sup_fwd_fp} and points in the same direction. Second, a ``free energy force'' 
\begin{align}
\label{sup_general_def_freeE_force}
 f^\mathcal{F}(L)
 &=
D 
      \  
      \frac{\partial}{\partial 
L}
 \log\left(
  \int_{L_\mathrm{ts}}^{L} dL' \;
   e^{-
  \int^{L'}
 \frac{2 f(L'')}{D} dL''
 }
    H(L')
 \right)
\\
 &=:
 D \frac{\partial}{\partial L} \log Z(L)
\end{align}
which reverses the combined entropy production of all sample paths of the force driven system. 
The partition function $Z(L)$ is here defined as the weighted integral over the Boltzmann factor for the sign inverted forward force $f(L)$
times the
sigmoidal function
\begin{align}
\label{sup_def_H}
H(L)
=
 1
 -
 \int_{L_\mathrm{ts}}^{L} P^{\mathrm{in}}(L') dL'
 \ .
\end{align}
For forward initial conditions  $P^\mathrm{in}(L) = \delta(L-L_f)$, the term $H(L)$ evaluates to a Heaviside step function. Above $L_f$ only the term $f(L)$, identical to the forward dynamics, contributes. Below $L_f$ the free energy force contributes fully.

For stochastic dynamics, the probability of the system to be found at a certain position does not only dependent on the potential, but also an the possible number of states it can assume, i.e.~on the entropy. When studied over all possible sample path lifetimes as implied by the construction of the TSA ensemble, the potential to be reverted is not only the integral over the force but also the entropic contribution summarized over all times, i.e. (and in analogy to stochastic thermodynamics\cite{seifert2005entropy}) the single particle free energy. To revert the forward force $f(L)$ and the effect of the stochastic forcing, the free energy force must be proportional to twice the force $f(L)$. Once to compensate for the forward dynamics and once to re-direct the dynamics on the free energy surface back to the forward initial conditions. The factor 2 can either be read off from the Boltzmann factor in Eq.~\eqref{sup_general_def_freeE_force}, or more intuitively, by substituting the Einstein relation $D=\frac{2 k_B T}{\gamma}$\cite{gardiner1985handbook}. $k_B$ would the Boltzmann constant, $T$ the temperature and $\gamma$ the friction coefficient. With this substitution it becomes apparent that the contribution due to the change in free energy comes with a factor of 2.

For an arbitrary input distribution $P^\mathrm{in}(L)$, the contribution of the free energy force to the drift vanishes gradually along a sigmoidal like curve with larger values of $L$. This implies that above the bulk of initial positions forward and TSA dynamics are indistinguishable in their dynamical law. Below however the free energy force reverses the forward dynamics and adds additional terms which we will study below.

While the drift and diffusion terms are straight forward to transfer, the sink is not represented by the SDE Eq.~\eqref{sup_tr_lv_tsa} and needs to be handled with care. For diffusion limited reaction diffusion systems\cite{wilemski1973general} it is known that a Fokker-Planck equation of the form
\begin{align}
\label{sup_react_diff_with_killing}
\frac{\partial}{\partial \tau}
P(L,\tau)
  =
-
k(L,\tau) P(L,\tau)
-
     \frac{\partial}{\partial L}
    a(L)   P(L,\tau) 
    + 
    \frac{D}{2} \frac{\partial^2}{\partial L^2}  
P(L,\tau)
\end{align}
transforms to a Langevin equation with Ito interpretation and killing measure $k(L,\tau)$\cite{holcman2005survival,schuss2015brownian}. The killing measure here defines the probability $p_k = k(L,\tau)d\tau$ of a sample path to be terminated at position $L$ at time $\tau$. It must be evaluated in each timestep and for each sample path.
Comparing terms, Eq.~\eqref{sup_react_diff_with_killing} and Eq.~\eqref{eq:revt_TSA_FP} tell us, that the killing measure for the reverse time TSA ensemble is given by 
\begin{align}
\label{sup_kill_tsa_fp}
 k(L,\tau)
 =\frac{
   P^\mathrm{in}(L)
 \rho_{L_\mathrm{ts}}(\tau|L)
 }{
 R_{L_\mathrm{ts}}(L,\tau;\tau_0)
 }
 \ .
\end{align}
It defines the fraction of at time $\tau$ and position $L$ still present sample paths that must be terminated. In general, it is surprisingly hard to correctly execute the killing measure on a single trajectory level when doing simulations. For example in the case of a delta-sink, i.e. a Robbin or radiation boundary condition, it yields the wrong result to simply terminate all sample paths that cross the sink within a finite simulation step $\Delta \tau$ with probability $p_k = k(L,\tau)\Delta\tau$. Instead, the sink crossing must be modeled as a diffusion process of duration $\Delta \tau$ and the case of not seen crossings within the simulation step $\Delta \tau$ must be included. For details we refer to Erban and Chapman\cite{erban2007reactive}.

For the killing measure of reverse time TSA ensembles these problems are in a sense less severe but come with a prize. Notably, the killing measure in Eq.~\eqref{sup_kill_tsa_fp} can not be executed on a trajectory by trajectory level as it depends on the full distribution of still present sample paths $R_{L_\mathrm{ts}}(L,\tau;\tau_0)$. This however suggests to simply do counting. We determine the fraction of still present sample paths $N_\tau$, that must be killed in the space-time interval $ [L,L+\Delta L]\times[\tau,\tau+\Delta \tau]$, 
i.e.~$\frac{P^\mathrm{in}(L) \rho_{L_\mathrm{ts}}(\tau|L) dL d\tau}{N_\tau}$
and divide it by the fraction of sample paths $\frac{n_{\tau,\Delta L}}{N_\tau}$, which are actually in the interval. $n_{\tau,\Delta L}$ denotes the bin count.
For ratios larger than one, we terminate all sample paths in the interval. For ratios smaller than one, we draw uniform random numbers and compare to the ratio. Formally this is called a metropolis criterion and we obtain the empirical killing measure
\begin{align}
\label{sup_kill_tsa_fp_emp}
 k(L,\tau) \Delta \tau 
 =\min 
 \left(1,
 \frac{
   P^\mathrm{in}(L) \Delta L
 \rho_{L_\mathrm{ts}}(\tau|L) \Delta \tau 
 }{
 n_{\tau,\Delta L} \Delta L
 }
 \right)
 \ .
\end{align}
Comparing the value of the killing measure for each trajectory at time $\tau$ to an iid random variable, this criterion allows to determine for each trajectory whether is must be terminated in the current step or not.

\subsection{Derivation of the reverse time TSA Fokker Planck equation}
\label{The_derivation_of_the_reverse_time_TSA_Fokker_Planck_equation}
Our derivation starts from the definition of the TSA ensemble for an arbitrary initial distribution $R_{l_\mathrm{ts}}(L,\tau;\tau_0)$ Eq.~\eqref{sup_rtsa_dist}, here defined in forward time $t$
 \begin{align}
\label{sup_prob_full_tr_ens_fwd}
R_{\widehat{L}_\mathrm{ts}}(\widehat{L},t;t_f)
&=
\int_{\widehat{L}_\mathrm{ts}}^\infty
d\widehat{L}_0
P^\mathrm{in}(\widehat{L}_0)
\int_{-\infty}^t
dt_0
\rho_{\widehat{L}_\mathrm{ts}}(t_f|\widehat{L}_0,t_0)
R_{\widehat{L}_\mathrm{ts}}(\widehat{L},t|\widehat{L}_0,t_0;t_f)
\ .
\intertext{Using the definition of TSA sub-ensembles Eq.~\eqref{sup_bridge_trans_bound} this expression simplifies to}
\label{sup_prob_full_tr_ens_fwd_simp}
&=
\int_{\widehat{L}_\mathrm{ts}}^\infty
d\widehat{L}_0
P^\mathrm{in}(\widehat{L}_0)
\int_{-\infty}^t
dt_0
\rho_{\widehat{L}_\mathrm{ts}}(t_f|\widehat{L},t) 
P(\widehat{L},t|\widehat{L}_0,t_0)
\ .
\end{align}
To derive the corresponding Fokker Planck equation we take the time derivative on both sides of Eq.~\eqref{sup_prob_full_tr_ens_fwd} and Eq.~\eqref{sup_prob_full_tr_ens_fwd_simp} and arrive at
\begin{align}
 \frac{\partial}{\partial t}
 &R_{\widehat{L}_\mathrm{ts}}(\widehat{L},t;t_f)
=
\int_{\widehat{L}_\mathrm{ts}}^\infty
d\widehat{L}_0
P^\mathrm{in}(\widehat{L}_0)
 \frac{\partial}{\partial t}
\int_{-\infty}^t dt_0 
\rho_{\widehat{L}_\mathrm{ts}}(t_f|\widehat{L},t) 
P(\widehat{L},t|\widehat{L}_0,t_0)
\ .
\intertext{Applying the Leibniz rule leads to}
\label{eq:Leibniz_rule_derivOfrtTSA}
&=
\int_{\widehat{L}_\mathrm{ts}}^\infty
d\widehat{L}_0
P^\mathrm{in}(\widehat{L}_0)
\left(
\rho_{\widehat{L}_\mathrm{ts}}(t_f|\widehat{L},t) 
 P(\widehat{L},t|\widehat{L}_0,t)
+
\int_{-\infty}^t dt_0 
 \frac{\partial}{\partial t}
 \left(
 \rho_{\widehat{L}_\mathrm{ts}}(t_f|\widehat{L},t)
P(\widehat{L},t|\widehat{L}_0,t_0)
\right)
\right)
\intertext{or equivalently}
&=
\int_{\widehat{L}_\mathrm{ts}}^\infty
d\widehat{L}_0
P^\mathrm{in}(\widehat{L}_0)
\left(
\rho_{\widehat{L}_\mathrm{ts}}(t_f|\widehat{L},t) 
 \delta(\widehat{L}-\widehat{L}_0)
+
\int_{-\infty}^t dt_0 
 \frac{\partial}{\partial t}
R_{\widehat{L}_\mathrm{ts}}(\widehat{L},t;t_f|\widehat{L}_0,t_0)
\right)
\ .
\intertext{
The first term evaluates to the hitting time distribution times the probability to be at the initial position $\widehat{L}_0$. 
The second term defines a double integral over the dynamics of a single contributing bridge process /sub-ensemble. We obtain
}
&=
\rho_{\widehat{L}_\mathrm{ts}}(t_f|\widehat{L},t) 
P^\mathrm{in}(\widehat{L})
+
\int_{\widehat{L}_\mathrm{ts}}^\infty
d\widehat{L}_0
P^\mathrm{in}(\widehat{L}_0)
\left(
\int_{-\infty}^t dt_0 
 \frac{\partial}{\partial t}
R_{\widehat{L}_\mathrm{ts}}(\widehat{L},t;t_f|\widehat{L}_0,t_0)
\right)
\ .
\end{align}
In the next step we substitute 
$ \frac{\partial}{\partial t}
R_{\widehat{L}_\mathrm{ts}}(\widehat{L},t;t_f|\widehat{L}_0,t_0)$
with the Fokker Planck equation of the non-normalized bridge process as derived above in this section. We substitute the dynamics in the form given by the intermediate result Eq.~\eqref{sup_rt_fp_full_dist_subst}, which allows us to handle the integral over $t_0$ and $L_0$ more easily. Note, that the transition probability to reach the final state $P(\widehat{L}_f,t_f|\widehat{L},t)$ in Eq.~\eqref{sup_rt_fp_full_dist_subst} was replaced by the hitting time probability
$\rho_{\widehat{L}_\mathrm{ts}}(t_f|\widehat{L},t)$,
and
the joint probability
$P(\widehat{L},t,\widehat{L}_f,t_f|\widehat{L}_0,t_0)$ 
of the two transitions 
by the respective joint probability
$R_{\widehat{L}_\mathrm{ts}}(\widehat{L},t;t_f|\widehat{L}_0,t_0)$,
to map the general bridge process to a bridge process which ends at a target state. The resulting evolution equation reads
\begin{align}
 &\frac{\partial}{\partial t}
 R_{\widehat{L}_\mathrm{ts}}(\widehat{L},t;t_f)
=
\rho_{\widehat{L}_\mathrm{ts}}(t_f|\widehat{L},t) P^\mathrm{in}(\widehat{L})
\notag \\
     &-
     \int_{\widehat{L}_\mathrm{ts}}^\infty
d\widehat{L}_0
P^\mathrm{in}(\widehat{L}_0)
      \int_{-\infty}^t dt_0
     \frac{\partial}{\partial \widehat{L}}
    \left[
      f(\widehat{L}) 
      R_{\widehat{L}_\mathrm{ts}}(\widehat{L},t;t_f|\widehat{L}_0,t_0) 
      - D \,
      \rho_{\widehat{L}_\mathrm{ts}}(t_f|\widehat{L},t) \frac{\partial}{\partial 
\widehat{L}}
P(\widehat{L},t|\widehat{L}_0,t_0)
    \right] 
\notag \\    
    &- 
\int_{\widehat{L}_\mathrm{ts}}^\infty
d\widehat{L}_0
P^\mathrm{in}(\widehat{L}_0)
      \int_{-\infty}^t dt_0
    \frac{D}{2} \frac{\partial^2}{\partial \widehat{L}^2}   
R_{\widehat{L}_\mathrm{ts}}(\widehat{L},t;t_f|\widehat{L}_0,t_0)
\ .
\intertext{
We recall that 
$R_{\widehat{L}_\mathrm{ts}}(\widehat{L},t;t_f|\widehat{L}_0,t_0)$
is defined as 
$ \rho_{\widehat{L}_\mathrm{ts}}(t_f|\widehat{L},t)P(\widehat{L},t|\widehat{L}_0,t_0)$ 
according to Eq.~\eqref{sup_bridge_trans_bound_notNorm}. This allows us to use the defining equation  of $R_{\widehat{L}_\mathrm{ts}}(\widehat{L},t;t_f)$ Eq.~\eqref{sup_prob_full_tr_ens_fwd_simp}, and we arrive at the simplified evolution equation
}
&=
\rho_{\widehat{L}_\mathrm{ts}}(t_f|\widehat{L},t) P^\mathrm{in}(\widehat{L})
\notag \\
     &-
     \frac{\partial}{\partial \widehat{L}}
    \left[
      f(\widehat{L}) 
      R_{\widehat{L}_\mathrm{ts}}(\widehat{L},t;t_f)
      - D \,
\rho_{\widehat{L}_\mathrm{ts}}(t_f|\widehat{L},t) \frac{\partial}{\partial 
\widehat{L}}
\int_{\widehat{L}_\mathrm{ts}}^\infty 
d\widehat{L}_0
P^\mathrm{in}(\widehat{L}_0)
      \int_{-\infty}^t dt_0
P(\widehat{L},t|\widehat{L}_0,t_0)
    \right] 
\notag \\    
    &- 
    \frac{D}{2} \frac{\partial^2}{\partial \widehat{L}^2} 
 R_{\widehat{L}_\mathrm{ts}}(\widehat{L},t;t_f)
\ .
\intertext{
To obtain the guiding force and recover $R_{\widehat{L}_\mathrm{ts}}(\widehat{L},t;t_f)$ in the second drift term as well, we
multiply  $\rho_{\widehat{L}_\mathrm{ts}}(t_f|\widehat{L},t)$  by 
$ \int_{\widehat{L}_\mathrm{ts}}^\infty d\widehat{L}_0 P^\mathrm{in}(\widehat{L}_0) \int_{-\infty}^t dt_0 P(\widehat{L},t|\widehat{L}_0,t_0)$ both in nominator and denominator
}
 &=
 \rho_{\widehat{L}_\mathrm{ts}}(t_f|\widehat{L},t) \delta(\widehat{L}-\widehat{L}_0) 
\notag \\
     &-\frac{\partial}{\partial \widehat{L}}
    \left[
      f(\widehat{L}) 
       R_{\widehat{L}_\mathrm{ts}}(\widehat{L},t;t_f)
\right.
\\
& \left.
      - 
\frac{
D \int_{\widehat{L}_\mathrm{ts}}^\infty d\widehat{L}_0 P^\mathrm{in}(\widehat{L}_0) \int_{-\infty}^t dt_0 P(\widehat{L},t|\widehat{L}_0,t_0)
\rho_{\widehat{L}_\mathrm{ts}}(t_f|\widehat{L},t)
}{
 \int_{\widehat{L}_\mathrm{ts}}^\infty d\widehat{L}_0 P^\mathrm{in}(\widehat{L}_0) \int_{-\infty}^t dt_0 P(\widehat{L},t|\widehat{L}_0,t_0)
}
\frac{\partial}{\partial 
\widehat{L}}
\int_{\widehat{L}_\mathrm{ts}}^\infty 
d\widehat{L}_0
P^\mathrm{in}(\widehat{L}_0)
      \int_{-\infty}^t dt_0
P(\widehat{L},t|\widehat{L}_0,t_0)
    \right] 
\notag \\    
    &- \frac{D}{2} \frac{\partial^2}{\partial \widehat{L}^2} 
 R_{\widehat{L}_\mathrm{ts}}(\widehat{L},t;t_f)
 \ .
\intertext{
The nominator then evaluates to $D R_{\widehat{L}_\mathrm{ts}}(\widehat{L},t;t_f)$,  using again Eq.~\eqref{sup_prob_full_tr_ens_fwd_simp}. The denominator together with the rest of the expression can then be simplified to the
form of a guiding force with $L_0$ and $t_0$ integrated out from the guiding probability. We arrive at our first intermediate result
}
\label{sup_full_tsa_pfwd}
&=
\rho_{\widehat{L}_\mathrm{ts}}(t_f|\widehat{L},t) P^\mathrm{in}(\widehat{L})
\notag \\
     &-
     \frac{\partial}{\partial \widehat{L}}
    \left[
      f(\widehat{L}) 
      - 
D
\frac{\partial}{\partial \widehat{L}}
\log 
\left(
\int_{\widehat{L}_\mathrm{ts}}^\infty 
d\widehat{L}_0
P^\mathrm{in}(\widehat{L}_0)
      \int_{-\infty}^t dt_0
P(\widehat{L},t|\widehat{L}_0,t_0)
\right)
    \right] 
    R_{\widehat{L}_\mathrm{ts}}(\widehat{L},t;t_f)
\notag \\    
    &- 
    \frac{D}{2} \frac{\partial^2}{\partial \widehat{L}^2} 
 R_{\widehat{L}_\mathrm{ts}}(\widehat{L},t;t_f)
%
\end{align}
in the derivation of the full reverse time TSA Fokker-Planck equation Eq.~\eqref{eq:revt_TSA_FP}.
Both this intermediate result and the dynamics for the free reverse time bridge Eq.~\eqref{sup_tr_fp} depend on the solution of the forward Fokker Planck $P(\widehat{L},t|\widehat{L}_0,t_0)$. Unfortunately, explicit solutions for $P(\widehat{L},t|\widehat{L}_0,t_0)$ are only known for very few cases. Interestingly, the above transformations have shown that it is not 
$P(\widehat{L},t|\widehat{L}_0,t_0)$
but
\begin{align}
\label{sup_cfb_density}
Q(\widehat{L})
:=
\lambda
\int_{\widehat{L}_\mathrm{ts}}^{\infty } d\widehat{L}_0 \; P^\mathrm{in}(\widehat{L}_0)
 \int_{-\infty}^t dt_0
P(\widehat{L},t|\widehat{L}_0,t_0)
\end{align}
we need an expression for. 
We introduce $\lambda$ as a normalization factor to treat $Q(\widehat{L})$ as a normalized distribution. The normalization is necessary as $P(\widehat{L},t|\widehat{L}_0,t_0)$ is implied with absorbing boundary conditions and decays with the hitting time distribution, as discussed above.

To derive an explicit expression for this integral we start
with the forward Fokker-Planck equation
\begin{align}
 \frac{\partial}{\partial t}  P(\widehat{L},t|\widehat{L}_0,t_0)
 = 
 - \frac{\partial}{\partial \widehat{L}} \left( f(\widehat{L}) P(\widehat{L},t|\widehat{L}_0,t_0) \right)
 + \frac{D}{2} \frac{\partial^2}{\partial \widehat{L}^2}
 P(\widehat{L},t|\widehat{L}_0,t_0)
\end{align}
and integrate out the dependence on the initial time $t_0$
\begin{align}
 \int_{-\infty}^t dt_0
 \frac{\partial}{\partial t}  P(\widehat{L},t|\widehat{L}_0,t_0)
 = 
  \int_{-\infty}^t dt_0
  \left(
 - \frac{\partial}{\partial \widehat{L}} \left( f(\widehat{L}) P(\widehat{L},t|\widehat{L}_0,t_0) \right)
 + \frac{D}{2} \frac{\partial^2}{\partial \widehat{L}^2} 
 P(\widehat{L},t|\widehat{L}_0,t_0)
 \right)
 \ .
\end{align}
To evaluate the left hand side we change to an integral over the distance $s=t-t_0$ and adapt the derivative accordingly. The resulting integral
\begin{align}
 \int_{-\infty}^t dt_0
 \frac{\partial}{\partial t}  P(\widehat{L},t|\widehat{L}_0,t_0)
 = 
 - \int_{\infty}^0 ds
 \frac{\partial}{\partial s}  P(\widehat{L},s|\widehat{L}_0,0) = -P(\widehat{L},0|\widehat{L}_0,0) 
 =
 -\delta(\widehat{L}-\widehat{L}_0) 
\end{align}
evaluates to minus the delta initial conditions of the forward process.
The $+\infty$ term vanishes due to the natural boundary condition we impose. We define the non-equilibrium stationary probability density with delta insertion probability at $\widehat{L}_0$
\begin{align}
Q(\widehat{L}|\widehat{L}_0):=
\lambda
\int_{-\infty}^t dt_0 P(\widehat{L},t|\widehat{L}_0,t_0) 
=
 - 
 \lambda
 \int_{\infty}^0 ds
   P(\widehat{L},s|\widehat{L}_0,0)
= Q(\widehat{L},\infty|\widehat{L}_0,0)
\ .
\end{align}
The generalization $Q(\widehat{L})$ Eq.~\eqref{sup_cfb_density} to arbitrary insertion probabilities is obtained by multiplication with $P^\mathrm{in}(\widehat{L}_0)$ and subsequent integration over $\widehat{L}_0$. 
The resulting ordinary differential equation then reads
\begin{align}
\label{ness_ode}
 - 
 \lambda
 P^\mathrm{in}(\widehat{L}) 
 &= 
 - \frac{\partial}{\partial \widehat{L}} \left( f(\widehat{L}) Q(\widehat{L}) \right)
 + \frac{D}{2} \frac{\partial^2}{\partial \widehat{L}^2} Q(\widehat{L})
 \notag \\
 Q(\widehat{L}_\mathrm{ts})&=0
 \ .
\end{align}
Integrating Eq.~\eqref{ness_ode} over the full interval $\widehat{L} \in [\widehat{L}_\mathrm{ts},\infty]$ and with
$
 \int_{\widehat{L}_\mathrm{ts} }^{\infty}
 P^\mathrm{in}(\widehat{L'}) \ d\widehat{L'} =1
 $
we find
\begin{align}
\lambda
 = 
 \frac{D}{2} \frac{\partial}{\partial \widehat{L}}  Q(\widehat{L}_\mathrm{ts}) 
\end{align}
For completeness note, that $\lambda$, as already discussed in Zhang et.~al.~\cite{zhang2011reconstructing}, defines a steady state probability flux at $\widehat{L}_\mathrm{ts}$.
Importantly equation Eq.~\eqref{ness_ode} can be solved exactly. For completness we here reproduce the result given in Zhang et.~al.~\cite{zhang2011reconstructing}.

To solve the ordinary differential equation Eq.~\eqref{ness_ode} for $Q(\widehat{L})$ we first integrate over $\widehat{L}$
\begin{align}
 -
 \left. f(\widehat{L'}) Q(\widehat{L'}) \right|_{\widehat{L'}_\mathrm{ts}}^{\widehat{L}}
 +
 \left.
 \frac{D}{2}
 \frac{\partial^2}{\partial \widehat{L}^2}
  Q(\widehat{L'})
  \right|_{\widehat{L}_\mathrm{ts}}^{\widehat{L}}
 =
 -
 \lambda
\int_{\widehat{L}_\mathrm{ts}}^{\widehat{L}} 
P^{\mathrm{in}}(\widehat{L'}) d\widehat{L'}
\end{align}
and obtain
\begin{equation}
\label{eq:ode_for_Q_const_D}
 -
 f(\widehat{L}) Q(\widehat{L})
 +
 \frac{D}{2}
 \frac{\partial}{\partial \widehat{L}}
  Q(\widehat{L})
 =
 \lambda
 \left(
 1
 - 
 \int_{\widehat{L}_\mathrm{ts}}^{\widehat{L}} 
 P^{\mathrm{in}}(\widehat{L'}) d\widehat{L'}
 \right)
 \ .
\end{equation}
Combining the terms on the left hand side into a single expression yields
\begin{align}
 \frac{D}{2} 
  e^{
  \int^{\widehat{L}}
 \frac{2 f(\widehat{L'})}{D} d\widehat{L'}
 }
 \frac{\partial}{\partial \widehat{L}}
 \left(
 Q(\widehat{L})
 e^{
  -
  \int^{\widehat{L}}
 \frac{2 f(\widehat{L'})}{D} d\widehat{L'}
 }
 \right)
 &=
 \lambda
 \left(
 1
 -
 \int_{\widehat{L}_\mathrm{ts}}^{\widehat{L}} P^{\mathrm{in}}(\widehat{L'}) d\widehat{L'}
 \right) 
 \ .
\end{align}
In the last step we again integrate from $\widehat{L}_\mathrm{ts}$ to $\widehat{L}$ and  solve for  $Q(\widehat{L})$ to obtain
\begin{align}
 Q(\widehat{L})
 &=
 \frac{2\lambda}{D}
 \;
   e^{
    \int^{\widehat{L}}
  \frac{2 f(\widehat{L'})}{D} d\widehat{L'}
  }
 \int_{\widehat{L}_\mathrm{ts}}^{\widehat{L}} d\widehat{L'} \;
   e^{-
  \int^{\widehat{L'}}
 \frac{2 f(\widehat{L''})}{D} d\widehat{L''}
 }
 \left(
 1
 -
 \int_{\widehat{L}_\mathrm{ts}}^{\widehat{L'}} P^{\mathrm{in}}(\widehat{L''}) d\widehat{L''}
 \right) 
  \ ,
\end{align}
which is the second intermediate result.

In the final step of the derivation of the full reverse time TSA dynamics we plug the obtained result for $Q(\widehat{L})$ into Eq.~\eqref{sup_full_tsa_pfwd}. 
Using the obtained result for $Q(\widehat{L})$, and $P^{\mathrm{in}}(\widehat{L})$ as normalized distribution, we find for the guiding force
\begin{align}
 D \frac{\partial}{\partial \widehat{L}} \log  Q(\widehat{L})
 =
 2 f(\widehat{L}) 
 +
 D \log\left(
  \int_{\widehat{L}_\mathrm{ts}}^{\widehat{L}} d\widehat{L'} \;
   e^{-
  \int^{\widehat{L'}}
 \frac{2 f(\widehat{L''})}{D} d\widehat{L''}
 }
 \left(
 1
 -
 \int_{\widehat{L}_\mathrm{ts}}^{\widehat{L'}} P^{\mathrm{in}}(\widehat{L''}) d\widehat{L''}
 \right) 
 \right)
 \ .
\end{align}
With this, we obtain the final reverse time Fokker Planck written in forward time
\begin{align}
&\frac{\partial}{\partial t}
R_{\widehat{L}_\mathrm{ts}}(\widehat{L},t;t_f;\widehat{L}_0)
  =
  P^\mathrm{in}(\widehat{L})
 \rho_{\widehat{L}_\mathrm{ts}}(t_f|\widehat{L},t) 
\notag \\
 &+
     \frac{\partial}{\partial \widehat{L}}
     \left(\!
    \left[
      f(\widehat{L}) 
      +D
      \  
      \frac{\partial}{\partial 
\widehat{L}}
 \log \! \left(
  \int_{\widehat{L}_\mathrm{ts}}^{\widehat{L}} d\widehat{L'} \;
   e^{-
  \int^{\widehat{L'}}
 \frac{2 f(\widehat{L''})}{D} d\widehat{L''}
 }
 \left(
 1
 -
 \int_{\widehat{L}_\mathrm{ts}}^{\widehat{L'}} P^{\mathrm{in}}(\widehat{L''}) d\widehat{L''}
 \right) 
 \right)
   \! \right] 
     R_{\widehat{L}_\mathrm{ts}}(\widehat{L},t;t_f;\widehat{L}_0)
     \!
     \right)
\notag \\    
    &- \frac{D}{2} \frac{\partial^2}{\partial \widehat{L}^2} 
R_{\widehat{L}_\mathrm{ts}}(\widehat{L},t;t_f;\widehat{L}_0)
\ .
\end{align}
To revert the time axis we set $\tau=t_f-t$, $L_f = \widehat{L}_0$ and obtain the full reverse time TSA Fokker-Planck equation as stated in Eq.~\eqref{eq:revt_TSA_FP}.

\section{Normalized TSA dynamics}
\label{the_renormalized_TSA_dynamics}
The TSA ensemble $R_{L_\mathrm{ts}}(L,\tau;\tau_0)$, derived in section \ref{time_to_completion_as_stochastic_variable}, 
is the exact mathematical description of an ensemble of target state aligned trajectories. In reverse time more and more trajectories reach the end of their lifetime and stop to contribute to the ensemble. At $\tau \to \infty$ the ensemble vanishes.
To calculate the moments of this ensemble, averages however must be taken with respect to a for each time $\tau$ normalized distribution. This normalized TSA ensemble then reads
\begin{align}
 \label{def_renorm_TSA}
 R^N_{L_\mathrm{ts}}(L,\tau;\tau_0)
 =
 \frac{
 R_{L_\mathrm{ts}}(L,\tau;\tau_0)
 }{
 \int_{L_\mathrm{ts}}^\infty 
 dL
 \;
 R_{L_\mathrm{ts}}(L,\tau;\tau_0)
 }
\end{align}

In this section we show that Eq.~\eqref{def_renorm_TSA} can not only be obtained subsequently by normalization, but also directly from a constitutive Fokker-Planck equation. We first observe that, using the definition of $R_{L_\mathrm{ts}}(L,\tau;\tau_0)$ as stated in Eq.~\eqref{sup_rtsa_dist} and integrating over $L$, the normalized TSA ensemble can be written as
\begin{align}
 \label{def_renorm_TSA_with_integrals}
 R^N_{L_\mathrm{ts}}(L,\tau;\tau_0)
 =
 \frac{
 R_{L_\mathrm{ts}}(L,\tau;\tau_0)
 }{
 \int_{L_\mathrm{ts}}^\infty P^\mathrm{in}(L_f)
 \int_\tau^\infty d\tau_f \rho_{L_\mathrm{ts}}(\tau_0|L_f,\tau_f)
 }
 \ .
\end{align}
Using this definition of the normalized TSA ensemble we can derive a constitutive Fokker-Planck equation of the form
\begin{align}
\label{eq:revt_TSA_FP_renorm}
&\frac{\partial}{\partial \tau}
R^N_{L_\mathrm{ts}}(L,\tau;\tau_0)
  =
  -
\frac{
\rho_{L_\mathrm{ts}}(\tau_0|L,\tau) 
P^\mathrm{in}(L)
-  
R^N_{L_\mathrm{ts}}(L,\tau;\tau_0)
\int_{L_\mathrm{ts}}^\infty
dL_f
P^\mathrm{in}(L_f)
\rho_{L_\mathrm{ts}}(\tau_0|L_f,\tau)
}{
\int_{L_\mathrm{ts}}^\infty
dL_f
P^\mathrm{in}(L_f)
\int_\tau^{\infty} d\tau_f 
\rho_{L_\mathrm{ts}}(\tau_0|L_f,\tau_f) 
}
\notag \\
 &-
     \frac{\partial}{\partial L}
     \left(
    \left[
       f(L) 
      +D
      \  
      \frac{\partial}{\partial 
L}
 \log\left(
  \int_{L_\mathrm{ts}}^{L} dL' \;
   e^{-
  \int^{L'}
 \frac{2 f(L'')}{D} dL''
 }
 \left(
 1
 -
 \int_{L_\mathrm{ts}}^{L'} P^{\mathrm{in}}(L'') dL''
 \right) 
 \right)
    \right] 
     R^N_{L_\mathrm{ts}}(L,\tau;\tau_0)
     \right)
\notag \\    
    &+ \frac{D}{2} \frac{\partial^2}{\partial L^2}  
R^N_{L_\mathrm{ts}}(L,\tau;\tau_0)
\ .
\end{align}
The detailed derivation is given below.
Importantly, the dynamics of the with $\tau$-decaying TSA ensemble Eq.~\eqref{eq:revt_TSA_FP} and the normalized ensemble Eq.~\eqref{eq:revt_TSA_FP_renorm} are identical. They differ however in the sink. While the un-normalized TSA dynamics looses probability mass according to their forward initial and lifetime statistics, is the normalized ensemble probability mass preserving. The same amount of trajectories that are terminated are simultaneously created. Sink and source however differ in their position. While the sink is positioned according to the forward initial distribution is the source proportional to the normalized TSA ensemble at time $\tau$ and the instantaneous change in the normalization. The whole expression is then evaluated with respect to the TSA ensemble normalization.

Under these considerations, the Langevin equations of both the not- and normalized ensemble are identical. While the killing measures are
identical in their location, the normalized ensemble features an additional source term that compensates for all terminated sample paths. Reverse time trajectories in this ensemble thus not only start at $\tau=\tau_0$, but over the whole time course of the reverse time dynamics.
In analogy to the discussion in section \ref{time_to_completion_as_stochastic_variable}, the killing measure reads
\begin{align}
  k(L,\tau)
 =\frac{
   P^\mathrm{in}(L)
 \rho_{L_\mathrm{ts}}(\tau_0|L,\tau)
 }{
 R_{L_\mathrm{ts}}(L,\tau;\tau_0)
 }
 \ .
\end{align}
Note, that the normalization of the sink and of the normalized ensemble cancel and only the un-normalized distribution $R_{L_\mathrm{ts}}(L,\tau;\tau_0)$ remains. In terms of the killing measure, normalized and un-normalized TSA Langevin dynamics are thus virtually identical.
For the normalized TSA ensemble additionally a source of the form
\begin{align}
s(L,\tau)
=
 \frac{
\int_{L_\mathrm{ts}}^\infty
dL_f
P^\mathrm{in}(L_f)
\rho_{L_\mathrm{ts}}(\tau_0|L_f,\tau)
}{
\int_{L_\mathrm{ts}}^\infty
dL_f
P^\mathrm{in}(L_f)
\int_\tau^{\infty} d\tau_f 
\rho_{L_\mathrm{ts}}(\tau_0|L_f,\tau_f) 
}
 \ .
\end{align}
occurs. We note, that the source is independent of the position $L$ and equals the instantaneous change of the normalization divided by the normalization.

The normalized TSA ensemble is clearly useful to calculate moments. From a conceptual perspective however an ensemble which creates sample paths  ``on the fly'' is clearly not the ensemble which matches the experimental protocol of target state alignment. It thus depends on the question we ask, whether the normalized ensemble could profitable be used. With the exception of the following derivation, in the rest of this chapter we always refer to the un-normalized case and indicate if a normalized version is used.

\subsection{Derivation of the normalized TSA dynamics}
\label{The_derivation_of_the_renormalized_TSA_dynamics}
In this section we derive the normalized TSA dynamics. We start from the definition of the normalized ensemble Eq.~\eqref{def_renorm_TSA} written in forward time
\begin{align}
 R^N_{\widehat{L}_\mathrm{ts}}(\widehat{L},t;t_f)
 &=
\frac{
R_{\widehat{L}_\mathrm{ts}}(\widehat{L},t;t_f)
}{
\int_{\widehat{L}_\mathrm{ts}}^\infty
d\widehat{L}_0
P^\mathrm{in}(\widehat{L}_0)
\int_{-\infty}^t dt_0 
\rho_{\widehat{L}_\mathrm{ts}}(t_f|\widehat{L}_0,t_0) 
}
\ .
 \notag \\
 \intertext{We then substitute the definition Eq.\eqref{sup_prob_full_tr_ens_fwd} of the un-normalized ensemble}
 &=
 \int_{\widehat{L}_\mathrm{ts}}^\infty
d\widehat{L}_0
P^\mathrm{in}(\widehat{L}_0)
\int_{-\infty}^t dt_0 
\frac{
\rho_{\widehat{L}_\mathrm{ts}}(t_f|\widehat{L},t) 
P(\widehat{L},t|\widehat{L}_0,t_0)
}{
\int_{\widehat{L}_\mathrm{ts}}^\infty
d\widehat{L}_0
P^\mathrm{in}(\widehat{L}_0)
\int_{-\infty}^t dt_0 
\rho_{\widehat{L}_\mathrm{ts}}(t_f|\widehat{L}_0,t_0) 
}
\ ,
\end{align}
to obtain our starting point expression.
In forward time the time-derivative of the normalized ensemble reads
\begin{align}
 &\frac{\partial}{\partial t}
 R^N_{\widehat{L}_\mathrm{ts}}(\widehat{L},t;t_f)
=
\frac{\partial}{\partial t}
\int_{\widehat{L}_\mathrm{ts}}^\infty
d\widehat{L}_0
P^\mathrm{in}(\widehat{L}_0)
\int_{-\infty}^t dt_0 
\frac{
\rho_{\widehat{L}_\mathrm{ts}}(t_f|\widehat{L},t) 
P(\widehat{L},t|\widehat{L}_0,t_0)
}{
\int_{\widehat{L}_\mathrm{ts}}^\infty
d\widehat{L}_0
P^\mathrm{in}(\widehat{L}_0)
\int_{-\infty}^t dt_0 
\rho_{\widehat{L}_\mathrm{ts}}(t_f|\widehat{L}_0,t_0) 
}
\ .
\intertext{Applying the Leibniz rule we find}
&=
\frac{
\int_{\widehat{L}_\mathrm{ts}}^\infty
d\widehat{L}_0
P^\mathrm{in}(\widehat{L}_0)
\rho_{\widehat{L}_\mathrm{ts}}(t_f|\widehat{L},t) 
 P(\widehat{L},t|\widehat{L}_0,t)
}{
\int_{\widehat{L}_\mathrm{ts}}^\infty
d\widehat{L}_0
P^\mathrm{in}(\widehat{L}_0)
\int_{-\infty}^t dt_0 
\rho_{\widehat{L}_\mathrm{ts}}(t_f|\widehat{L}_0,t_0) 
}
\notag \\
&+
\int_{\widehat{L}_\mathrm{ts}}^\infty
d\widehat{L}_0
P^\mathrm{in}(\widehat{L}_0)
\int_{-\infty}^t dt_0 
\frac{\partial}{\partial t}
\frac{
 \rho_{\widehat{L}_\mathrm{ts}}(t_f|\widehat{L},t)
P(\widehat{L},t|\widehat{L}_0,t_0)
}{
\int_{\widehat{L}_\mathrm{ts}}^\infty
d\widehat{L}_0
P^\mathrm{in}(\widehat{L}_0)
\int_{-\infty}^t dt_0 
\rho_{\widehat{L}_\mathrm{ts}}(t_f|\widehat{L}_0,t_0) 
}
\ .
\intertext{Recognizing $P(\widehat{L},t|\widehat{L}_0,t)$ in the first term as a delta function and evaluating the time derivative in the second term then leads to}
&=
\frac{
\int_{\widehat{L}_\mathrm{ts}}^\infty
d\widehat{L}_0
P^\mathrm{in}(\widehat{L}_0)
\rho_{\widehat{L}_\mathrm{ts}}(t_f|\widehat{L},t) 
 \delta(\widehat{L}-\widehat{L}_0)
}{
\int_{\widehat{L}_\mathrm{ts}}^\infty
d\widehat{L}_0
P^\mathrm{in}(\widehat{L}_0)
\int_{-\infty}^t dt_0 
\rho_{\widehat{L}_\mathrm{ts}}(t_f|\widehat{L}_0,t_0) 
}
\notag \\
&- 
\int_{\widehat{L}_\mathrm{ts}}^\infty
d\widehat{L}_0
P^\mathrm{in}(\widehat{L}_0)
\int_{-\infty}^t dt_0 
\frac{
 \rho_{\widehat{L}_\mathrm{ts}}(t_f|\widehat{L},t)
P(\widehat{L},t|\widehat{L}_0,t_0)
%
\int_{\widehat{L}_\mathrm{ts}}^\infty
d\widehat{L}_0
P^\mathrm{in}(\widehat{L}_0)
\rho_{\widehat{L}_\mathrm{ts}}(t_f|\widehat{L}_0,t) 
}{
\left(
\int_{\widehat{L}_\mathrm{ts}}^\infty
d\widehat{L}_0
P^\mathrm{in}(\widehat{L}_0)
\int_{-\infty}^t dt_0 
\rho_{\widehat{L}_\mathrm{ts}}(t_f|\widehat{L}_0,t_0) 
\right)^2
}
\notag \\
&+
\int_{\widehat{L}_\mathrm{ts}}^\infty
d\widehat{L}_0
P^\mathrm{in}(\widehat{L}_0)
\int_{-\infty}^t dt_0 
\frac{
\frac{\partial}{\partial t}
 \rho_{\widehat{L}_\mathrm{ts}}(t_f|\widehat{L},t)
P(\widehat{L},t|\widehat{L}_0,t_0)
}{
\int_{\widehat{L}_\mathrm{ts}}^\infty
d\widehat{L}_0
P^\mathrm{in}(\widehat{L}_0)
\int_{-\infty}^t dt_0 
\rho_{\widehat{L}_\mathrm{ts}}(t_f|\widehat{L}_0,t_0) 
}
\ .
\intertext{We next evaluate the delta-function in the first term. The resulting term can be identified with the sink in the non-normalized reverse time TSA, here divided by the TSA-normalization. The second term can be associated as a source that ensures that the reverse time TSA stays normalized. We substitute 
$R^N_{\widehat{L}_\mathrm{ts}}(\widehat{L},t;t_f)$ as defined in 
Eq.~\eqref{def_renorm_TSA_with_integrals} where possible.
The third term describes the dynamics of the TSA ensemble as in the non-normalized ensemble Eq.~\eqref{eq:Leibniz_rule_derivOfrtTSA}, divided by the normalization. The normalized reverse time SDE written in forward time $t$ then reads}
&=
\frac{
%
%
\rho_{\widehat{L}_\mathrm{ts}}(t_f|\widehat{L},t) 
P^\mathrm{in}(\widehat{L})
-  
R^N_{\widehat{L}_\mathrm{ts}}(\widehat{L},t;t_f)
\int_{\widehat{L}_\mathrm{ts}}^\infty
d\widehat{L}_0
P^\mathrm{in}(\widehat{L}_0)
\rho_{\widehat{L}_\mathrm{ts}}(t_f|\widehat{L}_0,t)
%
%
}{
\int_{\widehat{L}_\mathrm{ts}}^\infty
d\widehat{L}_0
P^\mathrm{in}(\widehat{L}_0)
\int_{-\infty}^t dt_0 
\rho_{\widehat{L}_\mathrm{ts}}(t_f|\widehat{L}_0,t_0) 
}
\notag \\
 &+
     \frac{\partial}{\partial \widehat{L}}
     \left(
    \left[
      f(\widehat{L}) 
      +D 
      \  
      \frac{\partial}{\partial 
\widehat{L}}
 \log\left(
  \int_{\widehat{L}_\mathrm{ts}}^{\widehat{L}} d\widehat{L}' \;
   e^{-
  \int^{\widehat{L}'}
 \frac{2 f(\widehat{L}'')}{D} d\widehat{L}''
 }
 \left(
 1
 -
 \int_{\widehat{L}_\mathrm{ts}}^{\widehat{L}'} P^{\mathrm{in}}(\widehat{L}'') d\widehat{L}''
 \right) 
 \right)
    \right] 
    R^N_{\widehat{L}_\mathrm{ts}}(\widehat{L},t;t_f)
     \right)
\notag \\    
    &- \frac{D}{2} \frac{\partial^2}{\partial \widehat{L}^2}
    R^N_{\widehat{L}_\mathrm{ts}}(\widehat{L},t;t_f)
\ .
\end{align}
Switching from forward time to the reverse time description we arrive at the final expression stated in Eq.~\eqref{eq:revt_TSA_FP_renorm}.

\section{Exactly solvable reverse time dynamics}
\label{Exactly_solvable_reverse_time_dynamics}
Our construction of the reverse time TSA ensemble in section \ref{time_to_completion_as_stochastic_variable} has revealed three conceptually different approaches to obtain information on its distribution and dynamics.
First, the TSA ensemble can be obtained from its definition Eq.~\eqref{sup_prob_full_tr_ens_fwd}. Given both the transition probability and the hitting time distribution of the forward problem are known we can construct the non-normalized bridge Eq.~\eqref{sup_bridge_trans_bound_notNorm}. Integration over all possible initial times and switching to the reverse time notation then leads to the reverse time TSA ensemble ending at $L_f$. If an initial distribution is known, it is easy to generalize the TSA ensemble to include a distribution of forward initial conditions. Below we discuss one example of this type.

The second method to obtain the reverse time TSA ensemble is based on sampling the reverse time dynamics. We derive an analytic expression for the reverse time sub-ensemble SDE Eq.~\eqref{sup_tr_lv} and synthesize sample paths with duration $T$ drawn from the forward lifetime time distribution. Three analytically tractable cases are shown below.

The third approach is based on directly solving the full reverse time TSA Fokker Planck equation Eq.~\eqref{eq:revt_TSA_FP}. We discuss one example.

\subsection{Constructing the TSA ensemble from its definition -- the random walk}
\label{Constructing_the_TSA_ensemble_from_its_definition}
TSA ensembles can be constructed from their defining equation Eq.~\eqref{sup_prob_full_tr_ens_fwd}, which demands to integrate over $t_0$ with respect to  the non-normalized bridge process.  We demonstrate this construction for the simple case of a target state aligned Gaussian random walk, or equivalently the diffusion process with $f(\widehat{L})=0$.
The transition probability of a diffusion process
with absorbing boundary conditions at $\widehat{L}_\mathrm{ts}=0$ is
\begin{align}
\label{sup_transProb_diffusion_absbound}
P(\widehat{L},t|\widehat{L}_0,t_0) 
=
 \frac{e^{-\frac{(\widehat{L}-\widehat{L}_0)^2}{2 D (t-t_0)}}-e^{-\frac{(\widehat{L}+\widehat{L}_0)^2}{2 D (t-t_0)}}}{ \sqrt{2 \pi D (t-t_0)}}
\end{align}
and can easily be obtained by using the technique of mirror charges. The lifetime (or hitting) time distribution is
\begin{align}
\label{eq:rw_hit_dist}
\rho_0(t_f|\widehat{L},t)=
 \frac{\widehat{L} e^{-\frac{\widehat{L}^2}{2 D (t_f-t)}}}{ \sqrt{2 \pi D (t_f-t)^3}}
\end{align}
and
follows directly from Eq.~\eqref{eq:def_hit_dist}.
Combining these two terms into the non-normalized bridge Eq.~\eqref{sup_bridge_trans_bound_notNorm}, and integrating over $t_0$ according to Eq.~\eqref{sup_prob_full_tr_ens_fwd},  we obtain
\begin{align}
\label{eq:tsa_rw_delta_in}
R_{0}(L,\tau;0|L_f)
=
 \frac{L (L+L_f -\left| L-L_f\right|) e^{-\frac{L^2}{2 D \tau }}}{\sqrt{2 \pi  D \tau
   ^3}}
   \ ,
\end{align}
here written in reverse time $\tau$ and $L_f=\widehat{L}_0$. The initial time of the reverse time process is set to zero.
It is the solution of the target state aligned  diffusion process with delta distributed forward initial probability at $P^\mathrm{in}(L)=\delta(L-L_f)$. Integrating over $L$ we obtain the cumulative of the hitting time distribution of the random walk.
The probability mass of the TSA ensemble changes proportional to the hitting time distribution Eq.~\eqref{eq:rw_hit_dist} and thus with the rate enforced by the sink in the full reverse time TSA Fokker-Planck equation Eq.~\eqref{eq:revt_TSA_FP}.

The Fokker-Planck equation for the reverse time TSA ensemble Eq.~\eqref{eq:revt_TSA_FP} can be used to directly verify this result by insertion. As the process under consideration is a random walk, for the forward force holds $f(L)=0$. 
We briefly discuss this substitution for the equivalent and easier to read TSA Langevin equation Eq.~\eqref{sup_tr_lv_tsa} and therein the free energy force Eq.~\eqref{sup_general_def_freeE_force}.
Substituting $f(L)=0$ into Eq.~\eqref{sup_general_def_freeE_force}, the exponential simplifies to one. With delta distributed forward initial conditions, $H(L)$ becomes a step function with jump at $L=L_f$. Below $L_f$, the step function $H(L)$ is one and the integral inside the logarithm evaluates to $L$. For $L>L_f$ we find $H(L)=0$ and thus no contribution of the free energy force to the TSA dynamics. For $L \neq L_f$, the resulting Langevin equation is
\begin{align}
\label{sup_tr_lv_rw_tsa_simp}
  dL(\tau) =
  \begin{cases}
\frac{D}{L} \;
       d\tau \
      + \;
      &\sqrt{D} \ d W_\tau
      \qquad \mathrm{for} \ \ L < L_f
      \\
      &\sqrt{D} \ d W_\tau
      \qquad \mathrm{for} \ \ L > L_f
    \ .
  \end{cases}
\end{align}
The drift term of the Fokker Planck equation is thus strongly simplified  and Eq.~\eqref{eq:tsa_rw_delta_in} can be checked by insertion. Note, that the TSA force law 
assumes the important form $\frac{D}{L}$ for $L < L_f$ that we will meet time and again in the following considerations. For $L = L_f$ the reverse time TSA Fokker-Planck Eq.~\eqref{eq:revt_TSA_FP} assumes a more complicated form as we must evaluate the delta initial conditions of the forward process $P^\mathrm{in}=\delta\left(L-L_f\right)$ inside the free energy force, which yields a non-analytical expression. To circumvent this problem, and to demonstrate the validity of Eq.~\eqref{eq:revt_TSA_FP} for non-delta distributed forward initial conditions, we mix the TSA ensemble solutions with delta insertion Eq.~\eqref{eq:tsa_rw_delta_in}, using a truncated Gaussian insertion probability
\begin{align}
P^\mathrm{in}(L_f)
=
 \frac{\sqrt{\frac{2}{\pi }} e^{-\frac{(L_\mathrm{in}-L_f)^2}{2 \sigma ^2}}}{\sigma 
   \left(\text{erf}\left(\frac{L_\mathrm{in}}{\sqrt{2} \sigma }\right)+1\right)}
\end{align}
normalized to the interval $L_f \in [0,\infty]$, and marginalize over $L_f$. We obtain the TSA reverse time ensemble 
\begin{align}
&R_{0}(L,\tau;0)
=
%
 e^{-\frac{L^2 \left(\frac{\sigma ^2}{D \tau
   }+1\right)+2 L_\mathrm{in}^2}{2 \sigma ^2}} 
%
   \notag \\
 &\times
 \frac{L e^{\frac{L^2+L_\mathrm{in}^2}{2 \sigma ^2}} \left(\sqrt{2 \pi } e^{\frac{L_\mathrm{in}^2}{2 \sigma ^2}}
   \left((L_\mathrm{in}-L) \text{erf}\left(\frac{L-L_\mathrm{in}}{\sqrt{2} \sigma }\right)+L_\mathrm{in}
   \text{erf}\left(\frac{L_\mathrm{in}}{\sqrt{2} \sigma }\right)+L\right)+2 \sigma \right)-2 \sigma  L
   e^{\frac{L_\mathrm{in} (2 L+L_\mathrm{in})}{2 \sigma ^2}} }{\pi  (D \tau )^{3/2}
   \left(\text{erf}\left(\frac{L_\mathrm{in}}{\sqrt{2} \sigma }\right)+1\right)}
\end{align}
for diffusive target search starting with Gaussian insertion probability. This expression satisifes Eq.~\eqref{eq:revt_TSA_FP} with $f(L)=0$.

\subsection{Sampling TSA ensembles via reverse time random bridges}
\label{Sampling_TSA_ensembles_from_reverse_time_bridges}
In this section, we discuss three different example cases for which the full TSA ensemble is explicitly constructed from fixed lifetime sub-ensembles. This can either be done by numerical integration of analytical obtainable sub-ensembles, or directly via sampling of reverse time bridge SDEs, where the lifetime is drawn from the hitting time distribution for each realization. We here discuss only the latter case for the three example cases of the Bessel, advected random walk and Ornstein-Uhlenbeck process.


\subsubsection{Bessel process}
\label{subsub_the_bessel_process}
\begin{figure}[ht]
\centerline{\includegraphics[width=0.8\linewidth]
{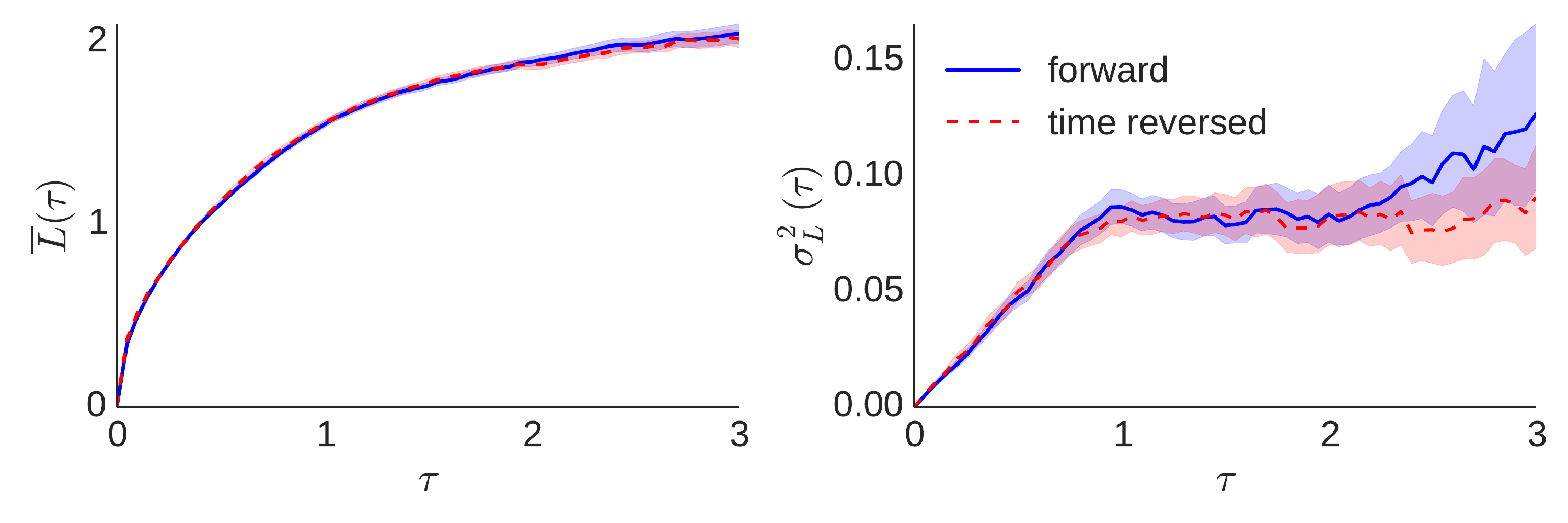}}
\caption{
\textbf{
Forward (blue) and time reversed dynamics (red) show excellent agreement with respect to the Bessel process}. Shown are the mean (Left) and variance (Right) of both 
processes with 95\% bootstrap confidence intervals. To exclude numerical 
inaccuracies due to rarely visited tails of 
Eq.~\eqref{sup_hit_bessel}, we directly sampled from the numerically 
obtained hitting time distribution of the forward process. Results were 
obtained using each 1000 realizations of the respective ensemble with 
parameter settings $\gamma = 1$, $D =0.2$ and $\widehat{L}_0=2$. 
The larger $\tau$, the broader the confidence intervals become, as less and 
less sample paths contribute to the averaging due to differences in their 
lengths $T_i$.
}\label{besselCompExact2}
\end{figure}

The dynamical law of the Bessel process reads\cite{bray2000random}
\begin{equation}
\label{sup_fwd_bessel_sde}
  d \widehat{L}(t) = - \frac{\gamma}{\widehat{L}}  \; dt + \sqrt{D} \; dW_t
 \ .
\end{equation}
Its solution with $\gamma >0$, 
absorbing boundary condition at $\widehat{L}_\mathrm{ts}=0$, and natural boundary 
conditions at $\widehat{L} \to \infty$ is known as Bessel process
and reads\cite{bray2000random}
\begin{equation}
\label{sup_fwd_bessel_trans}
P^{\mathrm{fw}}(\widehat{L},t|\widehat{L}_0,t_0=0) =
 \frac{\widehat{L}_0 
e^{-\frac{\widehat{L}^2+\widehat{L}_{0}^2}{2 D t}}
   \left(\frac{\widehat{L}
   }{\widehat{L}_0}\right)^{\frac{1}{2}-\frac{\gamma
   }{D}} I_{\frac{\gamma }{D}+\frac{1}{2}}\left(\frac{\widehat{L}
   \widehat{L}_0}{D t}\right)}{D t}
   \ .
\end{equation}
$I_\nu(z)$ is the modified Bessel 
function of the first kind, with series expansion
\begin{align}
\label{sup_mod_bess_first_kind}
 I_\nu(z) = \left(\frac{z}{2}\right)^\nu \sum_{k=0}^{\infty} \frac{\left(\frac{z^2}{2}\right)^k}{k! \Gamma(\nu+k+1)}
\end{align}
for real valued $\nu$ from Abramovitz \& Stegun Eq.~(AS 9.6.10)\cite{abramowitz1964handbook} . A derivation and the 
lifetime distribution
\begin{equation}
\label{sup_hit_bessel}
 \rho_0(T|\widehat{L}_0) =
 \frac{2^{-\frac{\gamma }{D}-\frac{1}{2}} 
e^{-\frac{\widehat{L}_0^2}{2
   D T}} \left(\frac{\widehat{L}_0^2}{D T}\right)^{\frac{\gamma
   }{D}+\frac{1}{2}}}{T \ \Gamma \left(\frac{\gamma
   }{D}+\frac{1}{2}\right)}
\end{equation}
can be found in\cite{bray2000random,martin2011first}. Knowing the 
forward probability 
distribution Eq.~\eqref{sup_fwd_bessel_trans}, the time reversed sub-ensemble SDE of this process can directly be obtained by evaluating the derivative of the logarithm of Eq.~\eqref{sup_fwd_bessel_trans} using the identity  
\begin{align}
\frac{d}{dz} \frac{I_{\nu+\frac{1}{2}}(z)}{\sqrt{z}}
= \frac{I_{\nu+\frac{3}{2}}(z)}{\sqrt{z}}
+ \frac{\nu I_{\nu+\frac{1}{2}}(z)}{z^{\frac{3}{2}}} 
\end{align}
from (AS 10.2.21)\cite{abramowitz1964handbook}.
We arrive at
\begin{equation}
\label{sup_tr_bessel_sde}
  dL(\tau) = 
  \left(
\frac{D}{L} - \frac{L}{(T - \tau)} + \frac{\gamma}{L}
 +
 \frac{L_f I_{\frac{\gamma
   }{D}+\frac{3}{2}}\left(\frac{L L_f}{D (T - 
\tau)}\right)}{(T - \tau)
   I_{\frac{\gamma }{D}+\frac{1}{2}}\left(\frac{L L_f}{D
   (T - \tau)}\right)}
   \right) d\tau
   + \sqrt{D} \; dW_\tau
   \ ,
\end{equation}
where the first term results from the power law term in Eq.~\eqref{sup_fwd_bessel_trans}, the second from the exponential and the last two from $I_\nu(z)$ using the stated identity. For the interpretation, recall that the initial conditions of the forward process are the final conditions of the reverse time process and thus $L_f =\widehat{L}_0$.
To obtain the full 
result for the time reversed dynamics, we first need to solve 
Eq.~\eqref{sup_tr_bessel_sde} for fixed $T$ and subsequently average over all 
$T$ according to the distribution of lifetimes Eq.~\eqref{sup_hit_bessel}.
In Fig.~\ref{besselCompExact2}, we compare two ways to synthesize this ensemble. One generated from 
Eq.~\eqref{sup_fwd_bessel_sde}, subsequently aligned and time reverted, the 
other directly generated using Eq.~\eqref{sup_tr_bessel_sde} and 
Eq.~\eqref{sup_hit_bessel}. Comparing statistics of both the 
mean and variance, as shown in Fig.~\ref{besselCompExact2}, confirms our 
calculations.

\subsubsection{Advected random walk} 
The advected random walk changes with constant drift velocity ($\gamma >0$) until the target state $\widehat{L}_\mathrm{ts}$ is reached. Its forward SDE is
\begin{equation}
\label{sup_fwd_advection}
 d \widehat{L}(t) = -\gamma \; dt + \sqrt{D} \; dW_t
 \ .
\end{equation}
We set the absorbing boundary to $\widehat{L}_\mathrm{ts} = 0$ and use natural boundary conditions at $\widehat{L} \to \infty$. The conditional 
probability for the forward process is
\begin{equation}
\label{sup_fwd_prob_drift}
P^{\mathrm{fw}}(\widehat{L},t|\widehat{L}_0,t_0=0) = 
\frac{1}{\sqrt{2 \pi D t}} 
\left(
e^{-\frac{(\widehat{L} + \gamma t -\widehat{L}_0)^2}{2 D 
t}}-e^{-\frac{(
   \widehat{L} +\gamma t + \widehat{L}_0)^2}{2 D t}+\frac{2 \gamma 
\widehat{L}_0}{D}}
\right)
   \ ,
\end{equation}
and the associated distribution of lifetimes is
\begin{align}
\label{sup_hit_drift}
 \rho_0(T|\widehat{L}_0) &=  \frac{\widehat{L}_0}{\sqrt{2 
\pi 
D T^3} } e^{-\frac{(\widehat{L}_0 - \gamma T)^2}{2 D 
T}}
\ ,
\end{align}
as given in Redner\cite{redner2001guide}.

\begin{figure}[ht]
\centerline{\includegraphics[width=0.8\linewidth]
{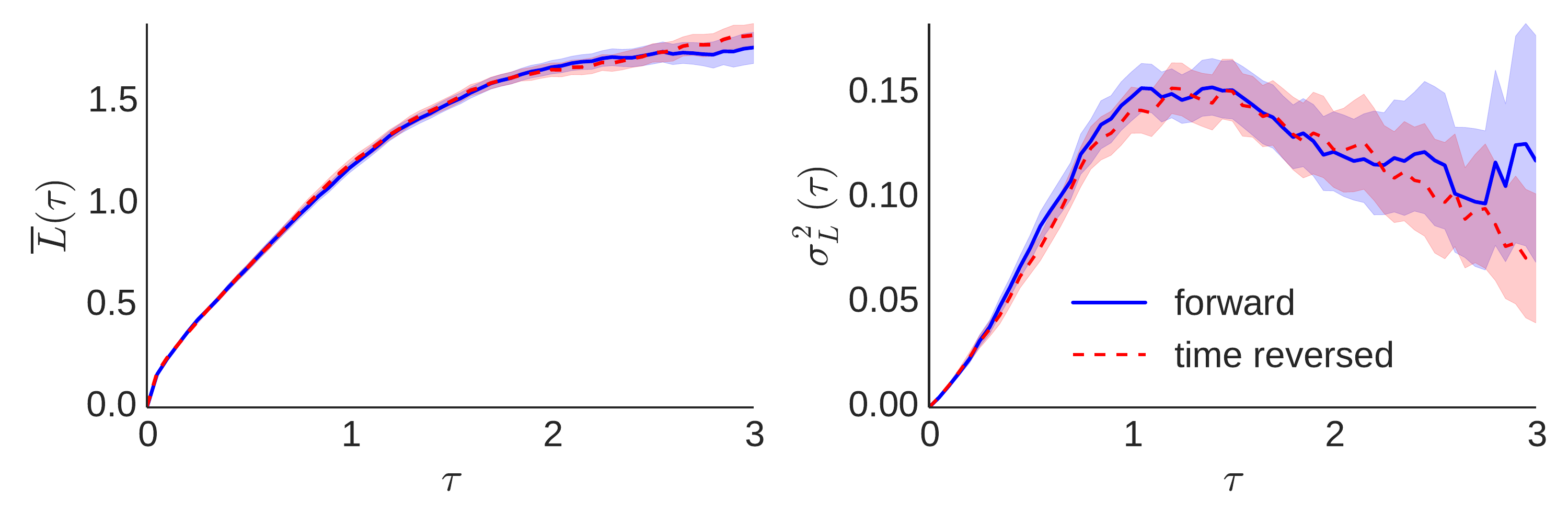}}
\caption{
\textbf{
Forward (blue) and time reversed dynamics (red) of the advected random walk show excellent agreement}. Shown are the mean (Left) and variance (Right) of both 
processes with 95\% bootstrap confidence intervals. To exclude numerical 
inaccuracies due to rarely visited tails of 
Eq.~\eqref{sup_hit_drift}, we directly sampled from the numerically 
obtained hitting time distribution of the forward process. Results were 
obtained using 1000 realizations of the respective ensemble with 
parameter settings $\gamma = 1$, $D =0.2$ and $\widehat{L}_0=2$. 
The larger $\tau$, the broader the confidence intervals become, as less and 
less sample paths contribute to the averaging due to differences in their 
lengths 
$T_i$.
}\label{driftCompExact2}
\end{figure}

Substituting Eq.~\eqref{sup_fwd_prob_drift} into the reverse time SDE for a sub-ensemble 
of fixed lifetime $T$ Eq.~\eqref{sup_tr_lv}, yields 
\begin{align}
\label{sup_tr_drift_sde}
  d L(\tau) = 
 - \frac{L - L_f \coth ( \frac{L 
L_f}{D (T-\tau)} )  }{ T-\tau }
 d \tau
 +
 \sqrt{D} \; dW_\tau
 \ .
\end{align}
Recall that the initial conditions of the forward process are the final conditions of the reverse time process and thus $L_f =\widehat{L}_0$.
Interestingly this time reversed SDE is independent of the drift velocity and thus identical to the random walk bridge. This puzzling finding is resolved when considering not only 
the sub-ensemble of fixed lifetime $T$, but the complete aligned and time reversed 
ensemble according to Eq.~\eqref{sup_prob_full_tr_ens}.
Averaging all sub-ensembles defined by Eq.~\eqref{sup_tr_drift_sde} over $T$ 
with respect to the distribution of lifetimes  
Eq.~\eqref{sup_hit_drift}, recovers the expected $\gamma$-dependence.
The $\gamma$-dependence is thus fully mediated by the lifetime distribution.
A numerical 
comparison of the forward ensemble (after time reversion and 
alignment), and the time reversed ensemble, defined by 
Eq.~\eqref{sup_tr_drift_sde} and Eq.~\eqref{sup_hit_drift},
is displayed in Fig.~\ref{driftCompExact2}.

\subsubsection{Ornstein-Uhlenbeck type process}
\begin{figure}[ht]
\centerline{\includegraphics[width=0.8\linewidth]
{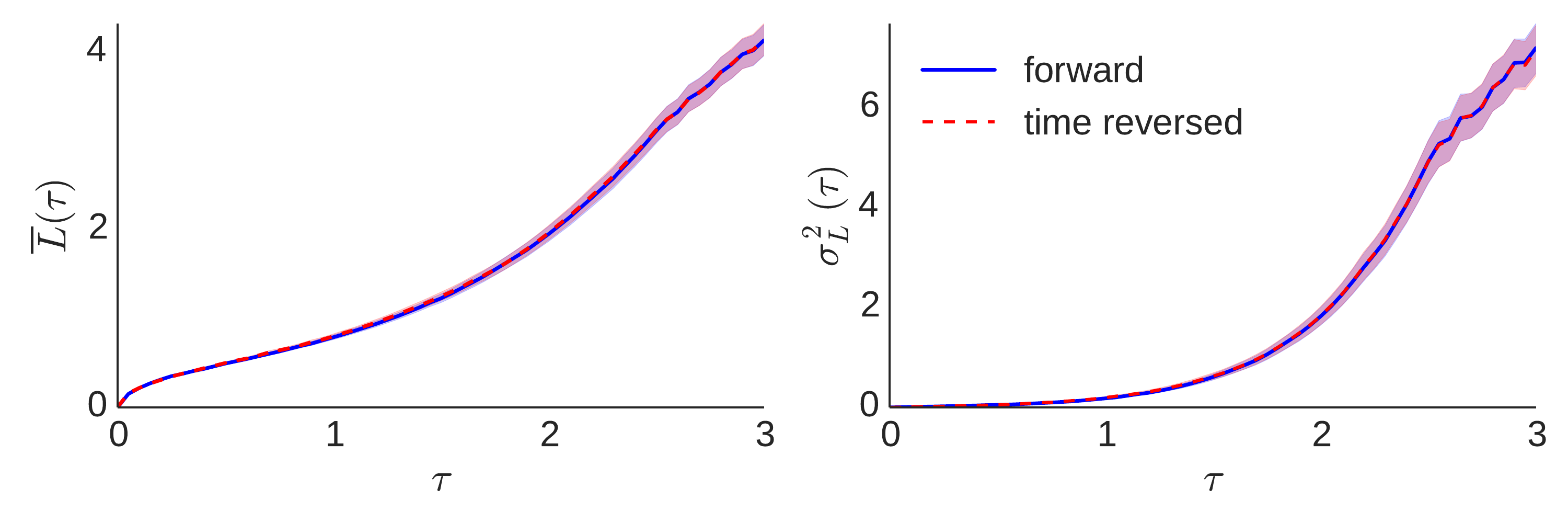}}
\caption{
\textbf{
Forward (blue) and time reversed dynamics (red) of Ornstein-Uhlenbeck type processes agree excellent}.
Shown are the mean (Left) and variance (Right) of both 
processes with 95\% bootstrap confidence intervals. To exclude numerical 
inaccuracies due to rarely visited tails of 
Eq.~\eqref{sup_hit_ou}, we directly sampled from the numerically 
obtained hitting time distribution of the forward process. Results were 
obtained using 1000 realizations of sample paths with 
parameter settings $\gamma = 1$, $D =0.2$ and $\widehat{L}_0=10$. Unlike for the previous examples, we used a larger value for $\widehat{L}_0$ to reveal the interesting concave convex shape of the mean, which, for too small $\widehat{L}_0$ values, is not detectable.
The larger $\tau$, the broader the confidence intervals become, as less and 
less sample paths contribute to the averaging due to differences in their 
lifetimes 
$T_i$.
}\label{ouCompExact2}
\end{figure}

The stochastic relaxation process describes dynamical processes with a linear restoring force. Its forward SDE is
\begin{equation}
\label{sup_fwd_ou_sde}
  d \widehat{L}(t) = -\gamma \widehat{L} \; dt + \sqrt{D} \; dW_t
 \ ,
\end{equation}
where $\gamma >0$.
The conditional probability density of this process with an absorbing 
boundary at $\widehat{L}_\mathrm{ts}=0$, can be constructed using the method of mirror charges. We obtain
\begin{align}
\label{sup_fwd_prob_ou}
P^{\mathrm{fw}}(\widehat{L},t|\widehat{L}_0,t_0=0) = 
\sqrt{
\frac{\gamma}{\pi D ( 1-e^{-2 \gamma  t} ) }
} 
\left(
 e^{-\frac{\gamma  \left(\widehat{L}-\widehat{L}_0 e^{- \gamma t
   }\right)^2}{D \left(1-e^{-2 \gamma  t}\right)}
   }
   -e^{
   -\frac{\gamma  \left(\widehat{L} + \widehat{L}_0 e^{- \gamma t 
   }\right)^2}{D \left(1-e^{-2 \gamma  t}\right)}
   }
\right)
\end{align}
and the associated distribution of lifetimes $T$
\begin{equation}
\label{sup_hit_ou}
 \rho_0(T|\widehat{L}_0) =
 \frac{2 D \widehat{L}_0 \left(\frac{\gamma }{D}\right)^{3/2} e^{2
   \gamma  T-\frac{\gamma  \widehat{L}_0^2}{D \left(e^{2 \gamma 
   T}-1\right)}}}{\sqrt{\pi } \left(e^{2 \gamma 
   T}-1\right)^{3/2}}
   \ ,
\end{equation}
given in Ricciardi and Sato\cite{ricciardi1988first}. Using 
the forward probability distribution from Eq.~\eqref{sup_fwd_prob_ou}, the 
time reversed SDE describing sub-ensembles of length $T$ reads
 \begin{align}
\label{sup_tr_ou_sde}
  dL(\tau) &= 
  \left(
  \gamma L
  +
D \frac{\partial}{\partial L}
\left(
 e^{-\frac{\gamma  \left(L-L_f e^{- \gamma (T-\tau)
   }\right)^2}{D \left(1-e^{-2 \gamma  (T-\tau)}\right)}
   }
   -e^{
   -\frac{\gamma  \left(L + L_f e^{- \gamma (T-\tau) 
   }\right)^2}{D \left(1-e^{-2 \gamma  (T-\tau)}\right)}
   }
\right)
\right) d\tau
 + \sqrt{D} \; dW_\tau
\notag\\
&=
  \left(
- \frac{\gamma L}{\tanh( (T -\tau) \gamma)} + \frac{L_f 
\gamma \coth(\frac{L 
L_f 
\gamma }{D \sinh( (T - \tau) \gamma)}) }{
  \sinh( (T - \tau) \gamma)} 
  \right) d\tau
 + \sqrt{D} \; dW_\tau
 \ .
\end{align}
Note, that the initial conditions of the forward process are now the final conditions of the reverse time process with $L_f=\widehat{L}_0$.
The full aligned and time reversed ensemble is then 
obtained by averaging over all possible sub-ensembles, defined by 
Eq.~\eqref{sup_tr_ou_sde}, with respect to the distribution of lifetimes 
given in  Eq.~\eqref{sup_hit_ou}. 
In Fig.~\ref{ouCompExact2}, we compare mean and variance obtained
by terminal alignment of Eq.~\eqref{sup_fwd_ou_sde}, and directly from Eq.~\eqref{sup_tr_ou_sde} sampling lifetimes from Eq.~\eqref{sup_hit_ou}. We find excellent agreement.


\subsection{Constructing TSA ensembles from their reverse time TSA Fokker-Planck equation -- the Bessel process}
\label{Constructing_TSA_ensembles_from_their_defining_Fokker-Planck_equation}
The third approach to obtain the density of complete reverse time TSA ensembles,  directly solves the defining Fokker-Planck Eq.~\eqref{eq:revt_TSA_FP} or SDE Eq.~\eqref{sup_tr_lv_tsa}. We here consider the case of the Bessel process for which this approach is analytically tractable. The forward process is defined by 
Eq.~\eqref{sup_fwd_bessel_sde} with initial condition $L_f=\widehat{L}_0$. The probability density which describes the forward process is given by Eq.~\eqref{sup_fwd_bessel_trans}. 
Defined as a TSA ensemble the SDE for the Bessel process reads
\begin{align}
\label{sup_tr_lv_bessel_tsa}
  dL(\tau) =
\left(
      - \frac{\gamma}{L}
      +D 
      \  
      \frac{\partial}{\partial 
L}
 \log\left(
  \int_{0}^{L} dL' \;
 L'^{\frac{2 \gamma}{D}}
 \;
 \Theta(\widehat{L}_0 - L')
 \right)  
 \right)
       d\tau
      +
      \sqrt{D} \ d W_\tau
      \ ,
\end{align}
where we substituted $f(L)= - \frac{\gamma}{L}$ into Eq.~\eqref{sup_tr_lv_tsa}. We assume $L_\mathrm{ts}=0$ and $L_f=\widehat{L}_0$. For $L \neq L_f$  this SDE simplifies to 
\begin{align}
\label{sup_tr_lv_bessel_tsa_simp}
  dL(\tau) =
  \begin{cases}
\frac{\gamma + D}{L} \;
       d\tau
      &+ \;
      \sqrt{D} \ d W_\tau
      \qquad \mathrm{for} \ \ L < L_f
      \\
- \frac{\gamma}{L} \;
       d\tau      
      &+ \;
      \sqrt{D} \ d W_\tau
      \qquad \mathrm{for} \ \ L > L_f
    \ ,
  \end{cases}
\end{align}
which both have known solutions in terms of their transition probability densities when defined on the complete interval $L\in [0,\infty]$ \cite{bray2000random}. These solutions can be used to piece wise solve the full SDE. 
We start with the solution of the process defined in Eq.~\eqref{sup_tr_lv_bessel_tsa_simp} for $L < L_f$, but assume for the moment, that we are interested in the solution on the complete interval $L \in [0,\infty]$. The transition probability, for arbitrary $L_0$, which describes this process, is then given as\cite{bray2000random}
\begin{align}
\label{sup_dist_fwd_bessel_bound_push}
P^{<}(L,\tau|L_0,\tau_0=0)
&=   
   \frac{L e^{-\frac{L^2+L_0^2}{2 D \tau}} \left(\frac{L}{L_0}\right)^{\frac{\gamma }{d}+\frac{1}{2}}
   I_{\frac{\gamma }{D}+\frac{1}{2}}\left(\frac{L L_0}{D \tau}\right)}{D \tau}
   \ .
\end{align}
Using the series expansion of $I_\nu(z)$ Eq.~\eqref{sup_mod_bess_first_kind} 
from Abramovitz \& Stegun (AS 9.9.10)\cite{abramowitz1964handbook}, we can evaluate this process for $L_0 \to 0$ and obtain
\begin{align}
\label{sup_dist_fwd_bessel_density}
P^{<}(L,\tau|L_0=0,\tau_0=0)
=
\frac{2^{-\frac{\gamma }{D}-\frac{1}{2}} (D \tau)^{-\frac{\gamma }{D}-\frac{3}{2}} e^{-\frac{L^2}{2 D \tau}}
   L^{\frac{2 \gamma }{D}+2}}{\Gamma \left(\frac{\gamma }{D}+\frac{3}{2}\right)}
   \ .
\end{align}
The needed contribution of the sub-process defined in Eq.~\eqref{sup_tr_lv_bessel_tsa_simp} ($L > L_f$) is more complicated to obtain, as its transition probability Eq.~\eqref{sup_fwd_bessel_trans} goes to zero for $L_0 \to L_\mathrm{ts}$. 
We first guess, that the $L$ dependent part of the solution of Eq.~\eqref{sup_tr_lv_bessel_tsa_simp} ($L > L_f$), now defined on $L\in [0,\infty]$, can be obtained by first normalizing the known solution of the Bessel process Eq.~\eqref{sup_fwd_bessel_trans}.  The resulting ensemble consists of sample paths which at time $t$ have not been absorbed at the target state but are generated by the process defined in Eq.~\eqref{sup_tr_lv_bessel_tsa_simp} ($L > L_f$).
We then take the limit $L_0 \to L_\mathrm{ts}$. The normalization is simply done by the cumulative of the hitting time distribution stated in Eq.~\eqref{sup_hit_bessel}. The seeked expression for $L_0 \to 0$ then reads
\begin{align}
\label{sup_bessel_renorm}
P^{>}(L,\tau|L_0=0,\tau_0=0)
\propto
\lim_{L_0 \to 0}
 \frac{L_0 e^{-\frac{L^2+L_0^2}{2 D \tau}} \left(\frac{L}{L_0}\right)^{\frac{1}{2}-\frac{\gamma
   }{D}} I_{\frac{\gamma }{D}+\frac{1}{2}}\left(\frac{L L_0}{D \tau}\right)}{D \tau \left(1-\frac{\Gamma
   \left(\frac{\gamma }{D}+\frac{1}{2},\frac{L_0^2}{2 D \tau}\right)}{\Gamma \left(\frac{\gamma
   }{D}+\frac{1}{2}\right)}\right)}
   =
   \frac{L e^{-\frac{L^2}{2 D \tau}}}{D \tau}
 \ ,
\end{align}
where ($>$) denotes the forward force law $-\frac{\gamma}{\widehat{L}}$. To find the limit $L_0 \to 0$ we again use the zero-th order of the series expansion for $I_\nu(z)$ stated in Eq.~\eqref{sup_mod_bess_first_kind}. 
With the help of Eq.(AS 6.5.3), Eq.(AS 6.5.4) and Eq.(AS 6.5.29) 
\cite{abramowitz1964handbook}, we replace the upper incomplete gamma 
function in the denominator of Eq.~\eqref{sup_bessel_renorm} by the series
\begin{align}
\Gamma (\nu,z)
=
\Gamma(\nu) - z^\nu \sum_{n=0}^{\infty} \frac{(-z)^n}{(\nu+n)n!}
\qquad 
\mathrm{for} \ |z| < \infty 
\ .
\end{align}
The zero-th order cancels the one in the denominator. The next order cancels all remaining $L_0$-terms of the nominator. 
The resulting term is proportional to the probability density of those sample paths which have survived until $\tau$ and started at $L_\mathrm{ts}=0$.

In the next step, we combine Eq.~\eqref{sup_dist_fwd_bessel_bound_push} and Eq.~\eqref{sup_bessel_renorm} into one expression which solves Eq.~\eqref{sup_tr_lv_bessel_tsa}.
Assuming continuity but not differentiability at $L = L_f$, we can easily connect the two solutions at 
$L_f$, as Eq.~\eqref{sup_bessel_renorm} is fully included in Eq.~\eqref{sup_dist_fwd_bessel_bound_push}. We obtain
\begin{align}
\label{sup_dist_rtTSA_bessel}
R_{0}(L,\tau;0|L_f)
=
 \begin{cases}
 \frac{2 (2 D \tau)^{-\frac{\gamma }{D}-\frac{3}{2}} e^{-\frac{L^2}{2 D \tau}}
   L^{\frac{2 \gamma }{D}+2}}{\Gamma \left(\frac{\gamma }{D}+\frac{3}{2}\right)} & L<L_f \\
 \frac{2(2 D \tau)^{-\frac{\gamma }{D}-\frac{3}{2}} e^{-\frac{L^2}{2 D \tau}}
   L_f^{\frac{2 \gamma }{D}+1} L }{\Gamma \left(\frac{\gamma }{D}+\frac{3}{2}\right)} & L\ge L_f
   \ .
\end{cases}
\end{align}
For  $L \neq L_f$, we confirm this solution by insertion into the Fokker-Planck equation corresponding to Eq.~\eqref{sup_tr_lv_bessel_tsa_simp}.
Furthermore, integrating $R_{0}(L,\tau;0|L_f)$ over $L$ yields one minus the cumulative of the forward hitting time Eq.~\eqref{sup_hit_bessel} and shows that $R_{0}(L,\tau;0|L_f)$ from Eq.~\eqref{sup_dist_rtTSA_bessel} decays with the rate given by the hitting time distribution as demanded by the reverse time TSA Fokker Planck equation Eq.~\eqref{eq:revt_TSA_FP}. 
As $L_\mathrm{ts}$, according to the Feller boundary classification\cite{martin2011first} is an entrance boundary and the other boundary is a natural boundary all lost probability mass must be due to the kink of $R_{0}(L,\tau;0|L_f)$ at $L_f$, i.e. due to the sink at $\delta(L-L_f)$. This confirms that Eq.~\eqref{sup_dist_rtTSA_bessel} is the full solution of the reverse time TSA Fokker-Planck associated with Eq.~\eqref{sup_tr_lv_bessel_tsa}.

To calculate mean and variance with respect to all not yet killed sample paths of the reverse-time ensemble,  we normalize Eq.~\eqref{sup_dist_rtTSA_bessel} to unity with respect to $L$ for all times $\tau$. We obtain for the mean
\begin{align}
 \overline{L}(\tau) &= \frac{A1 +A2}{B}
 \intertext{with}
A1 &= e^{\frac{L_f^2}{2 D \tau }} \left(\! \sqrt{2 \pi D \tau } L_f^{\frac{2 \gamma }{D}+1}
   \text{erfc}\left(\frac{L_f}{\sqrt{2 D \tau }}\right)+
   2 
   (2 D \tau
   )^{\frac{\gamma }{D}+1} \left(\!\Gamma \left(\frac{\gamma }{D}+2\right)-\Gamma \left(\frac{\gamma
   }{D}+2,\frac{L_f^2}{2 D \tau }\!\right)\!\right)\!\right)
   \notag \\
A2 &= 2 L_f^{\frac{2 (\gamma +D)}{D}}
\notag \\
\notag
B &=2
   L_f^{\frac{2 \gamma }{D}+1}
   + 
   2
   (2 D \tau )^{\frac{\gamma }{D}+\frac{1}{2}}
   e^{\frac{L_f^2}{2 D \tau }} \left(\Gamma \left(\frac{\gamma }{D}+\frac{3}{2}\right)-\Gamma
   \left(\frac{\gamma }{D}+\frac{3}{2},\frac{L_f^2}{2 D \tau }\right)\right)
\end{align}
and for the variance 
\begin{align}
\sigma^2_L(\tau)
=
 \frac{
  (2 D \tau )^{\frac{\gamma }{D}+\frac{3}{2}} e^{\frac{L_f^2}{2 D \tau
   }} \left(\Gamma \left(\frac{\gamma }{D}+\frac{5}{2}\right)-\Gamma \left(\frac{\gamma
   }{D}+\frac{5}{2},\frac{L_f^2}{2 D \tau }\right)\right)+L_f^{\frac{2 \gamma }{D}+1} \left(2 D \tau
   +L_f^2\right)
   }{
   L_f^{\frac{2 \gamma }{D}+1}
   +
   (2 D \tau
   )^{\frac{\gamma }{D}+\frac{1}{2}} e^{\frac{L_f^2}{2 D \tau }} \left(\Gamma \left(\frac{\gamma
   }{D}+\frac{3}{2}\right)-\Gamma \left(\frac{\gamma }{D}+\frac{3}{2},\frac{L_f^2}{2 D \tau
   }\right)\right)}
   -
   \overline{L}(\tau)^2
   \ .
\end{align}
Both mean and variance are depicted in Fig.~4b of the main text.


\section{TSA dynamics close to the target state}
\label{TSA_dynamics_close_to_the_target_state}
The dynamics valid for final stages of convergence to the target state can be characterized by a single SDE valid for all forward initial conditions. This claim is easily verified by inspecting Eq.~\eqref{sup_tr_lv_tsa}. For $L$ sufficiently below the bulk of the forward initial distribution $P^\mathrm{in}(L)$, the term 
$H(L)
=
 1
 -
 \int_{L_\mathrm{ts}}^{L} P^{\mathrm{in}}(L') dL'
$,
which describes the influence of the forward initial conditions on the reverse time TSA dynamics, evaluates to approximately one. If most of the sample paths have not yet reached the sink at $P^{\mathrm{in}}(L)$, this further implies that almost no sample path has terminated in reverse time. The killing measure $k(L,\tau)$ (Eq.~\eqref{sup_kill_tsa_fp}), i.e.~the product of hitting time and initial distribution normalized to the current number of still existing sample paths, therefore defines a sector of the $L-\tau$ plane, where, for $k(L,\tau) \approx 0$, TSA ensembles are -- up to exponentially small corrections -- governed by
\begin{align}
\label{sup_tr_lv_tsa_CFB}
  dL(\tau) =
\left(
      f(L) 
      +D 
      \  
      \frac{\partial}{\partial 
L}
 \log\left(
  \int_{L_\mathrm{ts}}^{L} dL' \;
   e^{-
  \int^{L'}
 \frac{2 f(L'')}{D} dL''
 } 
 \right)  
 \right)
       d\tau
      +
      \sqrt{D} \ d W_\tau
     \ .
\end{align}
This form of the reverse time TSA ensemble, valid for final stages of convergence to the target state, provides a powerful tool to infer forward dynamics irrespective of unknown initial conditions.

The reverse time SDE  for TSA ensembles close to completion Eq.~\eqref{sup_tr_lv_tsa_CFB} can be analytically expressed for a broad class of forward forces. Exemplary, we study the properties of power law like forward dynamics with force term
\begin{equation}
\label{sup_alpha_force_law}
 f(\widehat{L}) = - \gamma \widehat{L}^\alpha
 \ ,
\end{equation}
with $\gamma >0$, $\alpha \in \mathbb{R}$ and constant additive noise $D$. 
We derive analytical expressions for the free energy force using forward forces of the form Eq.~\eqref{sup_alpha_force_law}. We check their validity numerically and use the derived expressions to characterize general time-reversed ensembles near the target state.

The TSA reverse time dynamics for forward dynamics Eq.~\eqref{sup_alpha_force_law} are 
\begin{align}
\label{sup_tr_lv_ness_alpha_force}
dL(\tau)
=
 \left(
-\gamma  L^\alpha
+
 \frac{(D\alpha+D) \left(-\frac{2 \gamma }{D
   \alpha+D}\right)^{\frac{1}{\alpha+1}} e^{\frac{2 \gamma  L^{\alpha+1}}{D
   \alpha+D}}}{\Theta(\alpha+1)  \Gamma \left(\frac{1}{\alpha+1}\right)-\Gamma
   \left(\frac{1}{\alpha+1},-\frac{2 L^{\alpha+1} \gamma }{\alpha
   D+D}\right)}
\right) d\tau
   + \sqrt{D} \; dW_\tau
   \qquad \alpha \neq -1
   \ ,
\end{align}
where $\Gamma(z)$ is the gamma-function, $\Gamma \left( \nu,z \right)$ the upper incomplete gamma function and $\Theta(\nu)$ the Heaviside step function. The case for $\alpha=-1$ connecting $\alpha > 1$ and $\alpha<1$ is the Bessel process with reverse time TSA dynamics
\begin{align}
\label{sup_tr_lv_ness_bessel_force}
dL(\tau)
=
\frac{\gamma +D}{L}
 d\tau
   + \sqrt{D} \; dW_\tau
   \qquad \alpha =  -1
   \ .
\end{align}
A detailed derivation and an exemplary
comparison to the exact dynamics, already discussed in section \ref{subsub_the_bessel_process},
is provided below.


\subsection{TSA dynamics independent from forward initial conditions}
\label{deriving_ne_sde_approximation}
In this section we derive reverse time TSA dynamics under the assumption that the forward initial conditions are spatially and temporally well separated from the target state. We derive Eq.~\eqref{sup_tr_lv_ness_alpha_force} for a forward force of the form  $f(\widehat{L}) = - \gamma \widehat{L}^\alpha$, as stated in Eq.~\eqref{sup_alpha_force_law}.

We start from the definition of the free energy force as stated in Eq.~\eqref{sup_general_def_freeE_force} for general reverse time TSA dynamics with
\begin{align}
 f^\mathcal{F}(L)
 =
 D \frac{\partial}{\partial L}\log \widetilde{I}(L)
 \ ,
 \intertext{where the integral}
 \widetilde{I}(L) = \int_0^L e^{\frac{2 \Phi(L')}{D}} H(L') dL'
 \intertext{takes the form of a canonical partition function for the potential}
 \Phi(L) = - \int^L f(L') dL' 
 \ ,
\end{align}
 associated with the sign inverted forward force. The influence of the forward initial conditions on the TSA dynamics is captured by $H(L)$ Eq.~\eqref{sup_def_H}. It is a sigmoidal like function, which is one at the target state and decreases the closer $L$ is to the bulk of the forward initial condition. It approaches zero beyond. For well separated target states and forward initial conditions we can therefore safely set $H(L)=1$, close to $L_\mathrm{ts}$, and define 
$I(L)=\widetilde{I}(L)|_{H(L)=1}$.

The derivation of Eq.~\eqref{sup_tr_lv_ness_alpha_force} then starts with evaluating the potential
\begin{align}
\Phi(L)
=
\int^L \gamma  L^\alpha dL'
=
\frac{\gamma  L^{\alpha+1}}{\alpha+1} 
\qquad \qquad
&\ \alpha \neq -1 \ .
\end{align}
Substituting the resulting $\Phi(L)$ into $I(L)$ and with help of Eq.~(AS 6.5.25) from
Abramowitz \& Stegun\cite{abramowitz1964handbook}, yields
\begin{align}
 I(L)
=
\widetilde{C}(\alpha) \;
\Gamma \left(\frac{1}{\alpha+1},-\frac{2 \gamma  L^{\alpha+1}}{D
   (\alpha+1)}\right)
\Big|_0^L
\qquad \qquad
&\ \alpha \neq -1
\ ,
\end{align}
where $\Gamma \left(n,z \right)$ is the upper incomplete gamma function 
and 
$\widetilde{C}(\alpha)
:=
-\frac{1}{\alpha +1}  \left(-\frac{2 \gamma }{D(\alpha +1)}\right)^{-\frac{1}{\alpha +1}}
$ is a prefactor depending on $\alpha$ and independent of $L$. Depending on $\alpha$, the lower 
boundary assumes two different values. 
For $\alpha > -1$, the argument 
$z:=-\frac{2 \gamma  L^{\alpha+1}}{D(\alpha+1)}$ of the upper incomplete gamma function $\Gamma(\nu,z)$, goes to zero and  $\Gamma(\nu,z)$ reduces to $\Gamma(\nu)$. For $\alpha < -1$, the argument $z$ diverges ($z\to \infty$) as $L \to 0$ and $\Gamma(\nu,z) \to 0$, as can be read off from Eq.~(AS 6.5.32) in
Abramowitz \& Stegun\cite{abramowitz1964handbook}.
Combined into a single expression using the Heaviside step-function $\Theta(\nu)$ we arrive at
\begin{align}
  I(L)
 = \widetilde{C}(\alpha)
 \left(
  \Gamma \left(\frac{1}{\alpha+1},-\frac{2 \gamma  L^{\alpha+1}}{D
   (\alpha+1)}\right)
   -
   \Theta(\alpha+1)
   \Gamma \left(\frac{1}{\alpha+1} \right)
   \right)
   &\qquad \alpha \neq -1
   \ .
\end{align}
To obtain the final expression for the free energy force $f^\mathcal{F}$ we rewrite
\begin{align}
f^\mathcal{F}(L)
=
D \frac{\partial}{\partial L} \log I(L) 
=
\frac{D}{I(L)} \frac{\partial}{\partial L} I(L)
\ ,
\end{align}
which allows us to directly read off the sought expression, as  $\frac{\partial}{\partial L} I(L)$ is simply the integrand of $I(L)$, and obtain
\begin{align}
 \label{sup_freeEforce_fwd_powerlaw}
f^\mathcal{F}(L)
=
 \frac{(D\alpha+D) \left(-\frac{2 \gamma }{D
   \alpha+D}\right)^{\frac{1}{\alpha+1}} e^{\frac{2 \gamma  L^{\alpha+1}}{D
   \alpha+D}}}{\Theta(\alpha+1)  \Gamma \left(\frac{1}{\alpha+1}\right)-\Gamma
   \left(\frac{1}{\alpha+1},-\frac{2 L^{\alpha+1} \gamma }{\alpha
   D+D}\right)}
   \qquad \alpha \neq -1
   \ .
\end{align}
Substituting the resulting $f^\mathcal{F}(L)$ into the equation for the time 
reversed SDE Eq.~\eqref{sup_tr_lv_tsa_CFB} yields the final result as stated 
in Eq.~\eqref{sup_tr_lv_ness_alpha_force}.



\subsection{Comparing TSA dynamics with and without forward initial conditions}
\label{sup_sec_examples_ne}
To demonstrate the excellent agreement of the TSA approximation 
stated in Eq.~\eqref{sup_tr_lv_ness_alpha_force} and Eq.~\eqref{sup_tr_lv_ness_bessel_force} with the corresponding aligned 
forward simulations close to completion we use the three, from section \ref{Exactly_solvable_reverse_time_dynamics} known, analytically tractable 
example cases of constriction ($\alpha =-1$), advection ($\alpha =0$) and 
relaxation ($\alpha =1$).
From above, we further know, that the 
TSA-approximation of well separated initial and final conditions
breaks down, when a 
relevant fraction of time reversed sample 
paths reaches the level of the initial state of the forward process. To observe 
this 
phenomenon, we state the 
initial conditions of the forward process explicitly and adapt the plot range 
accordingly.

\subsubsection{Bessel process}
\label{constritionNESS}
\begin{figure}[ht]
\centerline{\includegraphics[width=0.8\linewidth]
{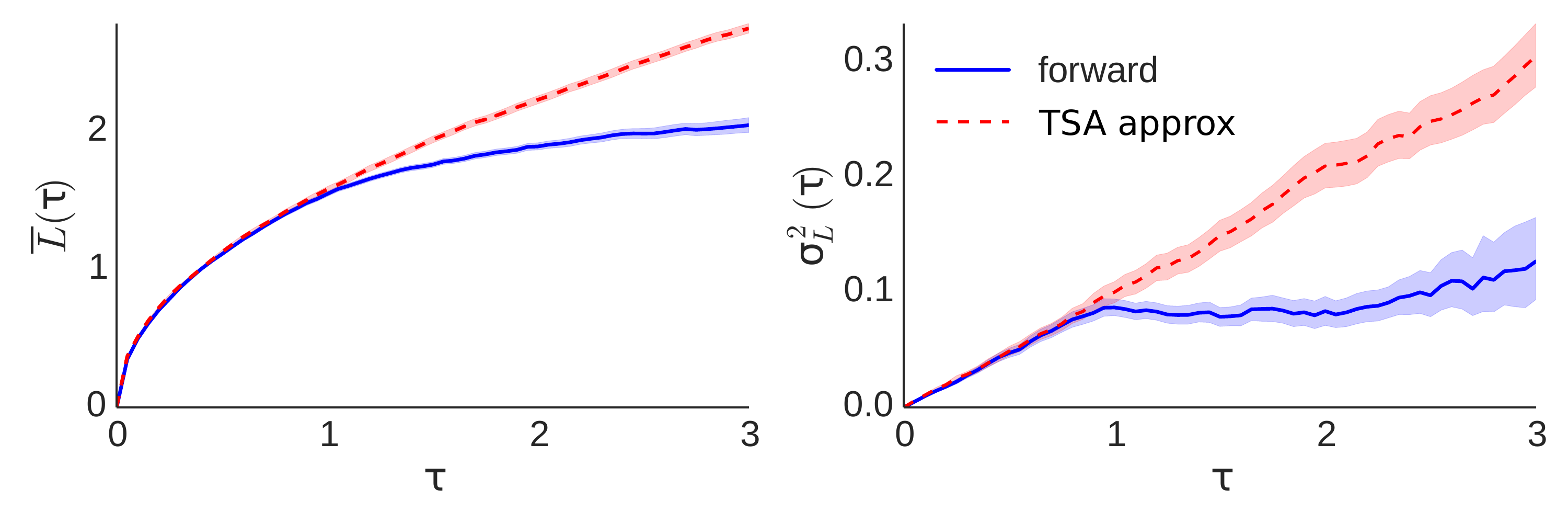}}
\caption{
\textbf{
Forward dynamics (blue) and the approximate time reversed TSA dynamics (red) of the Bessel process agree excellent close to the target state}.
Shown are the mean (Left) and variance (Right) of both 
processes  with 95\% bootstrap confidence intervals. Results were 
obtained using 1000 realizations for each of the respective ensemble with 
parameter settings $\gamma = 1$, $D =0.2$ and $\widehat{L}_\mathrm{in}=2$.
}\label{besselCompExactVsNess}
\end{figure}

The Bessel process within the TSA close to completion approximation is given by 
Eq.~\eqref{sup_tr_lv_ness_bessel_force}. Fig.~\ref{besselCompExactVsNess} 
shows the comparison between 
this approximation and the corresponding exact forward process starting at 
$\widehat{L}_\mathrm{in}=2$. Close to completion clearly both mean and 
variance are very well reproduced by 
Eq.~\eqref{sup_tr_lv_ness_bessel_force}.

Important for benchmark tests in section 
\ref{Moments_for_force_driven_TSA_dynamics}, Eq.~\eqref{sup_tr_lv_ness_bessel_force} can be solved analytically and results in the transition probability stated in Eq.~\eqref{sup_dist_fwd_bessel_bound_push}.

Specifying  Eq.~\eqref{sup_dist_fwd_bessel_bound_push} with respect to the initial conditions $(L_0,\tau_0)=(0,0)$ yields Eq.~\eqref{sup_dist_fwd_bessel_density},
which can be used to obtain expressions for mean
\begin{align}
\label{sup_ne_constriction_mean}
 \overline{L}(\tau)
 =
\frac{\sqrt{2} \sqrt{D \tau} \Gamma \left(\frac{\gamma
   }{D}+2\right)}{\Gamma \left(\frac{\gamma
   }{D}+\frac{3}{2}\right)} 
\end{align}
and
variance
\begin{align}
\label{sup_ne_constriction_var}
\sigma^2_L(\tau)
=
D \tau \left(\frac{2 \gamma }{D}-\frac{2 \Gamma \left(\frac{\gamma
   }{D}+2\right)^2}{\Gamma \left(\frac{\gamma
   }{D}+\frac{3}{2}\right)^2}+3\right)
\ .
\end{align}
Forming the joint probability distribution $P(L,\tau,L',\tau')$ from 
Eq.~\eqref{sup_dist_fwd_bessel_bound_push} and Eq.~\eqref{sup_dist_fwd_bessel_density} we obtain the two-time covariance function
\begin{align}
\label{sup_ne_constriction_cov}
C_L(\tau,\tau')
=
\frac{2 D \Gamma (\frac{\gamma}{D}+2)^2 \left(\sqrt{\tau'} 
\tau^{-\frac{\gamma}{D}-2} (\tau-\tau')^{\frac{\gamma}{D}+\frac{5}{2}} \,
_2F_1\left(\frac{\gamma}{D}+2,\frac{\gamma}{D}+2;\frac{\gamma}{D}+\frac{3}{2}
;\frac{\tau'}{\tau} \right)-\sqrt{\tau' \tau}\right)}{\Gamma
   \left(\frac{\gamma}{D}+\frac{3}{2}\right)^2} 
\end{align}
stated for $\tau > \tau'$, where $_2F_1(a,b;c;z)$ is the Gaussian hypergeometric function. The obtained mean Eq.~\eqref{sup_ne_constriction_mean}, variance Eq.~\eqref{sup_ne_constriction_var} and two-time covariance  Eq.~\eqref{sup_ne_constriction_cov} are later in this text used as a baseline reference for further derivations.

\subsubsection{Advected random walk}
\label{advectionNESS}
\begin{figure}[ht]
\centerline{\includegraphics[width=0.8\linewidth]
{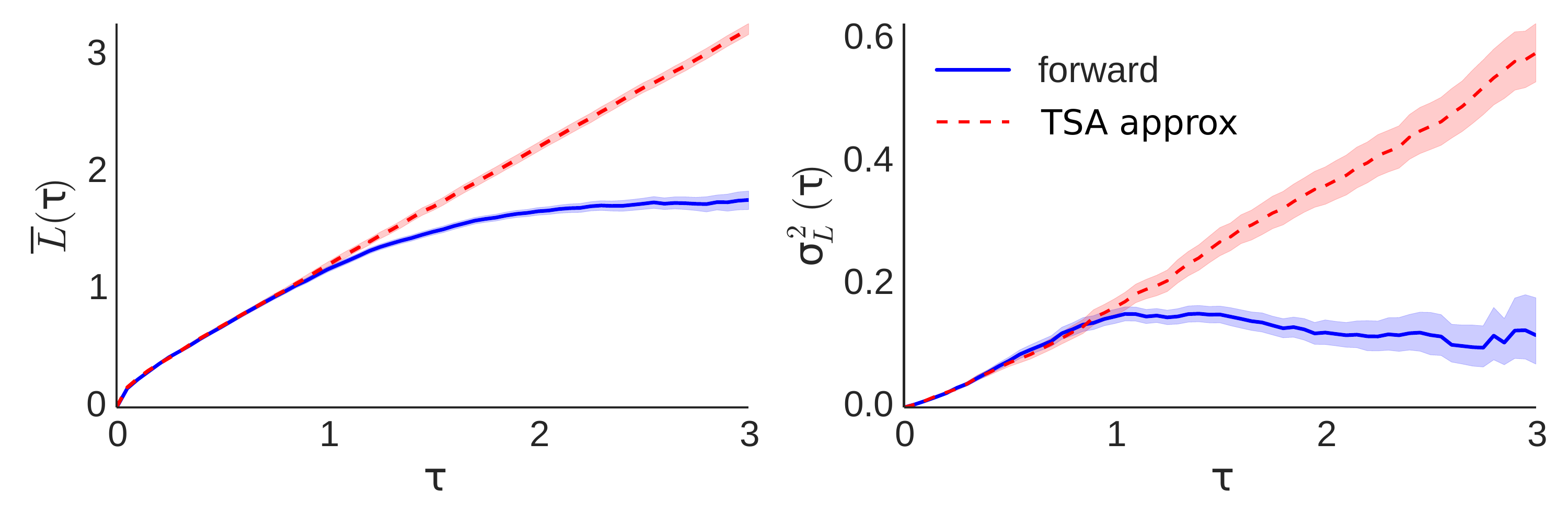}}
\caption{
\textbf{
Forward dynamics (blue) and the approximate time reversed TSA dynamics (red) of the advected random walk agree excellent close to the target state}.
Shown are the mean (Left) and variance (Right) of both 
processes with 95\% bootstrap confidence intervals. Results were 
obtained using each 1000 realizations of the respective ensemble with 
parameter settings $\gamma = 1$, $D =0.2$ and $\widehat{L}_\mathrm{in}=2$.
}\label{driftCompExactVsNess}
\end{figure}

Assuming advection and thus $\alpha =0$, the TSA approximation close to completion stated in 
Eq.~\eqref{sup_tr_lv_ness_alpha_force} simplifies to
\begin{align}
\label{sup_ne_advection}
d L =
\gamma
\left(
 \frac{2}{1-e^{-\frac{2 \gamma  L}{D}}} -1 
 \right) d\tau
 + \sqrt{D} \; dW_\tau
 \ .
\end{align}
Close to completion, the dynamics match both in mean and variance with the 
exact aligned forward simulation as displayed in 
Fig.~\ref{driftCompExactVsNess}.
With help of Eq.~\eqref{sup_ne_advection}, we can now explain the general 
behavior of aligned advection processes close to completion. For small $L$, 
Eq.~\eqref{sup_ne_advection} reduces to $\frac{D}{L}$ and induces a square root 
like behavior of the mean (see main text or 
Eq.~\eqref{sup_ne_constriction_mean}($\gamma=0)$). Further away from the absorbing
boundary, 
the time reversed force term simplifies to $\gamma$, yielding a linear behavior 
of the mean.
Then, eventually, the influence of the initial conditions 
($\widehat{L}_\mathrm{in}=2$) of the forward process can not any longer be 
neglected and the approximated and exact mean start to 
deviate.

\subsubsection{Ornstein-Uhlenbeck type process}
\label{relaxationNESS}
\begin{figure}[ht]
\centerline{\includegraphics[width=0.8\linewidth]
{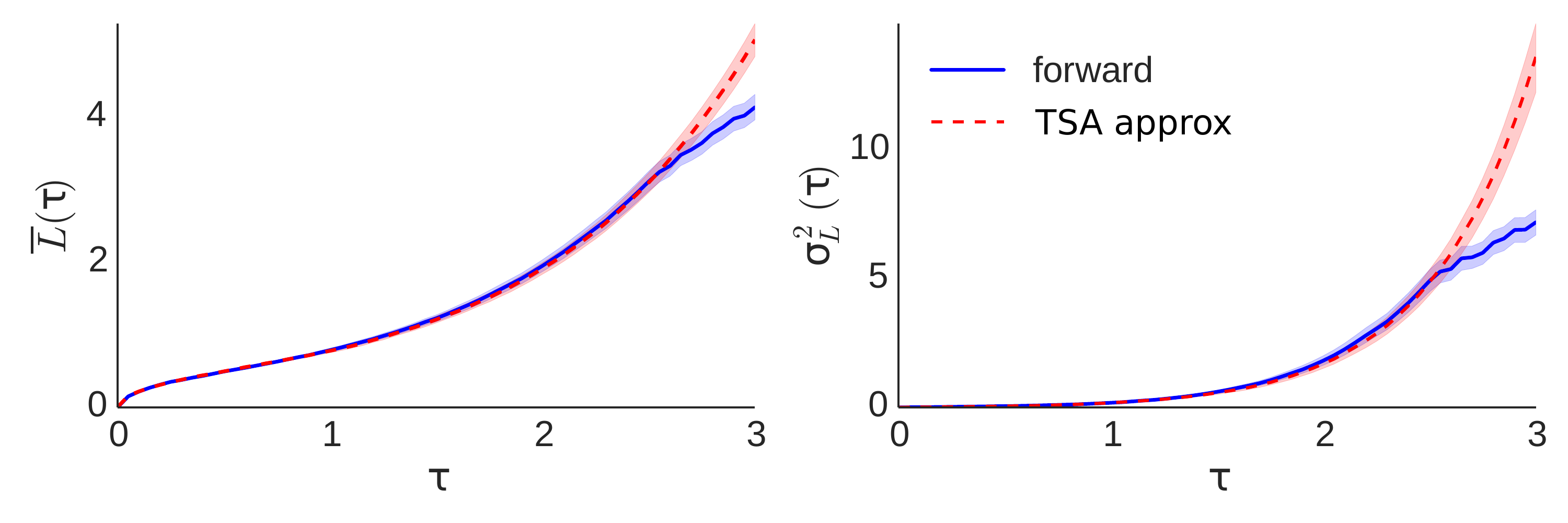}}
\caption{
\textbf{
Forward dynamics (blue) and the approximate time reversed TSA dynamics (red) of an Ornstein-Uhlenbeck type process agree excellent close to the target state}.
Shown are the mean (Left) and variance (Right) of both 
processes with 95\% bootstrap confidence intervals. Results were 
obtained using each 1000 realizations of the respective ensemble with 
parameter settings $\gamma = 1$, $D =0.2$ and $\widehat{L}_\mathrm{in}=10$.
}\label{ouCompExactVsNess}
\end{figure}

The TSA-approximation of well separated initial and final conditions Eq.~\eqref{sup_tr_lv_ness_alpha_force} reduces for the 
case 
of relaxation ($\alpha=1$) to the form
\begin{align}
d L = 
\left(
-\gamma  L
+
 \frac{\sqrt{\gamma  D}}{F\left(L \sqrt{\frac{\gamma
   }{D}}\right)}
 \right) d\tau
 + \sqrt{D} \; dW_\tau  
 \ ,
\end{align}
where $F(x)=e^{-x^2}\int_0^x e^{y^2} dy$ is the Dawson integral\cite{abramowitz1964handbook}.
In Fig.~\ref{ouCompExactVsNess},
we compare this approximation with the forward simulation. In particularly noteworthy here is the behavior of the 
mean, changing from a root like to a convex shape. We further observe that 
the approximate variance is a convex function, which seems to be a preserved 
feature of all processes with $\alpha >0$.


\section{Small $L$ and weak noise approximation -- Force and Noise driven TSA dynamics}
\label{Force_and_Noise_driven_TSA_dynamics}
Depending on the form of the force law $f(\widehat{L})$, dynamics close to the target state are either dominated by the deterministic drift or by random flucutations which eventually determine when a trajectory reaches the target state. To quantitatively characterize this effect we assume, that sufficiently close to the target state, most dynamics are well approximated by a powerlaw. Evaluating a small $L$-expansion of Eq.~\eqref{sup_tr_lv_ness_alpha_force} then yields separable contributions due to the forward force law and due to the noise. Evaluating the limit $L \to L_\mathrm{ts}$ allows to characterize which contribution dominates.

We start our derivation with the expansion of Eq.~\eqref{sup_tr_lv_ness_alpha_force} for small $L$. 
The behavior of the expansion depends on the argument of both the upper incomplete gamma function and the exponential
$ \left| \frac{2 \gamma L^{\alpha+1}}{D(\alpha+1)}\right|$
and whether it diverges or vanishes for $L \to 0$. If we only consider $L$ then its behavior is clear and we observe divergence for $\alpha<-1$ and convergence for $\alpha > -1$. If we, however, additionally demand that our results should not deteriorate in the limit of small noise, i.e. $L\to0$ and $D\to 0$, then the case $-1 < \alpha < 0$ switches its class and starts to diverge for $L \to 0$. A detailed argument and derivation is given below in this section. We here only state the results for the two regimes $\alpha \ge 0$ and $\alpha <0$.

For $\alpha \ge 0$ the reverse time SDE Eq.~\eqref{sup_tr_lv_ness_alpha_force} reduces to 
\begin{align}
 \label{sup_tr_lv_ness_alpha_pos_force}
  dL(\tau) = 
  \left(
  \frac{D}{L}
  - \gamma  \frac{ \alpha}{\alpha+2}  L^\alpha
  + \mathcal{O}\left(\frac{L^{2 \alpha +1}}{D}\right)
\right) d\tau
   + \sqrt{D} \; dW_\tau
   \qquad  \quad \alpha \ge 0
   \ .
\end{align}
Independent of the exact value of $\alpha$,
the force that drives the time reverse ensemble away from 
$\widehat{L}_\mathrm{ts}$ is thus always the pure free energy force $\frac{D}{L}$ as exhibited by the random walk case Eq.~\eqref{sup_tr_lv_rw_tsa_simp}. This finding has 
important 
implications for the reconstruction of $f(L)$ from TSA data. Close 
to $\widehat{L}_\mathrm{ts}$ all time reversed processes with $\alpha \ge 0$ mimic pure diffusion. 
The identification of genuine forces has thus to rely on dynamics sufficiently far away from $\widehat{L}_\mathrm{ts}$. For larger values of $L$, however, higher order terms can no longer be neglected. Their 
contributions mix and no clear universal approximation is achievable. We 
conclude, that a 
reliable reconstruction of the genuine force $f(L)$ for $\alpha > 0$ is only possible using the 
full time reversed expression as stated in 
Eq.~\eqref{sup_tr_lv_ness_alpha_force}.

For the domain $\alpha < 0$, and for $-1 < \alpha<0$ only under the assumption of small noise ($D \to 0$), Eq.~\eqref{sup_tr_lv_ness_alpha_force} reduces to
\begin{equation}
\label{sup_tr_lv_sn}
 dL(\tau) = 
 \left(
 \gamma L^\alpha - \frac{\alpha D}{L}
 + 
 \mathcal{O}\left(\frac{D^2}{L^{2+\alpha}}\right) 
 \right) d\tau
 +
\sqrt{D} \;dW_\tau 
\qquad \alpha < 0 
\ .
 \end{equation}
For $\alpha<-1$, and for $0<\alpha<-1$ given sufficiently small noise, the TSA dynamics close to the target state are fully governed by the sign inverted forward force law plus a contribution due to the noise similar to Eq.~\eqref{sup_tr_lv_ness_alpha_pos_force}, however, here reweight by a factor $\alpha$. We note that the case $\alpha=-1$, i.e.~the reverse time bessel process Eq.~\eqref{sup_tr_lv_ness_bessel_force}, is exactly included in Eq.~\eqref{sup_tr_lv_sn}. Dynamics with $\alpha <0$ should thus be distinguishable by inspecting their behavior near the target state.

In summary, Eq.~\eqref{sup_tr_lv_ness_alpha_pos_force} and Eq.~\eqref{sup_tr_lv_sn} together allow to distinquish between noise driven Eq.~\eqref{sup_tr_lv_ness_alpha_pos_force} and force driven Eq.~\eqref{sup_tr_lv_sn} TSA dynamics close to the target state. The noise driven case ($\alpha \ge 0$) close to the target state is indistinguishable from a pure TSA random walk with free energy force $\frac{D}{L}$. This process is a Bessel process and fully characterized by the transition probability Eq.~\eqref{sup_dist_fwd_bessel_bound_push}$(\gamma=0)$ and density Eq.~\eqref{sup_dist_fwd_bessel_density} $(\gamma=0)$. To form expectations about TSA ensemble data, we calculate then mean, variance and two-time correlation function below in this section.

The force driven case ($\alpha <0$) has a non negligible contribution due to the sign inverted forward force close to the target state. This sensitivity to the parameter of the forward dynamics close to $L_\mathrm{ts}$ motivates to seek for a small noise expansion around the sign inverted forward law. 
Further below, we demonstrate, that expressions for mean, variance and two-time correlation function can be obtain in the limit of small noise.


\subsection{The derivation of the small $L$ expansion}
\label{The_derivation_of_the_small_L_expansion}
The expansion of $f^\mathcal{F}(L)$ Eq.~\eqref{sup_freeEforce_fwd_powerlaw} depends on the limit properties of the argument 
$z=-\frac{2 L^{\alpha+1} \gamma }{\alpha D+D}$ of the rewritten free energy force
\begin{align}
 \label{sup_freeEforce_fwd_powerlaw_subst}
f^\mathcal{F}(L)
=
 \frac{C(D) e^{-z}}{\Theta(\frac{1}{\nu})  \Gamma \left(\nu\right)-\Gamma
   \left(\nu,z\right)}
   \qquad \alpha \neq -1
   \ 
\end{align}
and whether the limit of $|z|$, with respect to $D$ and $L$, rather approaches zero or infinity. For readability, we collect all non $z$-dependent terms into 
\begin{align}
C(D)= (D\alpha+D) \left(-\frac{2 \gamma }{D\alpha+D}\right)^{\frac{1}{\alpha+1}}
\end{align}
and set $\nu= \frac{1}{1+\alpha}$.
In the following three subsections we discuss the expansion of $f^\mathcal{F}(z)$ with respect to $z$, given small $L$, for the regimes $\alpha \in (-\infty,-1),(-1,\infty)$. We then reassess the reverse time forces in the interval $\alpha \in (-1,0)$ and with respect to its dependency on the size of the diffusion constant 
$D$.

\subsubsection{The case $|z| \to 0$}
For $\alpha >-1$ ($\nu>0$) the expansion variable $|z|$ decreases with decreasing $L$. For these cases the denominator of Eq.~\eqref{sup_freeEforce_fwd_powerlaw_subst} can be written as
\begin{align}
\Gamma(\nu) - \Gamma (\nu,z)
=
\Gamma(\nu) z^\nu e^{-z} \sum_{n=0}^{\infty} \frac{z^n}{\Gamma(\nu+n+1)}
\qquad 
\mathrm{for} \ |z| < \infty \ ,
\end{align}
using  Eq.~(AS 6.5.3), Eq.~(AS 6.5.4) and Eq.~(AS 6.5.29)\cite{abramowitz1964handbook}.
The free energy force $f^\mathcal{F}(L)$ is
\begin{align}
 f^\mathcal{F}(L)
 =\frac{C(D)}{
 \Gamma(\nu) z^\nu \sum_{n=0}^{\infty} \frac{z^n}{\Gamma(\nu+n+1)}
 }
\qquad 
\alpha > -1 \ ,
\end{align}
where we canceled the exponential function in nominator and denominator. Truncating the series in the denominator to include only the terms up to first order, we arrive at
\begin{align}
  f^\mathcal{F}(L)
 =\frac{\nu C(D) z^{-\nu}}{
1+ \frac{z}{\nu+1} + \mathcal{O}(z^2)
 }
\qquad 
\alpha > -1 \ .
\end{align}
We here used the recurrent relation $\Gamma(\nu+1)=\nu \Gamma(\nu)$ from Eq.~(AS 6.1.15) Abramowitz \& Stegun\cite{abramowitz1964handbook} to simplify the expression. In the last step we approximate the denominator to lowest order in $z$ and obtain
\begin{align}
   f^\mathcal{F}(L)
 =
 \nu C(D) z^{-\nu}
 \left(
1 - \frac{z}{\nu+1} + \mathcal{O}(z^2)
 \right)
\qquad 
\alpha > -1 \ .
\end{align}
Replacing the effective parameters and variable $z$ with the original parameters $\nu,\gamma,D$ and variable $L$, we arrive at the small $L$ expansion
\begin{align}
\label{sup_freeEforce_alph_g_minOne}
   f^\mathcal{F}(L)
 =
 \frac{D}{L}
 + \frac{2 \gamma  L^{\alpha }}{\alpha +2}
 + \mathcal{O}\left(\frac{L^{1+2\alpha}}{D}\right)
\qquad 
\alpha > -1 \ .
\end{align}
Substituting Eq.~\eqref{sup_freeEforce_alph_g_minOne} into Eq.~\eqref{sup_tr_lv_tsa_CFB} we arrive at the small $L$-expansion stated in Eq.~\eqref{sup_tr_lv_ness_alpha_pos_force}.


\subsubsection{The case $|z| \to \infty$}
\label{subsub_the_case_z_g_one}
For $\alpha < -1$ ($\nu <0$) the expansion variable $z$ goes to infinity for small $L$. With $\nu <0$ the Heaviside step function evaluates to zero and the denominator simplifies to $\Gamma(\nu,z)$ which we expand for large $z$, using
the known asymptotic expansion Eq.~(AS 6.5.32)\cite{abramowitz1964handbook}
\begin{align}
\label{sup_asymGammaFkt}
\Gamma \left( \nu, z \right)
=
z^{\nu-1}  e^{-z}  
\left(1 + \frac{\nu-1}{z}   
+ \mathcal{O}(z^{-2})
\right)
\qquad \mathrm{for} \ \left| z \right| \to \infty
\ .
\end{align}
We substitute this 
expansion into Eq.~\eqref{sup_freeEforce_fwd_powerlaw_subst} and arrive at
\begin{align}
\label{sup_freeEforce_z_g_one_expansionStepOne}
  f^\mathcal{F}(L)
 = -
 \frac{C(D) z^{1-\nu}}{
1+ \frac{\nu-1}{z} + \mathcal{O}(z^{-2})
 }
\qquad 
\alpha < -1 
\end{align}
after canceling the exponential terms. In a final step we approximate the denominator for $z\to \infty$, i.e.~  Taylor expand for small orders of $1/z$, and obtain
\begin{align}
   f^\mathcal{F}(L)
 =
 -
  C(D) z^{1-\nu}
 \left(
1 - \frac{\nu-1}{z} + \mathcal{O}(z^{-2})
 \right)
\qquad 
\alpha < -1 \ .
\end{align}
Re-substituting the original parameters and variable $L$, the final free energy force reads
\begin{align}
\label{sup_freeEforce_alph_sm_minOne}
   f^\mathcal{F}(L)
 =
 2 \gamma  L^{\alpha }
 - \frac{\alpha D}{L}
 + \mathcal{O}\left(\frac{D^2}{L^{2+\alpha}}\right)
\qquad 
\alpha < -1 \ .
\end{align}

\subsubsection{The $L$ and $D$ dependent range $-1 <\alpha < 0$}
\label{the_L_and_D_dependent_range}
For $\alpha$ in the range  $-1 <\alpha < 0$, the value of the expansion variable $z$ depends on the choice of the diffusion constant $D$. For large $D$ the limit behavior of $z$ is fully defined by $L$ and we find
\begin{align}
\lim_{L \to 0} |z|
= 
\lim_{L \to 0}
\left|
-\frac{2 L^{\alpha+1} \gamma }{\alpha D+D}
\right|
= 0
\qquad \mathrm{for} \quad  -1< \alpha < 0
\ .
\end{align}
If we additionally assume, that the expansion holds in the limit of small noise, the limiting behavior changes and we find 
\begin{align}
\lim_{L \to 0,D \to 0} |z|
= 
\lim_{L \to 0,D \to 0}
\left|
-\frac{2 L^{\alpha+1} \gamma }{\alpha D+D}
\right|
= \infty
\qquad \mathrm{for} \quad  -1< \alpha < 0
\ .
\end{align}
This implies, that for a given $L$, we can always find a small enough $D$ so that the expansion variable $z$ becomes large.
Thus, we also consider $|z| \to \infty$ in the regime $-1<\alpha<0$. Under this condition, the $\Theta(\alpha+1)$ dependent term in the denominator of Eq.~\eqref{sup_freeEforce_fwd_powerlaw_subst} does no longer vanish automatically. Eq.~\eqref{sup_freeEforce_z_g_one_expansionStepOne} then reads
\begin{align}
\label{sup_freeEforce_z_interm_expansionStepOne}
  f^\mathcal{F}(L)
 &= -
 \frac{C(D) z^{1-\nu}}{
m(z)
+
1+ \frac{\nu-1}{z} + \mathcal{O}(z^{-2})
 }
\qquad 
\alpha < -1 
\intertext{with}
m(z) &= - e^z z^{1-\nu} \Gamma(\nu)
\ .
\end{align}
In the following we argue, that $m(z)$ becomes negligible for large $|z|$. We show, that the limes of $m(z)$ with respect to $D \to 0$ is zero and conclude, that we can always find a $D$, which is small enough to neglect $m(z)$ in the expansion of $f^\mathcal{F}(L)$. For small $D$, 
Eq.~\eqref{sup_freeEforce_z_interm_expansionStepOne} therefore follows the same expansion as the case $\alpha <-1$ discussed before.

To proof these statements, we first recall, that $z=-\frac{2 L^{1+\alpha} \gamma}{(1+\alpha)D}$ is negative for $-1<\alpha<0$. The exponential thus decays with increasing $|z|$. With $\nu=\frac{1}{1+\alpha}$, the exponent $1-\nu$ is negative for $-1<\alpha<0$ and the power law $|z|^{1-\nu}$ decays with increasing $|z|$. For $|z|\to \infty$ we therefore find 
\begin{align}
\lim_{|z| \to \infty} |m(|z|)| 
= 0
\qquad \mathrm{for} \quad  -1< \alpha < 0
\ .
\end{align}

We next show, that $|m(z)|$ also approaches zero for $D \to 0$. We start by rewriting $m(z)$ in terms of $\alpha,\gamma,D$ and $L$ and obtain
\begin{align}
\label{sup_m_wrt_L}
 m(L)
 =
 \Gamma \left(\frac{1}{\alpha +1}\right) L^{\alpha } \left(-\frac{2 \gamma }{D(1+\alpha)}\right)^{\frac{\alpha }{\alpha +1}} e^{-\frac{2 \gamma  L^{\alpha +1}}{D(1+\alpha)}}
 \ .
\end{align}
To clarify the behavior of the first $D$-dependent term we note, that its absolute value can be written as
$
\left|
\left(-\frac{2 \gamma }{D(1+\alpha)}\right)^{\frac{\alpha }{\alpha +1}}
\right|
=
\left|
\frac{D(1-|\alpha|)}{2\gamma}
\right|^\frac{1-|\alpha|}{|\alpha|}
\ . 
$
With $|\alpha| \in (0,1)$, the exponent is greater than zero and the full term goes to zero for $D\to 0$.
The second $D$ dependent term in Eq.~\eqref{sup_m_wrt_L} is the exponential. For $D\to 0$ the exponent goes to $-\infty$ and the exponential vanishes. We can therefore write
\begin{align}
\lim_{D \to 0} |m(L)| 
= 0
\qquad \mathrm{for} \quad  -1< \alpha < 0
\ .
\end{align}
In the final step we show that $|m(L)|$ also vanishes in the simultaneous limit of $D\to 0$ and $L \to 0$. To perform this limit we set w.l.o.g.~$D=L$ and rewrite Eq.~\eqref{sup_m_wrt_L} accordingly. In the resulting equation
\begin{align}
\label{sup_m_wrt_L_smallLD}
 m(L)
 =
 \Gamma \left(\frac{1}{\alpha +1}\right) L^{\frac{\alpha^2}{1+\alpha} } \left(-\frac{2 \gamma }{1+\alpha}\right)^{\frac{\alpha }{\alpha +1}} e^{-\frac{2 \gamma  L^{\alpha}}{1+\alpha}}
 \ ,
\end{align}
we can directly read off that the powerlaw $L^{\frac{\alpha^2}{1+\alpha} }$ decays to zero for decreasing $L$, as does the exponential. We conclude, that
\begin{align}
\lim_{\substack{
L\to 0,D \to 0
\\
D\sim L
}
} |m(L)| 
= 0
\qquad \mathrm{for} \quad  -1< \alpha < 0
\end{align}
holds. It is thus always possible to find a small enough $D$,  where the large $z$ expansion of $f^{\mathcal{F}}(L)$, discussed above, holds also in the regime  $-1 <\alpha<0$.

We conclude this section by a remark. Reinspecting Eq.~\eqref{sup_m_wrt_L} we observe, that all $D$'s are multiplied by $1+\alpha$. For $\alpha$ decreasing from zero to minus one we therefore expect $m(L)$ to become smaller and smaller until it approaches zero for  $\alpha \to -1$ using the very same argument as for $D$. The large $|z|$ expansion will therefore hold for the larger $D$ values, the closer $\alpha$ is to $-1$.

\subsection{Moments and universal correlation function for noise driven TSA ensembles}
\label{moments_for_noise_driven_tsa_dynamics}
For $\alpha >0$, the SDE Eq.~\eqref{sup_tr_lv_ness_alpha_pos_force} with respect to the free energy force $f^\mathcal{F}(L)$ expanded in small $L$ tells us, that the dynamics close to the target state are purely governed by the force term $\frac{D}{L}$. From section \ref{Constructing_the_TSA_ensemble_from_its_definition}, or simply by inserting $f(L)=0$ into Eq.~\eqref{sup_tr_lv_tsa_CFB}, we know, that the SDE
\begin{align}
 \label{sup_tr_rw_approx}
  dL(\tau) =
        \frac{D}{L}
       d\tau
      +
      \sqrt{D} \ d W_\tau
\end{align}
describes the dynamics of a TSA random walk close to the target state and well separated from its forward initial conditions.
For completeness and easy use we here state the mean
\begin{align}
\label{sup_noiseDr_mean}
 \overline{L}(\tau)
 &=
\sqrt{\frac{8 D \tau}{\pi} }
\ ,
\intertext{variance}
\label{sup_noiseDr_var}
\sigma_L^2(\tau)
&=
\frac{(3 \pi -8) D \tau}{\pi }
\intertext{and two-time covariance}
\label{sup_noiseDr_cov}
 C_L(\tau,\tau')
   &=
\begin{cases} 
\frac{2 D \left(-4 \sqrt{\tau \tau'}+3 \sqrt{\tau (\tau'-\tau)}+(2 \tau+\tau') \sin
   ^{-1}\left(\sqrt{\frac{\tau}{\tau'}}\right)\right)}{\pi }
   \qquad \quad \, &\mathrm{for} \ \tau < \tau'
   \\
   \sigma_L^2(\tau)
   \qquad \qquad \qquad \qquad \qquad \qquad \qquad \qquad \qquad \, &\mathrm{for} \ \tau = \tau'
   \\
\frac{2 D \left(-4 \sqrt{\tau' \tau}+3 \sqrt{\tau' (\tau-\tau')}+(2 \tau'+\tau) \sin
   ^{-1}\left(\sqrt{\frac{\tau'}{\tau}}\right)\right)}{\pi }
   \qquad &\mathrm{for} \ \tau > \tau'   
   \ .
\end{cases}
\end{align}
They can be obtained either from the known transition probability Eq.~\eqref{sup_dist_fwd_bessel_bound_push} and density Eq.~\eqref{sup_dist_fwd_bessel_density} of the Bessel process with an entrance boundary at $L_\mathrm{ts}=0$ and in the limit $\gamma \to 0$, or from the moments stated in section \ref{subsub_the_bessel_process}.
Using the obtained variance and covariance, the correlation function
$
\mathrm{corr}_L(\tau,\tau') = C_L(\tau,\tau')/(\sigma_L(\tau) \sigma_L(\tau'))
$
reads 
\begin{align}
\label{sup_rt_rw_corr}
 \mathrm{corr}_L(\tau,\tau')
   &=
\begin{cases} 
\frac{2 \left(-4 \sqrt{\tau \tau'}+3 \sqrt{\tau (\tau'-\tau)}+(2 \tau+\tau') \sin
   ^{-1}\left(\sqrt{\frac{\tau}{\tau'}}\right)\right)}{(3 \pi -8) \sqrt{\tau \tau'} }
   \qquad \quad \, &\mathrm{for} \ \tau < \tau'
   \\
   1 \quad \  \
   \qquad \qquad \qquad \qquad \qquad \qquad \qquad \qquad \qquad \, &\mathrm{for} \ \tau = \tau'
   \\
\frac{2 \left(-4 \sqrt{\tau' \tau}+3 \sqrt{\tau' (\tau-\tau')}+(2 \tau'+\tau) \sin
   ^{-1}\left(\sqrt{\frac{\tau'}{\tau}}\right)\right)}{ (3 \pi -8) \sqrt{\tau' \tau}}
   \qquad &\mathrm{for} \ \tau > \tau'   
   \ .
\end{cases}   
\end{align}

Let us briefly discuss the obtained reverse time moments. For the mean, we observe that it grows with $\sqrt{\tau}$. It is thus similar to a forward random walk which starts at the boundary but never returns to it. Such a random walk can be constructed from Eq.~\eqref{sup_transProb_diffusion_absbound} (random walk with mirror charges), after normalization to one (for every $\tau$), and in the limit $\widehat{L}\to 0$. The result is the same we obtained as an intermediate step in the derivation of the exact TSA reverse time Bessel process Eq.~\eqref{sup_bessel_renorm} 
\begin{align}
 P(L,\tau)= 
 \frac{L e^{-\frac{L^2}{2 D \tau}}}{D \tau}
\end{align}
and is known as a Rayleigh distribution.
For the mean of this restricted random walk we find
\begin{align}
\overline{L}(\tau)
=
 \sqrt{\frac{\pi D \tau}{2}}
 \ ,
\end{align}
which shows the same $\tau$ dependence as the mean of the TSA random walk Eq.~\eqref{sup_noiseDr_mean} but with a different pre-factor. This observations motivates to also study the variance of the restricted random walk
\begin{align}
\sigma_L^2(\tau)
=
 \frac{1}{2} (4-\pi) D \tau
 \ .
\end{align}
We again find the same dependence on $\tau$ as in the variance of the TSA random walk but again a different pre-factor. This shows, that fitting mean or variance independently can lead to wrong conclusions. Comparing for example the coefficient of variation $CV = \frac{\sigma_L(\tau)}{\overline{L}(\tau)}$ we find the two different constants 
$CV_\mathrm{TSA-rw}=\sqrt{\frac{3 \pi }{8}-1} \approx 0.42 $
for the TSA random walk
and 
$CV_\mathrm{rest-rw}=\sqrt{\frac{4}{\pi }-1} \approx 0.52$ for the restricted random walk.
This shows the TSA random walk, as one might have assumed, is not the same as a restricted random walk.

While sample paths in both cases are generated by the same dynamics, they differ in the selection of sample paths which contribute to the ensemble. In the TSA ensemble, sample paths in principle have had the time to explore the full half plane $L \in [0,\infty]$ before they reach the target state. The fraction of sample paths which predominantly ``walk'' in one direction close to the target state is therefore enriched compared to a restricted random walk, which only neglects all sample paths which ever return to the initial conditions. Formally, its the difference between a TSA ensemble and a meander process\cite{majumdar2015effective}, which never returns to its initial position. For  mathematical details on meander processes we refer to Majumdar and Orland\cite{majumdar2015effective}.

\subsection{Moments for force driven TSA ensembles}
\label{Moments_for_force_driven_TSA_dynamics}
For $\alpha < -1$ and for $-1<\alpha<0$ under the additional constraint of small $D$, TSA dynamics are force driven close to the target state. This motivates to try a small noise expansion around the deterministic time reversed dynamics to obtain expression for the moments of the SDE Eq.~\eqref{sup_tr_lv_sn}. We adopt the approach defined for a standard SDE (see Gardiner chapter 6.2\cite{gardiner1985handbook}) to include order $D$ drift terms and find for the 
mean
\begin{align}
\label{sup_sn_mean}
 \overline{L}(\tau)
 &=
 \left( (1-\alpha) \gamma \tau \right)^{\frac{1}{1-\alpha}} 
 +
 D 
  \frac{(7 \alpha -3) ((1 - \alpha ) \gamma \tau)^{\frac{\alpha
   }{\alpha -1}}}{4 (3 \alpha -1) \gamma }
+ \mathcal{O}(D^{3/2})
\ ,
   \intertext{variance}
   \label{sup_sn_var}
\sigma_L^2(\tau)
&=
   D \frac{1 - \alpha}{1 - 3 \alpha} \tau
   + \mathcal{O}(D^2)
   \intertext{and covariance}
   \label{sup_sn_cov}
 C_L(\tau,\tau')
   &=
\begin{cases}   
\left(
\frac{\tau}{\tau'}
\right)^{\frac{\alpha }{\alpha -1}}
\; \sigma_{L}^2(\tau) \ +  \mathcal{O}(D^2)
   &\qquad \mathrm{for} \ \tau < \tau'
\\
\left(\frac{\tau'}{\tau}\right)^{\frac{\alpha }{\alpha -1}}
\sigma_{L}^2(\tau') + \mathcal{O}(D^2)
   &\qquad \mathrm{for} \ \tau > \tau'
\\
   \qquad \quad \ \ \
\sigma_{L}^2(\tau) \ +  \mathcal{O}(D^2)
   &\qquad \mathrm{for} \ \tau = \tau'
\end{cases}
\end{align}
by a perturbation theory calculation around the deterministic solution of Eq.~\eqref{sup_tr_lv_sn} in the Diffusion constant up 
to order $D$.
The correlation function 
$
\mathrm{corr}_L(\tau,\tau') = C_L(\tau,\tau')/(\sigma_L(\tau) \sigma_L(\tau'))
$
reads
\begin{align}
\label{sup_sn_corr}
 \mathrm{corr}_L(\tau,\tau')
   &=
\begin{cases}
\left(\frac{\tau}{\tau'}\right)^{\frac{3}{2} + \frac{1}{\alpha -1}}
   &\qquad \mathrm{for} \ \tau < \tau'
   \\
\left(\frac{\tau'}{\tau}\right)^{\frac{3}{2} + \frac{1}{\alpha -1}}
   &\qquad \mathrm{for} \ \tau > \tau'   
   \\
   1
   &\qquad \mathrm{for} \ \tau = \tau'
   \ .
\end{cases}
\end{align}
Note, that for all expressions 
(Eq.~\eqref{sup_sn_mean}-Eq.~\eqref{sup_sn_corr}) boundary conditions are already implicitly included due to the contribution of the free energy force.  
We further recall, that the expansion stated in Eq.~\eqref{sup_tr_lv_sn} was 
truncated at order $D$. Only terms up to this order are therefore 
considered in Eq.~\eqref{sup_sn_mean}, Eq.~\eqref{sup_sn_var} and 
Eq.~\eqref{sup_sn_cov}. Given this approximation,
we have show above in this section, that only the mean 
Eq.~\eqref{sup_sn_mean} carries a relevant contribution from the approximate 
free energy force $-\frac{\alpha D}{L}$.

The obtained expression can be used to infer power law forward forces $f(L) =-\gamma L^\alpha$ from TSA ensembles in the limit of small noise. For example, starting from the two-time correlation function, we read off the power law exponent $\alpha$. With $\alpha$ we can fit the variance and determine the diffusion constant $D$. The force strength $\gamma$ is obtained from the mean. 

For the small noise approximation to hold well, and
as a rule of thumb, the coefficient of variation 
$CV=\sigma_L(\tau) / \overline{L}(\tau)$, approximated as 
the lowest order small noise terms, should be well below one. For example for $\alpha=-1$ this implies $CV=\sqrt{D / (4 \gamma)}<1$. For all other $\alpha$-values we find $CV \sim \tau^{\frac{\alpha +1}{2 (\alpha -1)}}$. The approximation thus becomes better the larger $\tau$ given $-1<\alpha<0$, and for $\alpha<-1$ the smaller $\tau$. 
In the following, we will briefly comment on the expressions for 
$\overline{L}(\tau)$, $\sigma_L^2(\tau)$ and $C_L(\tau,\tau')$, and present a 
more detailed discussion and a step by step derivation below.

The mean $\overline{L}(\tau)$ (Eq.~\eqref{sup_sn_mean}) is given as the 
deterministic 
solution to Eq.~\eqref{sup_tr_lv_sn}($D\to0$) plus a (for all $\alpha < 0$) 
solely positive contribution of order $D$.
The case $\alpha < 0$ thus shows, that a purely 
deterministic approach to extract the mean from any experimentally obtained and 
terminally aligned data is bound to fail. As the mean, given by 
Eq.~\eqref{sup_sn_mean}, adds up a deterministic term and a deterministic positive contribution due to the noise, a 
pure deterministic approach overestimates the underlying forward force by the 
contribution of the noise. The difference between the deterministic 
approximation (order $D^0$ of Eq.~\eqref{sup_sn_mean}) and the correct analytic 
obtainable solution Eq.~\eqref{sup_ne_constriction_mean} is exemplary shown in 
\begin{figure}[ht]
\centerline{\includegraphics[width=0.8\linewidth]
{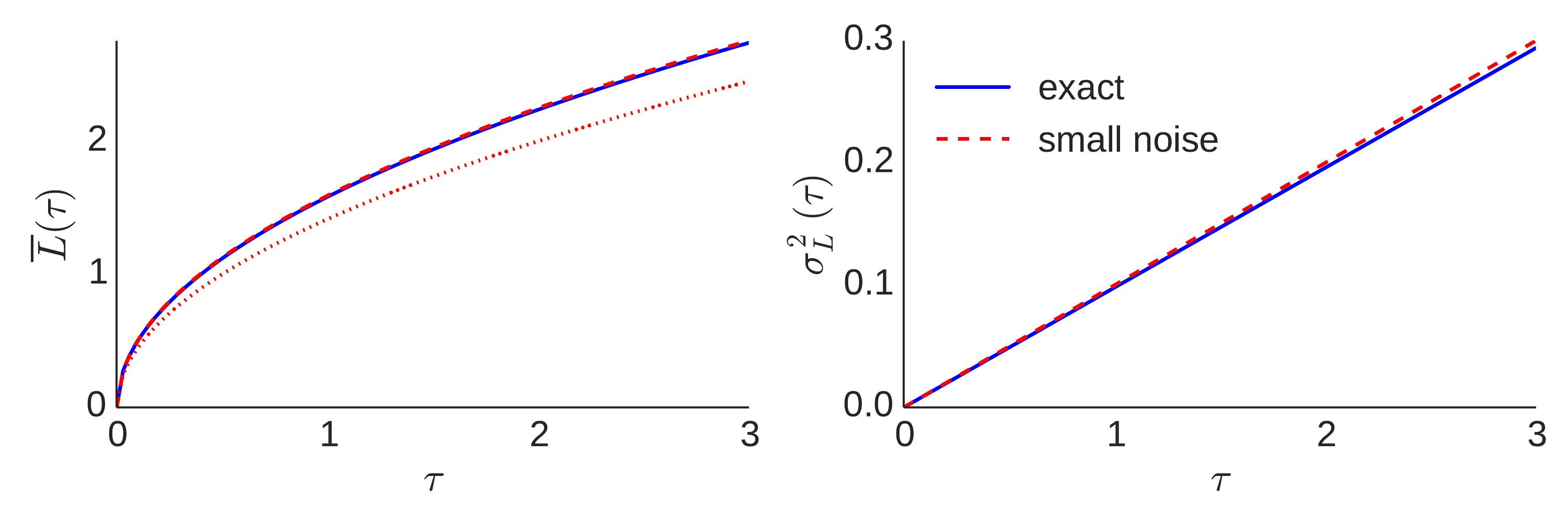}}
\caption{
\textbf{
The small noise approximation up to order $D$ is an excellent approximation of the exact dynamics}. 
Shown are
small noise expressions (red) for the mean (Left) 
Eq.~\eqref{sup_sn_mean} and variance (Right) Eq.~\eqref{sup_sn_var} in 
comparison to the exact analytic expressions (blue), as stated in 
Eq.~\eqref{sup_ne_constriction_mean} Eq.~\eqref{sup_ne_constriction_var} for 
the Bessel process. 
For the small noise expression of the mean we additionally show the curve up 
to order $D^0$ (dotted). Curves are obtained for the parameters $\alpha = -1$, 
$\gamma = 1$, $D=0.2$.
}\label{ExactTheoVsSNtheo}
\end{figure}
Fig.~\ref{ExactTheoVsSNtheo} (Left) for the case $\alpha = -1$. To be precise,
in Fig.~\ref{ExactTheoVsSNtheo} we compare the exact solution of the TSA Bessel process with forward initial conditions shifted to infinity, and the small noise approximation stated in Eq.~\eqref{sup_sn_mean} and Eq.~\eqref{sup_sn_var}. We find excellent agreement.

\begin{figure}[ht]
\centerline{\includegraphics[width=0.8\linewidth]
{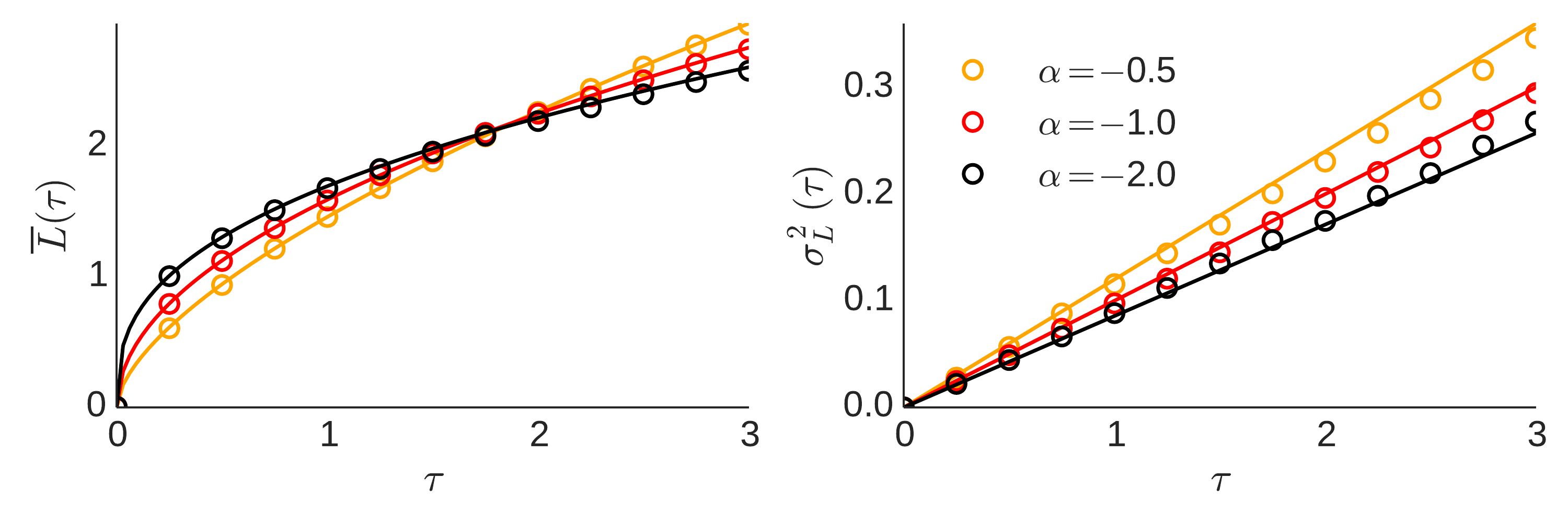}}
\caption{
\textbf{The small noise expansion is a very good approximation of its underlying TSA-SDE.}
Comparison between a numerical evaluation of Eq.~\eqref{sup_tr_lv_sn} 
(circles) and its approximation as analytic small noise expansion up to order 
$D^1$ (lines) as given in Eq.~\eqref{sup_sn_mean} and Eq.~\eqref{sup_sn_var}.
Shown are mean (Left) and variance (Right) 
for the cases $\alpha = -1/2$ (orange), $\alpha = -1$ (red) and 
$\alpha=-2$ (black). All other parameters are chosen identical with $\gamma = 
1$ 
and $D = 0.2$.
}\label{SNtheoVsSNsimul}
\end{figure}

The variance $\sigma_L^2(\tau)$, up to order $D$ (Eq.~\eqref{sup_sn_var}) is a 
$\gamma$-independent function linearly growing with $D \tau$. It thus shows the 
same dependence on $D \tau$ as the exact variance result
Eq.~\eqref{sup_ne_constriction_var} for the Bessel process SDE Eq.~\eqref{sup_tr_lv_sn} with 
$\alpha = -1$. Furthermore, if we assume $D/\gamma \to 0$, 
the variance for $\alpha = -1$, as stated in 
Eq.~\eqref{sup_ne_constriction_var} approaches 
the small noise solution from below until both assume the form $\frac{D 
\tau}{2}$. Taken together this 
indicates that in the limit of small noise, characterized by the $CV$, 
all relevant features of $\sigma_L^2(\tau)$ and $C_L(\tau,\tau')$,
are capture within an expansion up to order $D$.
To further test this for general $\alpha < 0$ we compare 
numerical solutions of Eq.~\eqref{sup_tr_lv_sn} with our analytic small noise 
expressions for mean Eq.~\eqref{sup_sn_mean} and variance 
Eq.~\eqref{sup_sn_var}. Results for $\alpha = -0.5$ and $\alpha = 
-2$ are displayed in Fig.~\ref{SNtheoVsSNsimul} and compared to the case $\alpha 
=-1$. 
All considered, we conclude that $\overline{L}(\tau)$ and $\sigma_L^2(\tau)$, as 
stated in Eq.~\eqref{sup_sn_mean} and Eq.~\eqref{sup_sn_var}, are 
sufficient to reconstruct $f(L)$ and $D$ for general $\alpha < 0$ and in the limit of small noise.

The two-time covariance function $C_L(\tau,\tau')$ is highly informative for force reconstruction. If enough sample paths are 
available ($\mathcal{O}(100) - \mathcal{O}(1000)$), the covariance stated in 
Eq.~\eqref{sup_sn_cov} is suitable to uniquely infer both the general form of the forward force $f(L)$ 
and the noise level.

\begin{figure}[ht]
\centerline{\includegraphics[width=0.95\linewidth]
{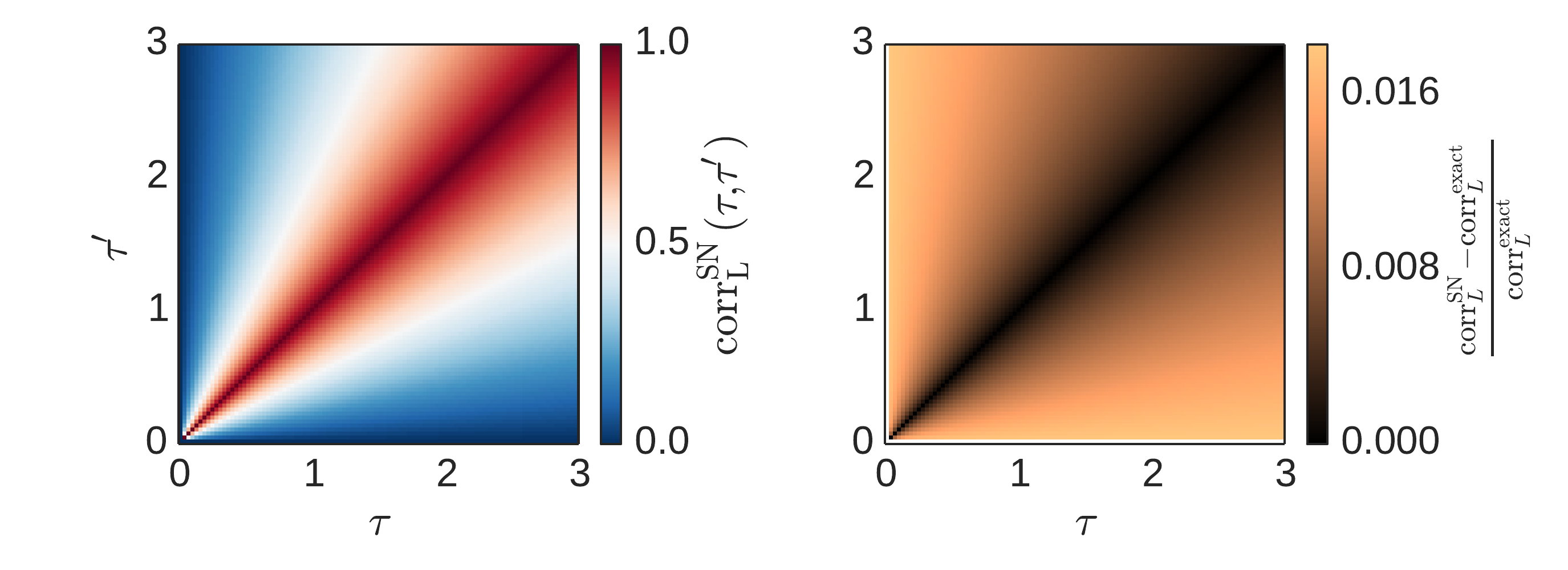}}
\caption{
\textbf{
The small noise approximation of the correlation function is an excellent approximation of the exact dynamics}. 
Shown is the small noise correlation function as given by 
Eq.~\eqref{sup_sn_corr} for the case $\alpha = -1$ (Left) and its deviations from 
the known 
exact correlation function of the Bessel process (Right) directly 
obtainable from Eq.~\eqref{sup_ne_constriction_var} and 
Eq.~\eqref{sup_ne_constriction_cov}.
}\label{compCorrBesselSN}
\end{figure}

The fraction $C_L(\tau,\tau')/(\sigma_L(\tau)\sigma_L(\tau)')$ represents the two 
time 
correlation function $\mathrm{corr}_L(\tau,\tau')$.
With its sole dependence on $\alpha$, $\mathrm{corr}_L(\tau,\tau')$ uniquely 
fixes the general form of the force law, irrespective of the reconstructed 
$\gamma$ and the noise level $D$. To check 
for the quality of Eq.~\eqref{sup_sn_corr}, we compare in 
Fig.~\ref{compCorrBesselSN} the case $\alpha = -1$ to its analytical 
counterpart obtained from Eq.~\eqref{sup_ne_constriction_var} and 
Eq.~\eqref{sup_ne_constriction_cov}. 
Using the same parameters as already in 
Fig.~\ref{ExactTheoVsSNtheo} we observe in Fig.~\ref{compCorrBesselSN} (Right) 
a 
maximal deviation of $2\%$ for large correlation times, and even smaller 
deviations for shorter correlation times.

\begin{figure}[ht]
\centerline{\includegraphics[width=0.8\linewidth]
{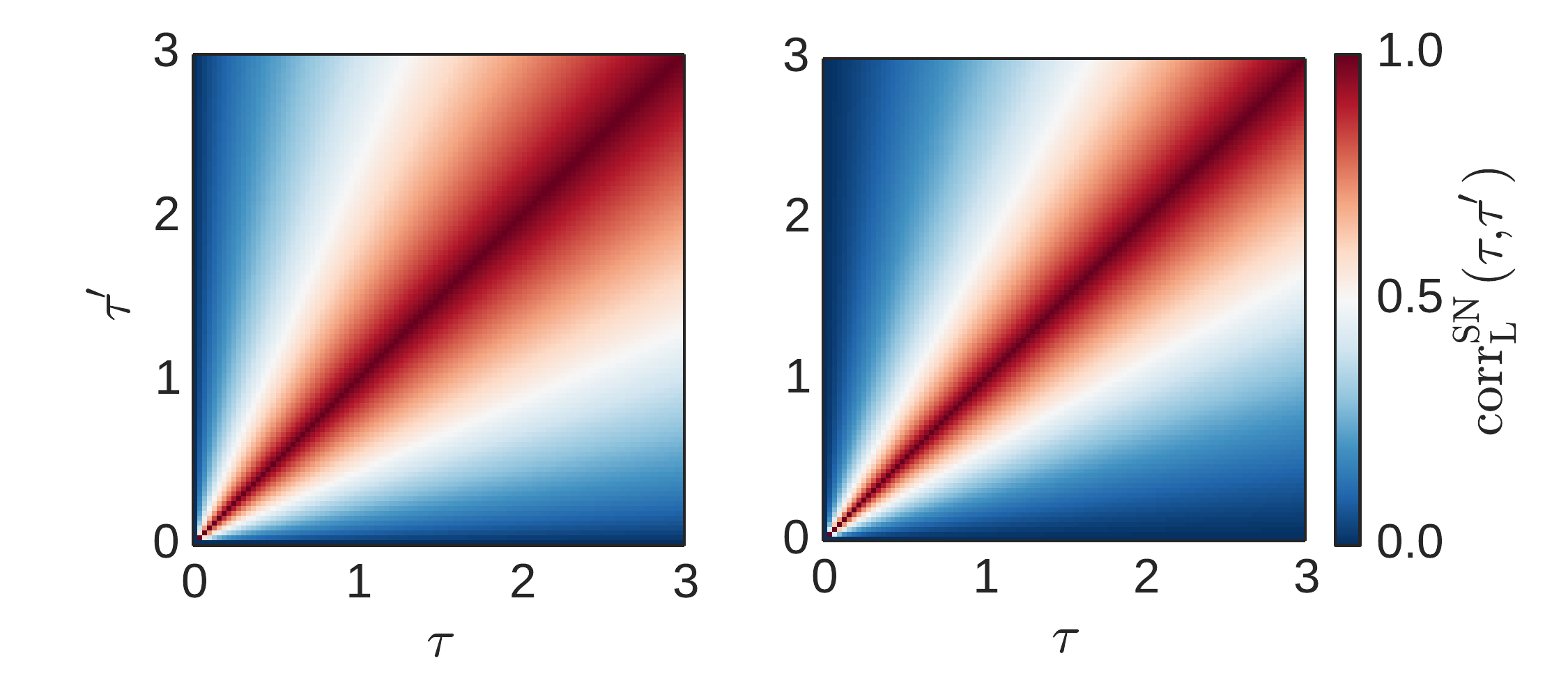}}
\caption{
\textbf{
Exemplary correlation functions obtained from the analytic small noise 
expression as stated in Eq.~\eqref{sup_sn_corr}}. Shown are the cases $\alpha=-1/2$ 
(Left) and $\alpha = -2$ (Right). The two remaining parameters are chosen 
identical with $\gamma =1$ and $D = 0.2$. For both we observe characteristic 
straight equi-value lines as demanded by Eq.~\eqref{sup_sn_corr}.
}\label{corrSNx2}
\end{figure}

Interestingly, the 2d representation of $\mathrm{corr}_L(\tau,\tau')$ as 
stated in Eq.~\eqref{sup_sn_corr} and displayed for the cases $\alpha = 
- 0.5$ and $\alpha = -2$ in Fig.~\ref{corrSNx2},
shows an easy to identify signature. 
Due to the sole dependence of Eq.~\eqref{sup_sn_corr} on the 
ratio of $\tau$ and $\tau'$ the 2d correlation function exhibits straight level lines ($\tau' = \mathrm{const} \cdot \tau$).
Given we are in the regime of small 
noise and do not observe this straight line structure in our experimental data, 
we can thus directly exclude all pure forward force laws of the form stated in 
Eq.~\eqref{sup_alpha_force_law} with $\alpha <0$. We 
further know from above, that processes 
with $\alpha \ge 0$ follow a 
force law $D/L$ close to completion. 
The respective correlation function can be calculated and is stated in 
Eq.~\eqref{sup_rt_rw_corr}. Connecting the two cases $\alpha <0$ and $\alpha>0$ we now have a theoretical understanding of the mean, variance and two-time correlation close to target states.





\subsubsection{Weak noise approximation of moments for force driven TSA ensembles}
\label{derive_small_noise_moments}
Starting from Eq.~\eqref{sup_tr_lv_sn}, we here derive the expressions Eq.~\eqref{sup_sn_mean}-Eq.~\eqref{sup_sn_corr} for mean $\overline{L}(\tau)$, variance  
$\sigma_L^2(\tau)$, covariance $C_L(\tau,\tau')$ and the correlation function 
$\mathrm{corr}_L(\tau,\tau')$. The calculation is valid in the domain 
$\alpha <0$, as discussed above, and in the limit 
of small noise. With some minor modifications we follow the weak noise approximation 
approach discussed in Gardiner (chapter 6.2)\cite{gardiner1985handbook}.

The derivation starts with rewriting Eq.~\eqref{sup_tr_lv_sn} in the 
general form
\begin{equation}
 dL = a(L) d\tau + \epsilon^2 b(L) d\tau + \epsilon \; d W_\tau
\end{equation}
with $\sqrt{D}$ substituted by the expansion parameter $\epsilon$. For small 
$\epsilon$ we expand
\begin{equation}
\label{sup_epsilon_expansion}
 L(\tau) = L_0(\tau) + \epsilon L_1(\tau) + \epsilon^2 L_2(\tau) + ...
\end{equation}
around the deterministic solution $L_0(\tau)$. We further assume that we can 
write
\begin{equation}
 a(L) = a(L_0 + \epsilon L_1 + \epsilon^2 L_2 + ...)
      = a_0(L_0) + \epsilon a_1(L_0,L_1) + \epsilon^2 a_2(L_0,L_1,L_2)
      + ... \ .
\end{equation}
The general expression for $a(L)$ then reads
\begin{equation}
 a(L) = a\left(L_0 + \sum_{m=1}^\infty \epsilon^m L_m \right)
 = \sum_{p=0}^\infty \frac{1}{p!} \frac{d^p a(L_0)}{d L_0} 
 \left(
 \sum_{p=0}^\infty \epsilon^m L_m
 \right)^p
\end{equation}
and after sorting terms we find for the first three terms
\begin{align}
 &a_0(L_0) = a(L_0) 
 \\
 &a_1(L_0,L_1) = L_1 \frac{d a(L_0)}{d L_0}
 \\
 \label{sup_a2}
 &a_2(L_0,L_1,L_2) = L_2 \frac{d a(L_0)}{d L_0} 
 + \frac{1}{2} L_1^2 \frac{d^2 a(L_0)}{d L_0^2}
 \ .
\end{align}
The same procedure can be applied to $b(L)$. Taken together the equations to 
lowest order in $\epsilon$ are
\begin{align}
\label{sup_dx0}
 &d L_0 = a(L_0) d\tau
 \\
\label{sup_dx1}
 &d L_1 = a_1(L_1,L_0) d\tau + d W_\tau
 \\
 \label{sup_dx2}
 &d L_2 = a_2(L_2,L_1,L_0) d\tau + b(L_0) d\tau
\end{align}
The first of these equations is the deterministic equation 
\begin{equation}
\label{sup_sn_Lzero}
 dL_0 = \gamma L_0^\alpha d\tau
 \ .
\end{equation}
As the process starts at the boundary at $L_0(0)=0$, its solution is simply 
given as
\begin{equation}
\label{sup_x0_res}
L_0 = \left( (1-\alpha) \gamma \tau \right)^{\frac{1}{1-\alpha}} 
\qquad \mathrm{with} \ \alpha <0
\ .
\end{equation}
The second equation Eq.~\eqref{sup_dx1} describes a generalized Ornstein-Uhlenbeck process
\begin{equation}
\label{sup_dx1_expl}
 d L_1 = k\left(L_0(\tau)\right) L_1 \; d\tau + d W_\tau
\end{equation}
with the time-dependent drift coefficient
\begin{equation}
 k\left(L_0\right) = \frac{d a(L_0)}{d L_0} = \gamma \frac{d L_0^\alpha}{d L_0}
	= \gamma \alpha L_0^{\alpha-1}
\ .
\end{equation}
Inserting the solution for $L_0(\tau)$ we find
\begin{equation}
\label{sup_k}
 k(L_0) = \gamma \alpha \left( (1-\alpha) \gamma \tau \right)^{-1}
 \ .
\end{equation}
Employing the initial conditions $L_1(0) = 0$, the solution of 
equation Eq.~\eqref{sup_dx1_expl} reads
\begin{equation}
\label{sup_x1_res}
 L_1(\tau) = \int_0^\tau  e^{
 \int_{\tau'}^\tau k(L_0(s)) ds
 }
 d W_{\tau'}
 = 
 \int_0^\tau
 \left(\frac{\tau}{\tau'}\right)^{-\frac{\alpha
   }{\alpha -1}}
   d W_{\tau'} \ .
\end{equation}
Together, Eq.~\eqref{sup_x0_res} for $L_0(\tau)$ and Eq.~\eqref{sup_x1_res} 
for $L_1(\tau)$ form the basis of all further calculations. Inserting these 
solutions into the expansion stated in Eq.~\eqref{sup_epsilon_expansion}
and taking the ensemble average we obtain expressions for mean
\begin{align}
\langle L(\tau) \rangle 
&= 
\langle L_0(\tau) \rangle 
+ \epsilon \langle L_1(\tau) \rangle + ... \ ,
\intertext{variance}
\label{sup_def_sn_var}
\sigma_L^2(\tau)
&=
\langle
\left(
L(\tau) - \langle L(\tau) \rangle
\right)^2
\rangle
=
\epsilon^2 \langle L_1^2(\tau) \rangle + ...
\intertext{and two the time covariance}
 C(\tau,\tau') &= 
\langle
\left(
L(\tau) - \langle L(\tau) \rangle
\right)
\left(
L(\tau') - \langle L(\tau') \rangle 
\right)
\rangle
=
\epsilon^2 \langle L_1(\tau) L_1(s) \rangle + ...
\end{align}
in lowest order of $\epsilon$ and valid for $\alpha <0$.
We first evaluate these expressions to lowest order, which for 
$\sigma_L^2(\tau)$ and $C_L(\tau,\tau')$ is already at order 
$\mathcal{O}(D)$. Corrections for the mean up to order $\mathcal{O}(D)$ are 
evaluated subsequently.
Note, that the average over a single Wiener integral, such as $\langle L_1(\tau) 
\rangle$ always evaluates to zero and therefore corrections to the mean will be 
of order $\epsilon^2$.

To 0th order, the solution for the mean is equivalent to the 
deterministic solution Eq.~\eqref{sup_x0_res}. For the variance, we have to 
calculate the second moment of equation Eq.~\eqref{sup_x1_res}, i.e.
\begin{align}
\label{sup_var_epsilon_expansion}
\sigma_L^2(\tau)
=
 D \langle L_1^2(\tau) \rangle 
 &=
  D 
  \int_0^\tau
  \int_0^\tau
   \left(\frac{\tau}{\tau'}\right)^{-\frac{ \alpha
   }{\alpha -1}}
 \left(\frac{\tau}{\tau''}\right)^{-\frac{ \alpha
   }{\alpha -1}}
   d W_{\tau'}
   d W_{\tau''}
   \notag \\
 &=  
 D \int_0^\tau
 \left(\frac{\tau}{\tau'}\right)^{-\frac{2 \alpha
   }{\alpha -1}}
   d \tau'
   =
   D \frac{1 - \alpha}{1 - 3 \alpha}\tau
   \qquad \mathrm{for} \ \alpha <0
   \ ,
\end{align}
where we replace $\epsilon^2$ by the original $D$. Note that $\langle L_1 
\rangle^2 = 0$ and thus drops out of the calculation. 
Similarly, we obtain the two-time covariance function
\begin{align}
\label{sup_2time_cov_epsilon_expansion}
 C(\tau,\tau') 
&=
D 
\int_0^{\tau} 
\int_0^{\tau'} 
  \left(\frac{\tau}{s}\right)^{-\frac{ \alpha
   }{\alpha -1}}
     \left(\frac{\tau'}{s'}\right)^{-\frac{ \alpha
   }{\alpha -1}}
   dW_s dW_{s'}
 \notag \\
 &= D \int_0^{\mathrm{min}(\tau,\tau')} 
  \left(\frac{\tau}{s}\right)^{-\frac{ \alpha
   }{\alpha -1}}
     \left(\frac{\tau'}{s}\right)^{-\frac{ \alpha
   }{\alpha -1}}
   d s
\notag
   \\
   &=
\begin{cases}   
   D \frac{(\alpha -1) \tau^{\frac{1}{\alpha -1}+2} \tau'^{-\frac{\alpha
   }{\alpha -1}}}{3 \alpha -1} 
   &\qquad \mathrm{for} \ \tau < \tau'
   \\
   D \frac{(\alpha -1) \tau'^{\frac{1}{\alpha -1}+2} \tau^{-\frac{\alpha
   }{\alpha -1}}}{3 \alpha -1} 
   &\qquad \mathrm{for} \ \tau > \tau' 
   \\
   D \langle L_1^2(\tau) \rangle
   &\qquad \mathrm{for} \ \tau = \tau'
\end{cases}
\qquad \mathrm{and} \ \alpha <0
\end{align}
again in terms of $D$.
Using the variance Eq.~\eqref{sup_var_epsilon_expansion}, we 
can 
rewrite this result in the final form stated in Eq.~\eqref{sup_sn_var}.
The two-time correlation function also directly follows from 
Eq.~\eqref{sup_2time_cov_epsilon_expansion}. Using the definition
\begin{align}
\mathrm{Corr}(\tau,\tau') := 
\frac{C(\tau,s)}{\sigma_L(\tau) \sigma_L(\tau')}
\end{align}
we obtain the final form stated in Eq.~\eqref{sup_sn_corr}.

In a last step we calculate the first $D$-dependent contribution to the mean, 
characterizing deviations from the deterministic solution. As contributions due 
to $L_1(\tau)$ average out, terms of order $\epsilon^2$ become important
and thus the contribution $-\alpha D / L$  of the free energy 
force as stated in Eq.~\eqref{sup_tr_lv_sn}. Using the series expansion 
of $L(\tau)$ in orders of $\epsilon$ as stated in 
Eq.~\eqref{sup_epsilon_expansion}, 
we next evaluate the contribution of $L_2(\tau)$ to the mean. This is most 
simply 
achieved by directly evaluating an averaged Eq.~\eqref{sup_dx2}. 
Using the specifications due to Eq.~\eqref{sup_a2} we then obtain a linear 
differential equation
\begin{equation}
 \frac{d \langle L_2 \rangle}{d\tau}
 =
 \frac{d a(L_0)}{dL_0} \langle L_2 \rangle
 + \frac{1}{2} \frac{d^2 a(L_0)}{d L_0^2} \langle L_1^2 \rangle
 -\frac{\alpha}{L_0}
\end{equation}
with the solution
\begin{equation}
\label{L_2_avg_homog}
\langle L_2(\tau) \rangle
=
 \frac{(7 \alpha -3) ((1 - \alpha) \gamma  \tau)^{\frac{\alpha
   }{\alpha -1}}}{4 (3 \alpha -1) \gamma }
\qquad \mathrm{for} \ \alpha <0
\ .
\end{equation}
The final result Eq.~\eqref{sup_sn_mean} is obtained from
\begin{align}
 \overline{L}(\tau) = L_0(\tau) + D \langle L_2(\tau) \rangle \ .
\end{align}


\section{A model of cytokinetic ring constriction with distinct effective force laws}
To assess whether TSA ensemble analysis can distinguish different regimes of directed dynamics we here study the exemplary process of Cytokinesis. 
Cytokinesis is the process of separation of a mother cell into two daughter cells at the end of the cell cycle. For animal cells this separation is driven by a contractile ring formed from actin filaments, myosin and associated proteins. 
We used a model that describes the contraction and molecular turnover of force generating acto-myosin bundles, i.e.~an assembly of filaments and motor complexes\cite{zumdieck2007stress}.
Two coupled differential equations define the model. The first describes how the balance of viscous and elastic forces determines the ring diameter and 
leads to ring constriction or dilation. The viscous force is written as $-\xi^{-1}\dot{R}$, where $\dot{R}$ denotes the time derivative of the ring radius and $\xi$ an effective friction coefficient. The elastic response of the cell upon cell deformation derives from a harmonic energy term  
\begin{equation}
E(R)=\frac{K}{2}(R-R_0)^2
\end{equation}
with elastic modulus $K>0$ and equilibrium radius $R_0$.
Combined with a term $\Sigma$ for the contractile stress, we obtain a force balance equation of the form
\begin{equation}
\label{eq:ring_dynamics}
 \dot{R}=-\xi \left(
 \frac{\partial E}{\partial R}
 + 2 \pi \Sigma
 \right) \ .
\end{equation}
The dynamics of the filament concentration $c = N/(2 \pi R)$ with $N$ filaments is defined along a ring of length $2 \pi R$. Without turnover the ring perimeter therefore changes with $R \propto 1/c$. Including turnover we obtain the material balance equation
\begin{align}
\label{eq:material_balance}
\dot{c}=k_p -k_d c -\frac{\dot{R}}{R} c
\end{align}
with filament polymerization and depolimerization rates $k_p$ and $k_d$. 
The contractile stress $\Sigma=A N_b c^2$ models the active force generation of actin filaments and myosin motors.
$A$ is the (positively defined) effective material coefficient, $N_b$ the number of distinct filament bundles and $c$, the number density of filaments per unit length along a bundle.  The contractile force is proportional to $c^2$ to account for the fact that actin filaments need a partner to exert forces. 
Together Eq.~\eqref{eq:ring_dynamics} and Eq.~\eqref{eq:material_balance} define an elementary model of cytokinetic ring constriction.

Zumdieck et.~al.~ used parameters from independent measurements curated from the literature\cite{zumdieck2007stress}. Comparing numerical solutions of the model, using these parameters
with time laps recordings of the first-division constriction dynamics of C.~elegans oocytes, Zumdieck et.~al.~found good agreement between model and measurement.

To allow for different stochiometrics of the potential force generating molecules such as actin and myosin, and to include a stochastic component, we slightly generalized this model dynamics
\begin{align}
\label{cytokin_zumdiek_1}
 d\widehat{L}(t) &= - \left( \gamma_1^\lambda (\widehat{L}(t) - \widehat{L}_m) + \gamma_2^\lambda \widehat{c}(t)^n \right) dt
  + \sqrt{D} \  d W_t
 \\
\label{cytokin_zumdiek_2}
 d \widehat{c}(t) &= \left(
 k_\mathrm{on}^\lambda -k_\mathrm{off}^\lambda \widehat{c}(t)
 \right) dt
  + 
 \frac{d\widehat{L}(t)}{\widehat{L}(t)} \widehat{c}(t)
 \ .
\end{align}
We define $\widehat{L}:=R$, $\gamma_1^\lambda:=K \xi$, $\widehat{L}_m:=R_0$, $\gamma_2^\lambda:=2 \pi A N_b$, $k_\mathrm{on}^\lambda:=k_p$ and $k_\mathrm{off}^\lambda:=k_d$. 
To allow for single and pairwise interactions of the force generating molecules we replace the exponent of the concentration by a parameter $n:=\{1,2\}$. In this version, the index $\lambda:=\{m,c\}$ further accounts for the possibility to switch between two different regimes. A maintenance ($m$) phase where a fixed ring perimeter is stabilized, and a constriction ($c$) phase where the ring diameter shrinks until cell separation. For simplicity the switching between the two regimes  $m \to c$ is modeled as a discrete shift in parameter space 
$(\gamma_1^m,\gamma_2^m,k_\mathrm{on}^m,k_\mathrm{off}^m) \to
(\gamma_1^c,\gamma_2^c,k_\mathrm{on}^c,k_\mathrm{off}^c) 
$. 
Both generalizations, stochiometrics of myosin and stochasticity, were already suggested by the original authors.~\cite{zumdieck2007stress}.

Below we show that this generalized model exhibits three qualitatively different biophysically plausible regimes, which can be approximated with a single effective SDE of the form 
\begin{align}
\label{sup_effective_cytokin_langevin}
 d\widehat{L}(t) = - \gamma_\mathrm{eff} \widehat{L}^\alpha + \sqrt{D} \; dW_t
\end{align}
with $\alpha=0,-1,-2$. These three effective dynamics are recovered from Eq.~\eqref{cytokin_zumdiek_1} and Eq.~\eqref{cytokin_zumdiek_2}, for the biological meaningful limits of fast turnover, stalled myosin turnover, and stalled actin turnover.

\subsection{The fast turnover limit}
In the limit of fast turnover ($k_\mathrm{on}^\lambda \gg 1$, $k_\mathrm{off}^\lambda \gg 1$) and $k_\mathrm{on}^\lambda / k_\mathrm{off}^\lambda=\mathrm{const.}$, the above model reduces to 1d dynamics of the ring perimeter $\widehat{L}$ which, during the contraction phase, is driven by a constant force $\gamma_\mathrm{eff}$. 
In this limit $\widehat{c}(t)$ relaxes very fast to its equilibrium concentration $k_\mathrm{on}^\lambda / k_\mathrm{off}^\lambda$. The remaining SDE for the ring perimeter $\widehat{L}$ is
\begin{align}
\label{high_turnover_1d}
d\widehat{L}(t) &= 
- 
\left( \gamma_1^\lambda \left( \widehat{L} - \widehat{L}_m \right)
+
\gamma_2^\lambda \left(\frac{k_\mathrm{on}^\lambda}{k_\mathrm{off}^\lambda}\right)^n
%
\right) dt
+ \sqrt{D} \;  d W_t
 \ .
\end{align}
Assuming a very ``soft'' cell shape and a strong constriction force we set $\gamma_1^c =0$ and find an equation of the form of Eq.~\eqref{sup_effective_cytokin_langevin}. 
This is the force to deform the entire cell.
An analytic expression for the reverse time force close to completion can be obtained using Eq.~\eqref{sup_tr_lv_tsa_CFB}. It turns out that the effect of $\gamma_1 > 0$ on the moments of the aligned reverse time ensemble close to completion is manure. Only the mean changes marginally and rises slightly slower than in the case with $\gamma_1 = 0$.

\subsection{Stalled turnover limits}
In the limit of stalled turnover 
($k_\mathrm{on}^\lambda  \approx 0$, $k_\mathrm{off}^\lambda \approx 0$)
all force generating molecules are trapped in the ring over the time scale of the constriction phase. We thus set $k_\mathrm{on}^c$ and $k_\mathrm{off}^c$ to zero. The concentration dynamics from Eq.~\eqref{cytokin_zumdiek_2} and during constriction then reduces to 
\begin{equation}
 \frac{d(\widehat{c}(t) \widehat{L}(t))}{dt}=\dot{\widehat{c}}(t)\widehat{L}(t) + \widehat{c}(t)\dot{\widehat{L}}(t) = 0 \ ,
\end{equation}
implying a fixed number of force generating molecules on the ring. In this limit
\begin{equation}
 \widehat{c}(t) \widehat{L}(t) = \mathrm{const.}
\end{equation}
As this is maintained during constriction, the constant value is inherited from the value of this product at the end of the maintenance phase, the switching time $t_\mathrm{sw}$. We can therefore replace $\widehat{c}(t)$ in the perimeter dynamics Eq.~\eqref{cytokin_zumdiek_1}, and obtain
\begin{align}
\label{stalled_turnover_1d}
 d\widehat{L}(t) = - \left( \gamma_1^c (\widehat{L} - \widehat{L}_m) 
 + \gamma_2^c \left(\frac{\overline{ \widehat{c}(t_\mathrm{sw})} \overline{\widehat{L}(t_\mathrm{sw})} }{\widehat{L}(t)} \right)^n \right) dt
  + \sqrt{D} \  d W_t
\end{align}
with $\overline{\widehat{c}(t_\mathrm{sw}) }$ approximated by the ratio of $k_\mathrm{on}^m/k_\mathrm{off}^m$ and
\begin{align}
\overline{\widehat{L}(t_\mathrm{sw})}= \widehat{L}_m - \frac{\gamma_2^m}{\gamma_1^m} \overline{\widehat{c}(t_\mathrm{sw})} \ ,
\end{align}
where $\overline{\widehat{c}(t_\mathrm{sw})}$ and $\overline{\widehat{L}(t_\mathrm{sw})}$ are the average values during the maintenance phase.
For the example displayed in the main text, we again assume dominance of the constriction force ($\gamma_1^c=0$) during the constriction phase. 
Note that, as the concentration of force generating molecules diverges, this is a particular mild assumption here.
The effective dynamics than terminates either $\propto 1/\widehat{L}$ or $ \propto 1/\widehat{L}^2$ 
depending on the model variants ($n=1$: myosin; $n=2$: actin) for the force generating molecule.

\subsection{Simulations}
Sample paths for the three different constriction scenarios with constriction forces constant ($\alpha=0$), proportional to $\propto 1/\widehat{L}$ ($\alpha=-1$) or $ \propto 1/\widehat{L}^2$ ($\alpha=-2$) were generated from Eq.~\eqref{cytokin_zumdiek_1} and Eq.~\eqref{cytokin_zumdiek_2}. 
All simulations were started from the equilibrium distribution of the maintenance phase. After a time $t_\mathrm{sw}$ (chosen larger than the window used for inference), the dynamics for each sample path was switched to the constriction regime.
The case corresponding to $\alpha=-1$ is provided in Fig.~6 of the main text. Here we additionally display the two cases with $\alpha=0,-2$ (see Fig.~\ref{fig:meanVarCytokinX2}). Tab.\ref{cytokin_parameter} lists the parameter settings for all simulations.
%
\begin{table} [h!]
\resizebox{\textwidth}{!}{%
\centering
\begin{tabular}{LCCCCCC}
\toprule
\multicolumn{1}{l}{Parameters} &
\multicolumn{2}{c}{Simulation 1 ($\alpha=0$)}    &
\multicolumn{2}{c}{Simulation 2 ($\alpha=-1$)}    &
\multicolumn{2}{c}{Simulation 3 ($\alpha=-2$)}    \\ 
\cmidrule(lr){2-3}
\cmidrule(lr){4-5}
\cmidrule(lr){6-7}

&
\multicolumn{1}{c}{maintainance} &
\multicolumn{1}{c}{constriction}     &
\multicolumn{1}{c}{maintainance} &
\multicolumn{1}{c}{constriction}     &
\multicolumn{1}{c}{maintainance} &
\multicolumn{1}{c}{constriction}     \\
\midrule

n & 1 & 1 & 1 & 1 & 2 & 2 \\
k_\mathrm{on}\, / \mu\mathrm{m}\, s^{-1} & 400 & 2000 & 0.5 & 0 & 0.7 & 0\\
k_\mathrm{off}\, /  s^{-1}  & 10 & 20 &  0.01 & 0 & 0.1 & 0\\
\gamma_1\, /  s^{-1}  & 0.075 & 0 & 0.075 & 0 & 0.075 & 0 \\
\gamma_2\, / \mu\mathrm{m}^{1+n}\, s^{-1}  & 2 \pi \cdot 3.6 \cdot 10^{-5} & 2 \pi \cdot 3.6 \cdot 10^{-5} & 2 \pi \cdot 3.6 \cdot 10^{-5} & 2 \pi \cdot 3.6 \cdot 10^{-5} & 2 \pi \cdot 3.6 \cdot 10^{-5} & 2 \pi \cdot 3.6 \cdot 10^{-5}\\
\widehat{L}_m\, / \mu\mathrm{m} & 14.5 & 14.5 & 14.5 & 14.5 & 14.5 & 14.5 \\
D\, / \mu\mathrm{m}^2\, s^{-1}  & 0.04 & 0.04 & 0.04 & 0.04 & 0.04 & 0.04\\
\bottomrule
\end{tabular}}
\caption{
\textbf{
Parameter settings used for 3 different sample path ensembles created from Eq.~\eqref{cytokin_zumdiek_1} and Eq.~\eqref{cytokin_zumdiek_2}.} 
Simulation 2 is discussed in the main text.
}\label{cytokin_parameter}
\end{table}
For integration we used a standard Euler scheme with time step $\Delta t = 0.05s$.

\subsection{Maximum likelihood path ensemble inference}
To infer the most likely constriction scenario hidden in each of the forward simulations as defined in Table~\ref{cytokin_parameter}, we maximize the path likelihood of the respective ensemble after alignment and time reversion. 
We implement the approach discussed in section \ref{reverse_time_ensemble_path_inference}. We infer the reverse time force from the TSA ensemble and reconstruct the underlying forward force. We assumed the forward force law is of power law form $f(\widehat{L})=-\gamma \widehat{L}^\alpha$. We can therefore directly read off the forward force $f(L)$ from the inferred reverse time force $f^\mathrm{TSA}(L)$.

For each of the three simulations defined in Table \ref{cytokin_parameter} the inference procedure is identical. We here discuss Simulation 2 (with true effective $\alpha=-1$) shown in the main text. The idea is to treat the simulation data as black box, as experimental data would be, and determine which forward force law $f(\widehat{L})$ is most suitable to describe the constriction dynamics.
In principle the maximal included lifetime of the reverse time trajectories $N \Delta \tau$ should be optimized as well. Starting from very short trajectories for which contributions of the maintenance phase are negligible, and increasing the length until
distinctions from a pure ensemble become detectable, then
marks the transition from constriction to maintenance. We here, for simplicity, directly choose $N$ based on visual inspection of the aligned reverse time ensemble.

There are two options to find the best model. First, to simultaneously optimize for $\alpha,\gamma_\mathrm{eff},D$. Second to to choose one $\alpha$ from the three above described mechanisms, assuming it contains the correct model, and infer the best fitting $\gamma_\mathrm{eff},D$. With the first approach we find the best model with its maximum likelihood parameters.
With the latter we find the best fitting parameters for each of the candidates ($\alpha=0,-1,-2$). This then allows to quantitatively compare the three different models either in terms of their maximum likelihood values or, in terms of their moments which we each calculate from simulations of the respective reverse time SDE Eq.~\eqref{sup_tr_lv_ness_alpha_force} using the newly inferred maximum likelihood parameter $\gamma_\mathrm{MAP},D_\mathrm{MAP}$. 

\begin{figure}
\centering
\includegraphics[width=.8\linewidth]{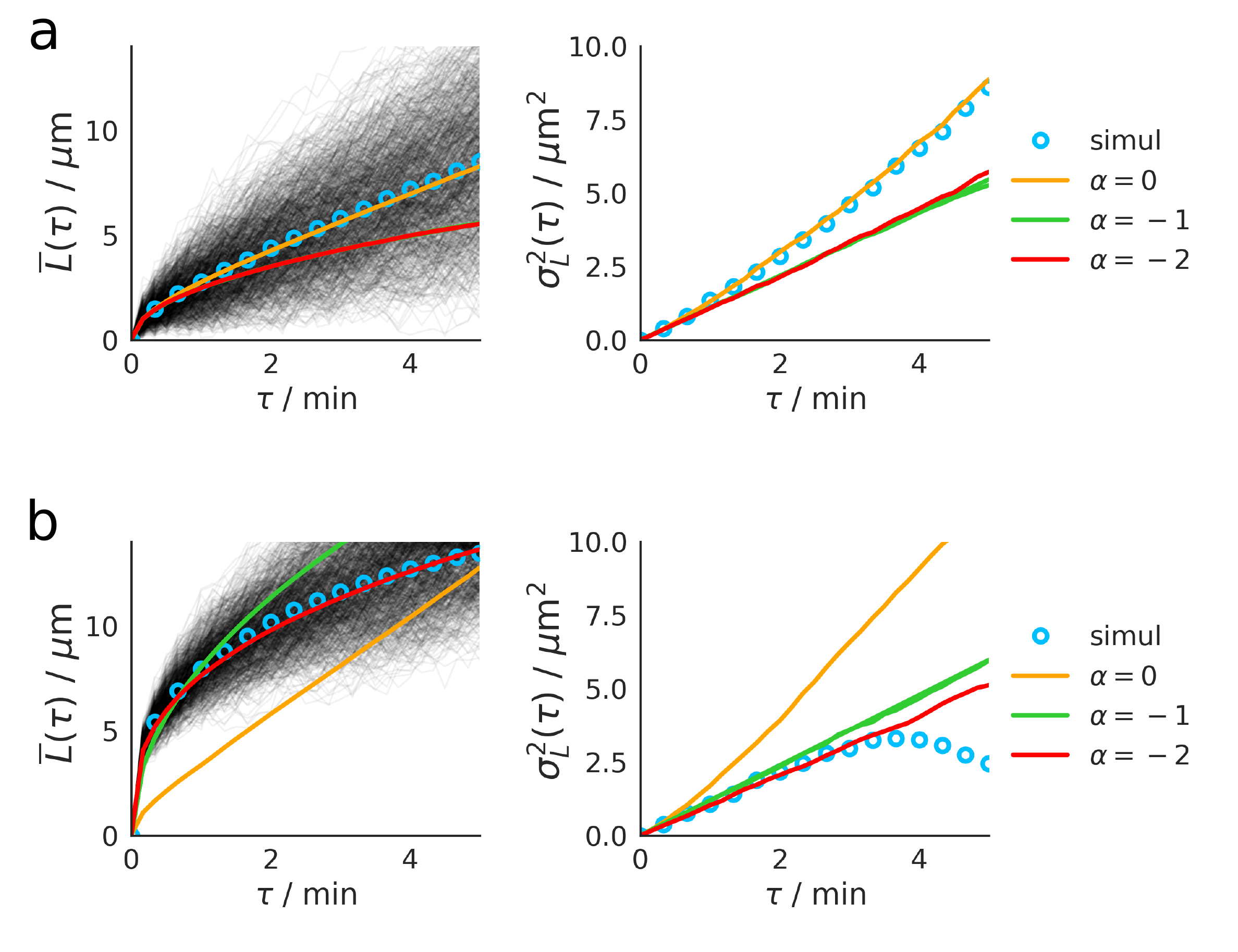}
	\caption{
	\textbf{
    The dynamical law leading to constriction can unambiguously be inferred from realizations of the Zumdieck model, using path integral maximum likelihood inference Eq.~\eqref{loglike}.}
    We visually confirm the inferred underlying force law $f(\widehat{L})$ by numerically evaluating mean and variance for the reverse-time dynamics under the assumption of well seaprated initial and final conditions, for each of the three inferred maximum likelihood parameter sets $(\gamma_{\mathrm{ML}}^\alpha,D_{\mathrm{ML}}^\alpha)$
    \textbf{(a):}
    Given the correct underlying force law is $f(\widehat{L})=-\gamma$ ($\alpha=0$),
    only the case $\alpha=0$ (yellow line) predicts both the mean and variance of the Zumdieck model (blue circles).
    \textbf{(b):}
    Given the correct underlying force law is $f(\widehat{L})=-\frac{\gamma}{\widehat{L}^2}$ ($\alpha=-2$),
    only the case $\alpha=-2$ (red line) predicts both the mean and variance of the Zumdieck model (blue circles).
    }
    \label{fig:meanVarCytokinX2}
\end{figure}

For Simulation 2 (with true effective $\alpha=-1$) the results of this comparison are shown in Fig.~6 of the main text. 
For the ``true'' case $\alpha=-1$, the full terminal aligned simulation and the inferred dynamics perfectly coincide in their moments. For $\alpha=0,-2$ they strongly deviate.
A similar observation holds for Simulation 1 and 3 shown in Fig.~\ref{fig:meanVarCytokinX2}. For both, the correct dynamics are recovered while the deviant models can not be matched in their statistics.
In total this shows that our maximum likelihood path ensemble inference approach for reverse time analysis allows to reliable infer the correct underlying mechanism while rejecting deviant models. The theory of TSA ensemble inference thus provides a reliable framework for the inference of effective dynamics. In the following sections we will explore generalizations to this baseline framework.

\section{Reverse-time ensemble path inference}
\label{reverse_time_ensemble_path_inference}
The reverse time TSA ensemble approach aims at the inference of the underlying forward dynamics from dynamics with identifiable dynamics only close to the target state. 
We here propose to directly infer the reverse time force $f^\mathrm{TSA}(L)$ from TSA sample paths using a path-ensemble-likelihood inference scheme. 

The construction idea of this likelihood inference scheme follows classical maximum likelihood and Bayesian path inference approaches\cite{el2015inferencemap,masson2014mapping,tuerkcan2012bayesian,masson2015pitfalls,dargatz2010bayesian,durham2002numerical,hurn2007seeing}. We first define the transition probability for discrete sample paths starting at $L_\mathrm{ts}$ at time $\tau=0$ and ending at $\tau_f$ for simplicity. Sample points are taken with equal spacing $\Delta \tau$. We considering sample paths of duration $T=N \Delta \tau$. The transition probability for individual realizations of such a Markovian process reads
\begin{align}
 P(Lf,\tau_f|L_\mathrm{ts},0)
 =
 \prod_{i=0}^{N-1}
  P(L_{i+1},\tau_{i+1}|L_{i},\tau_i)
  \ .
\end{align}
Assuming individual, statistically independent sample paths, the total likelihood of a finite size sample $n_\mathrm{ens}$ of sample paths is the product of individual transition probabilities. Taking the logarithm the log likelihood function is
\begin{align}
\log \mathcal{L}
 =
 \sum_{j=1}^{n_\mathrm{ens}}
 \sum_{i=0}^{N-1}
 \log
  P(L^{(j)}_{i+1},\tau_{i+1}|L^{(j)}_{i},\tau_i)
\end{align}
which can be used for maximum likelihood inference of model parameter. The exact transition probability $P(L_{i+1},\tau_{i+1}|L_{i},\tau_i)$ is not available analytically for most stochastic processes. 
A simple approximation for $P(L_{i+1},\tau_{i+1}|L_{i},\tau_i)$ useful for closely sampled trajectories is of Gaussian form
\begin{align}
P(L_{i+1},\tau_{i+1}|L_{i},\tau_i)
=
 \frac{1}{\sqrt{2 \pi D \Delta \tau} } 
 e^{
 - \frac{
 \left(
 L_{i+1} - L_{i} - f^\mathrm{TSA}(L_{i}) \Delta \tau
 \right)^2
 }
 {
 2 D \Delta \tau
 }
 }
\ .
\end{align}
This is equivalent to a first order stochastic simulation of an SDE in Ito interpretation, where the change $dL$ with respect to the deterministic force times $\Delta \tau$ plus a random step with zero mean and variance $D \Delta \tau$ is evaluated.
The log-likelihood under the approximation of Gaussian transition probabilities thus reads
\begin{align}
\label{loglike}
 \log \mathcal{L} =
 \sum_{j=1}^{n_\mathrm{ens}} \sum_{i=0}^{N-1}
 \left(
 -\frac{ \left(L^{(j)}_{i+1}-L^{(j)}_{i} -f^\mathrm{TSA}(L^{(j)}_i) \Delta \tau \; \right)^2}{2 D \Delta \tau} 
 -  \frac{1}{2}\log(2 \pi D \Delta \tau) 
 \right)
 \ .
\end{align}
This approximation becomes exact for arbitrary small time steps, and deviations occur as the time-steps become bigger. It is therefore recommended to check after optimization, if the residual increments after subtraction of the deterministic part are actually Gaussian. If not, more sophisticated path inference techniques can be used\cite{ozaki1985statistical,kessler1997estimation,elerian1998note}, which can cope with ``almost'' Gaussian transition probabilities\cite{ait2002maximum,ait2008closed} or even arbitrary step size\cite{beskos2006exact}.

For forward power law forces of the form $f(L)=-\gamma L^\alpha$, and close to the target state, the corresponding reverse time forces $f^\mathrm{TSA}(L)=f(L) + f^\mathcal{F}(L)$ can be stated exactly using Eq.~\eqref{sup_tr_lv_ness_alpha_force} for the free energy force $f^\mathcal{F}(L)$. 
Using $f^\mathcal{F}(L)$, in the form as defined in Eq.~\eqref{sup_tr_lv_ness_alpha_force}, the log-likelihood is not easily evaluated numerically for small $L$. As both $\Gamma(\nu)$, $\Gamma(\nu,z)$ and $\exp(z)$ need to be evaluated numerically and separately, slight inconsistencies can lead to numerical instabilities. We therefore transformed $f^\mathcal{F}(L)$ to forms in which only one special function is evaluated numerically. This is feasible when treating $\alpha<-1$ and $\alpha>-1$ separately. The full free energy force then splits into three cases
\begin{align}
\label{eq:numerical_stable_freeEforce}
 f^\mathcal{F}(L)
 =
 \begin{cases}
\frac{D}{-\frac{2 L^{\alpha+1} \gamma }{\alpha D+D}
    \, _1F_1(1; 1+ \frac{1}{1+\alpha} ; -\frac{2 L^{\alpha+1} \gamma }{\alpha D+D} )}
\qquad &\mathrm{for} \ \alpha > -1 + \epsilon
\\
\frac{2 \gamma}{L} - \frac{\alpha D}{L}
\qquad &\mathrm{for} \ -1 + \epsilon < \alpha < -1 - \epsilon
\\
\frac{- D (1+ \alpha)}{-\frac{2 L^{\alpha+1} \gamma }{\alpha D+D}
    \, U(1; 1+ \frac{1}{1+\alpha} ; -\frac{2 L^{\alpha+1} \gamma }{\alpha D+D} )}
    \qquad &\mathrm{for} \ -1 - \epsilon < \alpha < -1 - \epsilon 
    \ .
 \end{cases}
\end{align}
In our experience this representation with $\epsilon=0.05$ worked robustly. Here $\, _1F_1(a;b;z)$ is the Kummer confluent hypergeometric function, $U(a;b;z)$ is Tricomi's confluent hypergeometric function. To obtain this representation we used the substitutions
\begin{align}
\Gamma(a) -\Gamma(a,z)
=
a^{-1} z^a e^{-z} 
_1F_1(1;1+a;z)
\end{align}
which can be derived from
from Abramovitz \& Stegun (AS 6.5.3) and (AS 6.5.12)\cite{abramowitz1964handbook},
and 
\begin{align}
 \Gamma(a,z)
 =
 z^a U(1;1+a;z) e^{-z}
\end{align}
which follows
from Abramovitz \& Stegun (AS 13.1.29) and (AS 13.6.28)\cite{abramowitz1964handbook}.
We point out, that $f^\mathrm{TSA}(L(\tau_0))$ diverges for $L(\tau_0)=L_\mathrm{ts}$. We therefore leave the target state data point out and start the inference with the next or second next state.

The strength of the path ensemble inference formalism compared to optimizing predicted
moments such as mean and variance is, that the optimization is performed with respect to the full likelihood. No considerations about sufficient statistics or how to weigh the different moments relative to each other in a loss function are required. 
By directly inferring the underlying reverse time SDE, mean, variance etc.~become predictions.
The comparison between observed and predicted moments can then be used to test the inference and accept or reject the model.



\bibliography{literature_SI}